\documentclass{aa}  

\usepackage{graphicx}
\usepackage{txfonts}
\usepackage{caption}
\usepackage{natbib}
\bibpunct{(}{)}{;}{a}{}{,} 

\begin{document} 
\title{Tracing the ISM magnetic field morphology: The potential of multi-wavelength polarization measurements}
\author{S. Reissl\inst{\ref{inst1}} \and S. Wolf\inst{\ref{inst1}} \and D. Seifried\inst{\ref{inst2}}}

\institute{\centering Institut für Theoretische Physik und Astrophysik, Christian-Albrechts-Universit\"{a}t zu Kiel, Leibnizstraße 15, 24098 Kiel, Germany \\
								{sreissl@astrophysik.uni-kiel.de} \\
								{wolf@astrophysik.uni-kiel.de}\label{inst1}
\and
					 \centering Hamburger Sternwarte, Universität Hamburg, Gojenbergsweg 112, 21029 Hamburg, Germany \\
								{dseifried@hs.uni-hamburg.de}\label{inst2}
}
						
\abstract
   {}
{We present a case study to demonstrate the potential of multi-wavelength polarization measurements. The aim is to investigate the effects that dichroic polarization and thermal re-emission have on tracing the magnetic field in the interstellar medium (ISM). Furthermore, we analyze the crucial influence of imperfectly aligned compact dust grains on the resulting synthetic continuum polarization maps.}
   {We developed an extended version of the well-known 3D Monte-Carlo radiation transport code MC3D for multi-wavelength polarization simulations running on an adaptive grid. We investigated the interplay between radiation, magnetic fields and dust grains. Our results were produced by post-processing both ideal density distributions and sophisticated  magnetohydrodynamic (MHD) collapse simulations with radiative transfer simulations. We derived spatially resolved maps of intensity, optical depth, and linear and circular polarization at various inclination angles and scales in a wavelength range from $7\ \rm{\mu m}$ to $1\ \rm{mm}$.}
   {We predict unique patterns in linear and circular polarization maps for different types of density distributions and magnetic field morphologies for test setups and sophisticated MHD collapse simulations. We show that alignment processes of interstellar dust grains can significantly influence the resulting synthetic polarization maps. Multi-wavelength polarization measurements allow one to predict the morphology of the magnetic field inside the ISM. The interpretation of polarization measurements of complex structures still remains ambiguous because of the large variety of the predominant parameters in the ISM.}
 {}

   \maketitle
%

\section{Introduction}
Magnetic fields in the interstellar medium (ISM) affect astrophysical processes in various environments and scales in a crucial way. Especially the collapse of molecular clouds and the driving forces behind star formation are topics of ongoing research \citep{1999osps.conf..305M,2000ApJ...530..277E, 2007ARA&A..45..565M}. So far, vital questions about the role of magnetic fields remain unanswered. \\
The current paradigms of what drives star formation (magnetically supported \citep{1999osps.conf..305M} or turbulent models \citep{2004ASPC..322..299K} make different  predictions of the magnetic field geometry. In models preferring magnetic support, gas falls parallel to the field lines. The magnetic field lines are frozen in to the matter, mass-dragged field lines form an hourglass morphology \citep{2005ApJ...631..411N}. In general, the magnetic field lines are expected to be smooth, with a regular structure. \\
The turbulent model, in turn, predicts a magnetic field that is too weak to resist the deformation by the local turbulent ISM environment. In this case, the magnetic field lines have an irregular and chaotic structure. Such irregularities should be easily recognizable in polarization measurements because of a correlation between polarization patterns and the density distribution.\\
To test the role of magnetic fields in the star formation process proper knowledge of the magnetic field strength and field morphology involved are required. Different methods are available for determining the magnetic field strength, such as the Chandrasekhar-Fermi method \citep[][]{1953ApJ...118..113C, 2004ApJ...600..279C}, Zeeman measurements \citep[e.g.][]{2008arXiv0808.1150C} and the field morphology by taking advantage of the polarization effects of non-spherical aligned dust particles inside the ISM \citep[e.g.][]{1999ApJ...525L.109G,2004MNRAS.352.1347L,2010ApJ...717.1262T}. However, from an observational point of view, the difficulty is that the reliability of polarization measurements and their interpretation depends on a wide range of physical parameters that are still discussed.\\
So far, knowledge about magnetic field morphologies in star formation regions has been accumulated by measurements of thermal dust re-emission by aligned dust grains at far-infrared (FIR) to sub-millimeter (sub-mm) wavelength \citep[e.g][]{2011AA...535A..44F}. Although the polarization of light by interstellar dust was discovered in the middle of the last century \citep{1949Natur.163..283H}, its potential as a tool for tracing the magnetic field in the ISM is still high. Especially circular polarization is an often neglected source of additional information. In the mid-/far-infrared wavelength range where extinction is not dominated by scattering one can expect significant insights into the magnetic field structures by polarization measurements in combination with the ongoing research in grain alignment mechanisms. Existing theories of grain alignment \citep[e.g.][]{1951ApJ...114..206D,2007JQSRT.106..225L} agree that rotating non-spherical dust grains align with their shorter axis parallel to the magnetic field direction. As a result, previously unpolarized light will be polarized by thermal dust remission and dichroic extinction perpendicular and parallel, respectively, to the magnetic field lines. This makes multi-wavelength observation on multiple scales a powerful tool completing our understanding of the fundamental physics of star formation processes. \\
The paper is arranged in the following order: In Sect. \ref{sq:pol} in detail we present the theoretical basis for the considered mechanisms of dichroic polarization. In Sect. \ref{sect:dust} the parameters of the applied dust grain model are described. We introduce the considered mechanisms of grain alignment in Sect. \ref{sq:PDG} and \ref{sq:IDG}. To investigate the effects of polarization and alignment we apply these mechanisms in Sect. \ref{setupIDEAL} to simple setups of various density and temperature distributions with analytically modeled field configurations. In Sect. \ref{setupMHD} we extend the investigation to more realistic setups including sophisticated MHD collapse simulations. We discuss and summarize the effects observed in the resulting synthetic polarization maps in Sect. \ref{disc} and demonstrate that the magnetic field morphology can be traced from the complexity of the involved density distribution.
\section{Basics}
\subsection{Dichroic polarization}
\label{sq:pol}
As the source of polarization in the ISM we consider the effects of dichroic extinction and re-emission of light by compact elongated dust grains aligned with the magnetic field lines. Additional contributions to linear and circular polarization from scattering at optical/near-infrared wavelengths are not considered. In the presence of magnetic fields dust particles align with their minor axis parallel to the field direction. Previously unpolarized light is polarized by passing non-spherical particles because of different cross sections that are parallel and perpendicular to the particles alignment axis. This results in linear and circular polarization. Our approach is to analyze the change in the Stokes vector that is caused by aligned grains along the line of sight. The polarization effects of extinction and re-emission are treated separately. By applying the Stokes vector formalism to the radiative transfer equation we obtain a system of equations in matrix form for the extinction:

\begin{equation}
\frac{\rm{d}}{dl}\vec{S}=- n_{\rm{d}}(s)\hat{C}\vec{S}.
\label{eq:extinction}
\end{equation}
The Stokes vector $\vec{S} = \left( I , Q , U , V \right)^T$ changes along a distinct pathway $dl$ that depends on the number density $n_{\rm{d}}(s)$ of the dust grains and their cross-section matrix $\hat{C}$ (see Appendix \ref{apA}). The parameter $I$ corresponds to intensity, $\pm Q$ and $\pm U$ to a linear polarization, and $\pm V$ indicates counter-clockwise and clockwise circular polarization, respectively. Here the $+Q$ - system is defined to be in the $EW$ - direction on the plane sky.\\
The temperature $T_{\rm{d}}$  of the dust grains leads to an additional contribution to intensity and linear polarization as a result of its thermal re-emission. However, the characteristic of the re-emitted light depends on the magnetic field direction, the absorption properties, and the geometry of the dust particle. The angle of highest re-emission is perpendicular to the magnetic field. For the source term of the radiative transfer equation we consider the dust grains to be perfect blackbody radiators:

\begin{equation}
dI = n_{\rm{d}}(s) C_{\rm{abs}}B_{\rm{\lambda}}(T_{\rm{d}})dl,
\label{eq:reemissionI}
\end{equation}

\begin{equation}
dQ = n_{\rm{d}}(s) \Delta C_{\rm{abs}}B_{\rm{\lambda}}(T_{\rm{d}})\cos(2\varphi(s)) dl,
\label{eq:reemissionQ}
\end{equation}

\begin{equation}
dU = -n_{\rm{d}}(s) \Delta C_{\rm{abs}}B_{\rm{\lambda}}(T_{\rm{d}})\sin(2\varphi(s)) dl.
\label{eq:reemissionU}
\end{equation} Here, $B_{\rm{\lambda}}(T_{\rm{d}})$ is the Planck function and $C_{\rm{abs}}$ the cross section for absorption. For the definition of the angle $ \varphi$ see Fig. \ref{Fig:Angle}. The V parameter remains unaffected by thermal re-emission. Here, the Stokes vector and the dust grains are considered to be in the same frame of reference. Below we discuss the problems of separate coordinate systems and their transformation in detail. \\
For non-spherical particles the cross sections for absorption, extinction, and birefringence vary with respect to the incident angle and polarization angle of the passing light rays. Subsequently, dichroic extinction and thermal re-emission mechanisms lead to linear as well as circular polarization, depending on the number density $n_{\rm{d}}$, the topology of the magnetic field $\vec{B}$, and the dust temperature $T_{\rm{d}}$. Therefore, the polarized light carries with it the information of the projected configuration of the magnetic field and the local dust parameter along the line of sight. \\
The degree of linear and circular polarization depends on optical depth and temperature. In low optical depth regions the dominating factor of linear polarization is the dust grain temperature because of thermal re-emission. In the intermediate region where the ISM starts to become optically thick, dichroic extinction can lead to high values of the $Q$, $U$ and $V$ parameter. However, with increasing optical depth, $U$ and $V$ fast approach zero and the light remains polarized in $Q$, while the intensity $I$ decreases. The exact degree of polarization depends on the interplay of all the physical parameters inside the ISM along the line of sight.

\subsection{Dust model}
\label{sect:dust}
Interstellar dust grains are a composite of materials of distinct size distributions and complex shapes. The problem of modeling the cross-section behavior depends on the particle shape, size, material, wavelength, and the angles of incident radiation. The composition and computation of the optical properties of these dust grains are still a matter of debate.\\
We calculated the optical properties of the dust according to the dust model by \cite{1977ApJ...217..425M}. The MRN model, named after its authors, consists of a compositions of silicate and carbon distributed with a power-law distribution of $n(a)\propto a^{-3.5}$ in a range of effective radii $a_i$ with sharp cut-offs at $a_{min}=5nm$ and $a_{max}=250nm$. However, the MRN model is missing the extinction features in the FUV and UV regions because of absent polycyclic aromatic hydrocarbons (PAHs) at the small-size end of the distribution. \\
\cite{1997A&A...323..566L} proposed a dust model with a silicate – PAH mixture that satisfies the interstellar extinction, polarization, and abundance constraints. This model required hypothetical optical constants to fit the extinction in the far-UV and is of no practical use. \cite{2001ApJ...551..807D} and \cite{2001ApJ...554..778L} presented a  model  in agreement with observations by using optical properties of the PAHs based on laboratory data. But this model requires an unrealistic amount of certain elements inside the ISM, which is inconsistent with solar abundances.\\
Currently, there is no dust model available that simultaneously fits all extinction, re-emission and, abundance constraints. However, in the desired range of wavelengths from $7\rm{\mu m}$ to $1\ \rm{mm}$, the MRN dust model still agrees with observations of the ISM extinction curve. Furthermore, because of its power law-size distribution, the MRN model predicts a larger amount of small and partly aligned dust species to dominate the extinction. The degree of polarization is determined by large and rare grains with a higher alignment efficiency. This is consistent with the findings of \cite{1951ApJ...114..206D} and \cite{1980A&A....88..194H} which makes the MRN model an excellent candidate for our investigations. The application of different dust models would primarily only result in a shift in the characteristic features of graphite and silicate depending on composition or in additional absorption and emission lines if we were to choose a dust model with a PAH component.\\
For each wavelength we calculated 150 absorption and extinction efficiencies logarithmically distributed over the range of effective radii. We uses a composition of $62.5 \%$ silicate and $37.5 \%$ graphite. The graphite component is highly anisotropic in nature. Therefore, the refractive index depends on the orientation of light relative to the symmetry axis of the graphite. To take into account the anisotropic structure, we applied the $\frac{1}{3} - \frac{2}{3}$ approximation as presented in \cite{1993ApJ...414..632D}. We assumed both materials to have similar size distributions and alignment behaviors.\\
The corresponding cross sections are defined as the product of an efficiency factor and the geometric cross section $C_{\lambda} = \pi a^2 Q_{\lambda}$, where $a$ is the radius of a spherical grain of equivalent volume.  The mass of the particles does therefore not depend on the particle shape.\\
We calculated the efficiencies for extinction of the grains using the discrete dipole approximation (DDA). The idea behind this approximation is to model the shape of the dust grains with a distinct number of dipoles and material parameter as it is implemented in the well-tested program DDSCAT 7.2 \citep{2012arXiv1202.3424D}. We used the refraction indices of silicate and graphite as published in \citep{2000AAS...197.4207W}.\\
We weighted the efficiencies for each wavelength according to the power-law distribution to derive the average cross sections over all effective radii:

\begin{equation}
\overline{C_{\lambda}}= \frac{\sum_i{\pi a_i^2 n(a_i) Q_{\lambda}}}{\sum_i{n(a_i)}}.
\label{eq:weight}
\end{equation} We assumed a gas-to-dust ratio of $100:1$.

\subsection{Perfect grain alignment}
\label{sq:PDG}
\begin{figure}[]   
 \centering
	\includegraphics[width=0.4 \textwidth]{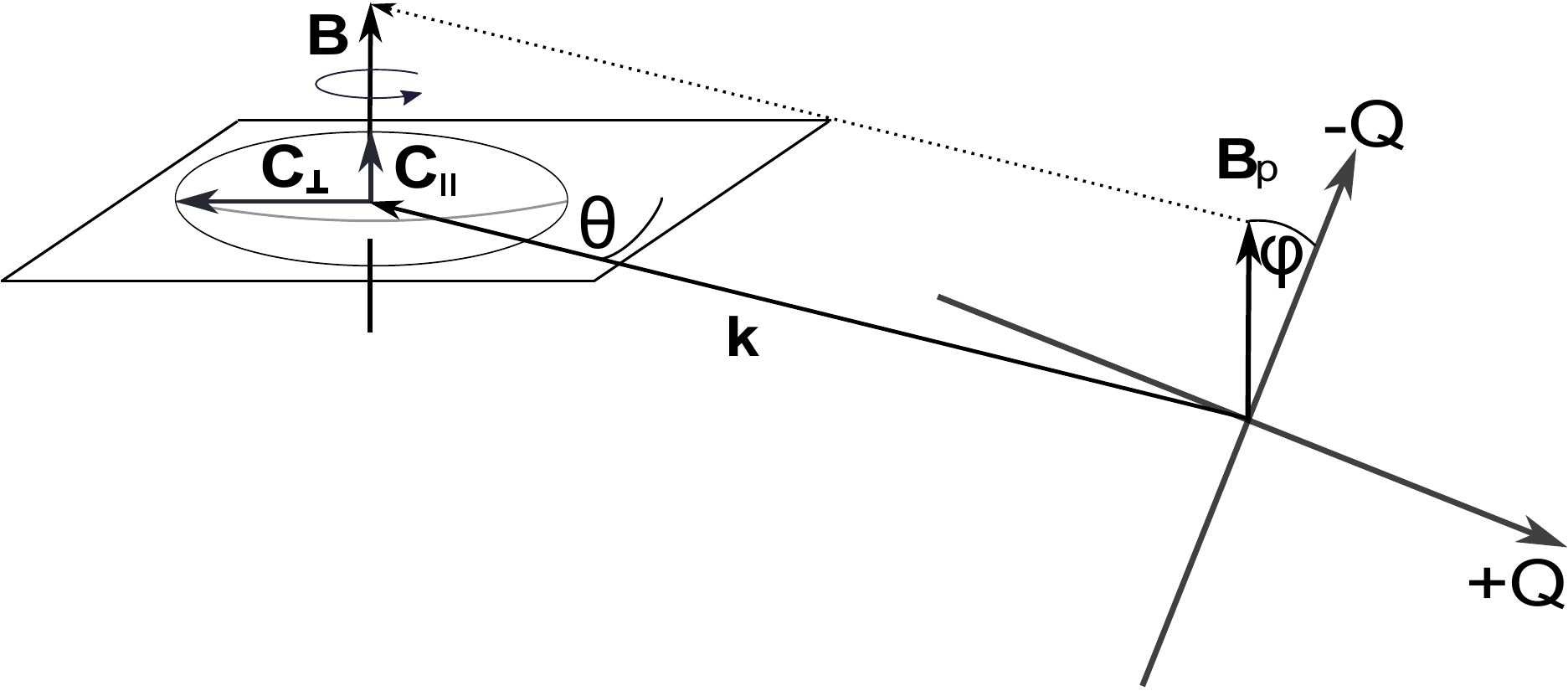}
	\caption{\small Geometrical configuration for oblate dust grains with their shorter axis perfectly aligned (PA) with the direction of the magnetic field \textbf{B}. The angle $\varphi$ is defined by the projection of \textbf{B} into the Q – coordinate system of the Stokes vector and the inclination angle $\vartheta$ is made by the direction of propagation \textbf{k} and the plane of rotation perpendicular to \textbf{B}.}
	\label{Fig:Angle}
\end{figure}

In the case of perfect grain alignment (PA) the cross sections are only dependent on the two angles defined by the direction of the incident light and the orientation of the coordinate system of the Stokes vector with respect to the direction of the  magnetic field $\vec{B}$. \\
In the general case, the aligned grain and the incident light are not in the same frame of reference. To rotate the Stokes parameter relative to the coordinate system of the dust grain, $Q$ and $U$ of the Stokes vector have to be adjusted. The $I$ and $V$ components remain invariant under rotation. We defined the angle $\varphi$ to be the one between the magnetic field vector projected into the coordinate system of the Stokes vector and the $-Q$ direction (see Fig. \ref{Fig:Angle}). To rotate the Stokes vector by an arbitrary angle we applied the rotation matrix $R$ and $R^{-1}$:

\begin{equation}
R = \begin{pmatrix} 0 & 0 & 0 & 0 \\ 0 & \cos(2\varphi) & -\sin(2\varphi) & 0\\ 0 & \sin(2\varphi) & \cos(2\varphi) & 0 \\ 0 & 0 & 0 & 0 \end{pmatrix}.
\label{eq:transformation}
\end{equation}
The second angle, $\Theta$, is defined by the direction of wave propagation $\vec{k}$  of the incident light and the plane perpendicular to the magnetic field vector. To correct the cross-sections that depend on the tilt one finds \citep{1968nim..book..221G}

\begin{equation}
\overline{C_{\lambda}}(\Theta)= \frac{1}{2} \left[ \overline{C}_{\lambda,\bot}+\overline{C}_{\lambda,||}+ \left(\overline{C}_{\lambda,\bot}-\overline{C}_{\lambda,||}\right) \cos^2(\Theta) \right],
\label{eq:Qsum}
\end{equation}

\begin{equation}
\Delta \overline{C_{\lambda}}(\Theta)=  \frac{\overline{C}_{\lambda,\bot}-\overline{C}_{\lambda,||}}{2}  \sin^2(\Theta).
\label{eq:Qdiff}
\end{equation}
However, perfect alignment of the dust grains does not take into consideration any deviation due to the local physical conditions of the ISM and remains an ideal approximation.
	
\subsection{Imperfect grain alignment}
\label{sq:IDG}

\begin{figure}[]  
\centering 
	\includegraphics[width=0.4 \textwidth]{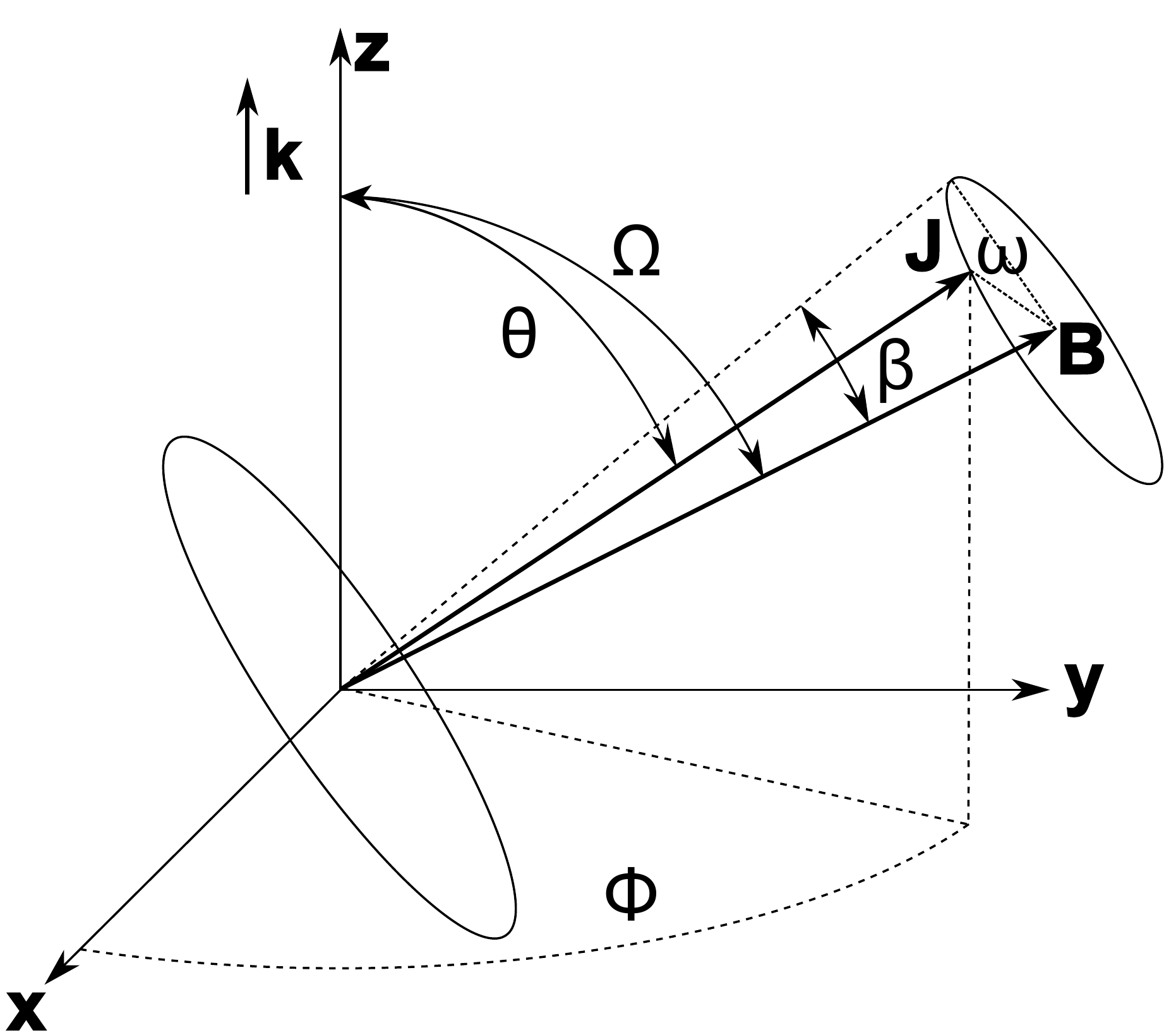}
	\caption{\small Geometrical configuration of a rotating oblate dust grain with angular momentum \textbf{J} no longer in alignment with the magnetic field \textbf{B}. The precession of \textbf{J} around \textbf{B} is defined by the precession angle $\omega$ and the cone angle $\beta$.
The direction of propagation \textbf{k} is parallel to the z-axis. The inclination angle $\theta$ is not constant but varies with $\omega$, the angle $\Omega$ between \textbf{k}, and magnetic field, and $\beta$.}
	\label{Fig:IDG}
\end{figure}

The perfect alignment of dust grains is hindered by gas-grain collisions. Depending on the local conditions in the ISM, dust grains are partially misaligned. However, a consistent theory of grain alignment is still lacking. The proposed mechanism is based on alignment e.g. due to paramagnetic relaxation \citep{1951ApJ...114..206D} or radiative torques (RAT), the grain interaction with the radiation field \cite{2007AAS...21113807H}. However, an analytical alignment distribution function for the RAT alignment of dust grains is still missing. The required magnetic field strength for the RAT alignment to be efficient is still debated as well \citep[see ][]{2009ApJ...704.1204H}.\\
Therefore, we implemented the classical imperfect Davis-Greenstein (IDG) mechanism to model the impact of imperfectly aligned dust grains on our radiative transfer and polarization simulations. We assumed silicate and graphite grains to be equally affected by IDG. For imperfectly aligned dust grains the cross sections depend on the angular distribution \citep{1967ApJ...147..943J}
\begin{equation}
f(\xi(a),\beta)=\frac{\xi(a)\times sin(\beta)}{\left( \xi^2(a)cos^2(\beta)+sin^2(\beta)\right)^{1.5}}
\label{eq:coneangledist}
\end{equation}
of the cone angle $\beta$ between the angular momentum vector of the dust grain and the magnetic field vector (see Fig. \ref{Fig:IDG}). The angle between the direction of propagation and the magnetic field is $\Omega$. The quantity $\Theta$ is no longer constant but depends on $\beta$, $\Omega$ and the precession angle $\omega$. The angular distribution of $\beta$ is determined by the grain alignment parameter:
\begin{equation}
\xi^2(a) = \frac{a+\delta_{\rm{0}}\times \frac{T_{\rm{d}}}{T_{\rm{g}}}}{a+\delta_{\rm{0}}}.
\label{eq:xi}
\end{equation}
Here, $a$ is the effective radius, $T_{\rm{d}}$ the dust temperature, and $T_{\rm{g}}$ the kinetic gas temperature. The grain alignment parameter depends on the ratio $T_{\rm{d}}/T_{\rm{g}}$ and $\delta_{\rm{0}}$ and acts as a threshold between perfectly aligned and imperfectly aligned dust grains. For grains with an effective radius below this threshold $\delta_{\rm{0}}$, the cone angle distribution $f(\xi(a),\beta)$ behaves like a delta function and the shorter axis parallel to the particle is aligned to the magnetic field direction. With decreasing $\delta_{\rm{0}}$  more dust grain species become disaligned and behave like spherical particles. Subsequently, if kinetic gas temperature $T_{\rm{g}}$ and dust temperature $T_{\rm{d}}$ have the same value, no alignment can take place. With the dimensionless material constant $\kappa = 2.5 \times 10^{-12}$ \citep{1951ApJ...114..206D}, the magnetic field strength $B$, and the number density of the gas $n_H$, one can derive the parameter

\begin{equation}
\delta_{\rm{0}} = 8.23\times 10^{31} \frac{\kappa B^2}{n_{\rm{g}} T_{\rm{d}} \sqrt{T_{\rm{g}}}} [\rm{m}].
\label{eq:delta}
\end{equation}
Values of $\delta_{\rm{0}}$ up to a factor of 100 higher can be obtained by including superparamagnetic relaxation due to ferromagnetic inclusions or dust grains surrounded by an ice mantle \citep{1967ApJ...147..943J,1978ApJ...219L.129D,1994MNRAS.268...40S}. With an increasing effective radius, $\xi^2(a)$ approaches unity for lower values of $a$. Thus, particles with an effective radius of $a>\delta_{\rm{0}}$ hardly contribute to the net polarization while their extinction remains unaffected.\\
The cross sections can be calculated by averaging over size distribution $n(a)$, precession angle $\omega$,  and cone angle distribution $f\left( \xi(a) ,\beta\right)$. For oblate dust particles one obtains

\begin{equation}
\left\langle C\right\rangle_{\lambda} = \left(\frac{2}{\pi}\right)^2 \frac{ \sum_i  \int_0^{\frac{\pi}{2}} \int_0^{\frac{\pi}{2}} \pi a_i^2 n(a_i) Q_{\lambda}(\Theta) f(\xi(a_i),\beta) d\omega d\beta }{\sum_i n(a_i)},
\label{eq:IDGsum}
\end{equation}

\begin{equation}
\left\langle \Delta C\right\rangle_{\lambda} = \frac{2}{\pi^2}\frac{ \sum_i  \int_0^{\pi} \int_0^{\frac{\pi}{2}} \pi a_i^2 n(a_i) \Delta Q_{\lambda}(\Theta) f(\xi(a_i),\beta) \cos(2\Phi) d\omega d\beta }{\sum_i n(a_i)}.
\label{eq:IDGdiff}
\end{equation}
Since polarization depends on the difference of cross sections for wobbling dust grains, certain orientations cancel each other out. Therefore, $\left\langle \Delta C\right\rangle_{\lambda}$ must be weighted with $\cos(2\Phi)$ where the factor can be expressed as a function of $\beta$, $\Theta$ and $\omega$ \citep[for details see][]{1980A&A....88..194H,2010MNRAS.404..265D}.

		\begin{figure*}
   \centering
	 \includegraphics[width=1.0\textwidth]{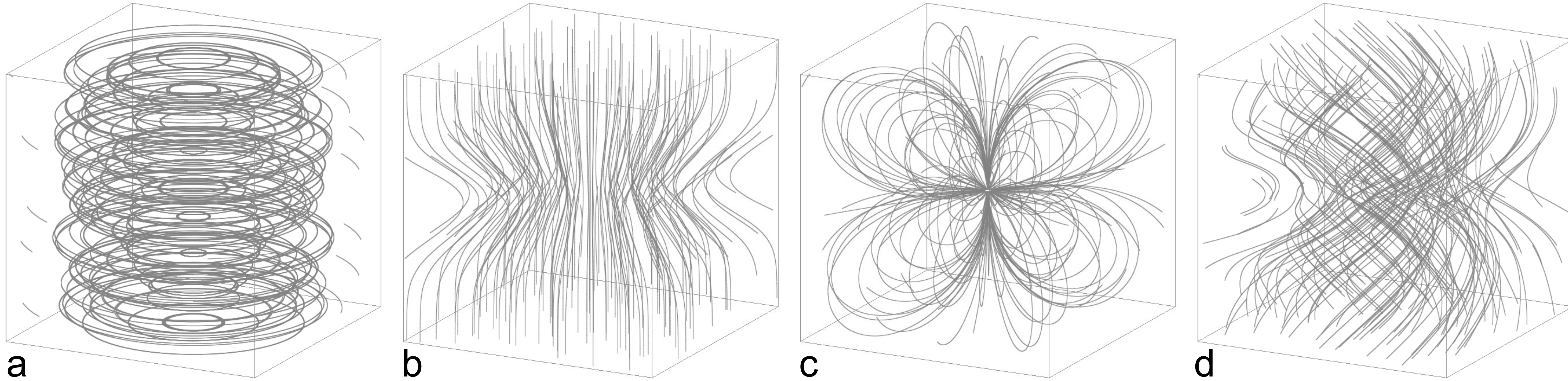}
	
   \caption{\small 3D plots for the toroidal (panel \textbf{a}, Eq. \ref{eq:Btor}), hourglass (panel \textbf{b}, Eq. \ref{eq:Bhour}), quadrupole (panel \textbf{c}, Eq. \ref{eq:Bquad}), and helical (panel \textbf{d}, superposition of Eq. \ref{eq:Btor} and Eq. \ref{eq:Bhour}) magnetic field morphologies.}

\label{fig:morph}
\end{figure*}
\begin{table*}[]
\centering
    \begin{tabular}{|l|c|c|c|c|c|c|}
    \hline
		Setup & Align. & $\lambda$ $[\rm{\mu m}]$ & Temp. $T_{\rm{g}}$[\rm{K}]& Temp. $T_{\rm{d}}$[\rm{K}] & Morph. & Mag. field [\rm{T}]\\ \hline		    \hline	
$\rm{BE_{toro}}$ & PA &	811 & X & 15 &	torodial	&X \\ \hline	

$\rm{BE_{quad1}}$ & PA & 811 & X& 5 - 15	& quadrupole	&  X \\ \hline	
$\rm{BE_{quad2}}$ & IDG & 811 & 10 - 25 & 5 - 15	& quadrupole.	&  $2.0\times 10^{-8}$ \\ \hline	

$\rm{BE_{hour1}}$ & PA & 811 & X& 5 - 15	& hourglass	&  X \\ \hline	
$\rm{BE_{hour2}}$ & IDG & 811 & 10 - 25 & 5 - 15	& hourglass	&  $2.0\times 10^{-8}$ \\ \hline	

$\rm{BE_{helic1}}$ & PA & 811 & X& 5 - 15	& hour. \& torod.	&  X \\ \hline	
$\rm{BE_{helic2}}$ & IDG & 811 & 10 - 25 & 5 - 15	& hour. \& torod.	&  $2.0\times 10^{-8}$ \\ \hline	

$\rm{MHD_{sim1}}$ &	PA & $1 - 10^3$& X & $0 - 280$	& irregular & X \\ \hline	
$\rm{MHD_{sim2}}$ &	IDG & $1 - 10^3$& $0 - 2000$ & $0 - 280$	& irregular & $8.8\times 10^{-10} - 9.4\times 10^{-5}$ \\ \hline	
\end{tabular}
		
\caption{\small Physical parameters for the presented ideal Bonnor – Ebert sphere setups and MHD setups. The characteristic radius for the Bonnor – Ebert sphere is $R_c= 1100\ \rm{AU}$ with an central number density of $n_0=10^{13}\ \rm{m^{-3}}$. For the PA alignment not all parameters are required (see Sect. \ref{sq:PDG}). Parameters irrelevant for the radiative transport calculations are marked with an $X$.}
\label{tab:1}
 \end{table*}

\section{Applications}
We aim to model the underlying processes that lead to linear and circular polarization of previously unpolarized light to reveal the morphology of the magnetic field in the ISM. We developed an extended version of the well-established 3D Monte Carlo radiative transfer code MC3D \citep{ 2003CoPhC.150...99W} and implement the equations of Sect. \ref{sq:pol} for dichroic absorption and re-emission as well as the grain alignment mechanisms of Sects. \ref{sq:PDG} and \ref{sq:IDG} as an optional feature. To increase performance and reduce the amount of data, we applied an adaptive Cartesian grid with an octree-structure. The grid automatically refines the cell structure in regions with higher density distribution and optical depth, respectively, inside the model space. The dust temperature is predetermined by the model setup.\\
The probability for a scattering event is completely determined by the ratio of $ C_{\rm{sca}} / C_{\rm{ext}}$. Since we operate in a regime of wavelengths of $7\ \rm{\mu m} - 1\ \rm{mm}$ where this ratio strongly decreases from $1.3\times 10^{-3} $ to $7.9\times10^{-9} $, the contribution of scattering is marginal even in optically thick regions.
Hence, we applied an algorithm purely based on ray-tracing in contrast to a more time consuming Monte-Carlo - scattering algorithm to calculate the polarization maps. Each grid cell contains a unique set of parameters. As a first step the cross sections for all cells in the model space were re-calculated for all required angles and wavelength. Then the positive and negative intensity contributions were added up along given paths. The reference frame of the Stokes parameter is determined by the first point of interaction with the ISM inside the model space. Finally, all Stokes vectors were rotated into the coordinate system of the model space.\\
According to the Stokes formalism, linear polarization is determined by the degree of linear polarization $P_{\rm{l}}$ and its position angle in the plane of the sky:

\begin{equation}
P_{\rm{l}} = \frac{\sqrt{Q^2+U^2}}{I},
\label{eq:Pl}
\end{equation}

\begin{equation}
\tan(2\chi)=\frac{Q}{U}.
\label{eq:X}
\end{equation} The degree of circular polarization $P_{\rm{c}}$ is defined to be
\begin{equation}
P_{\rm{c}} = \frac{V}{I}.
\label{eq:Pc}
\end{equation} The grain temperature only weakly depends on the effective radius \citep{ 1980A&A....88..194H} and on elongation for an aspect ratio of $s \leq 2$ \citep{ 1999A&A...349L..25V}. Hence, we applied a constant temperature to the entire ensemble of dust grains weighted over the effective radii. Since our main focus is not on the grain parameter itself, we applied a fixed ratio of $s = 2$ for the oblate dust grains to all setups. Higher aspect ratios would result in higher peak values and more contrast in the maps of linear and circular polarization.

\begin{figure}[]
	\begin{minipage}[c]{0.49\linewidth}
			\begin{center}
				\includegraphics[width=1.0\textwidth]{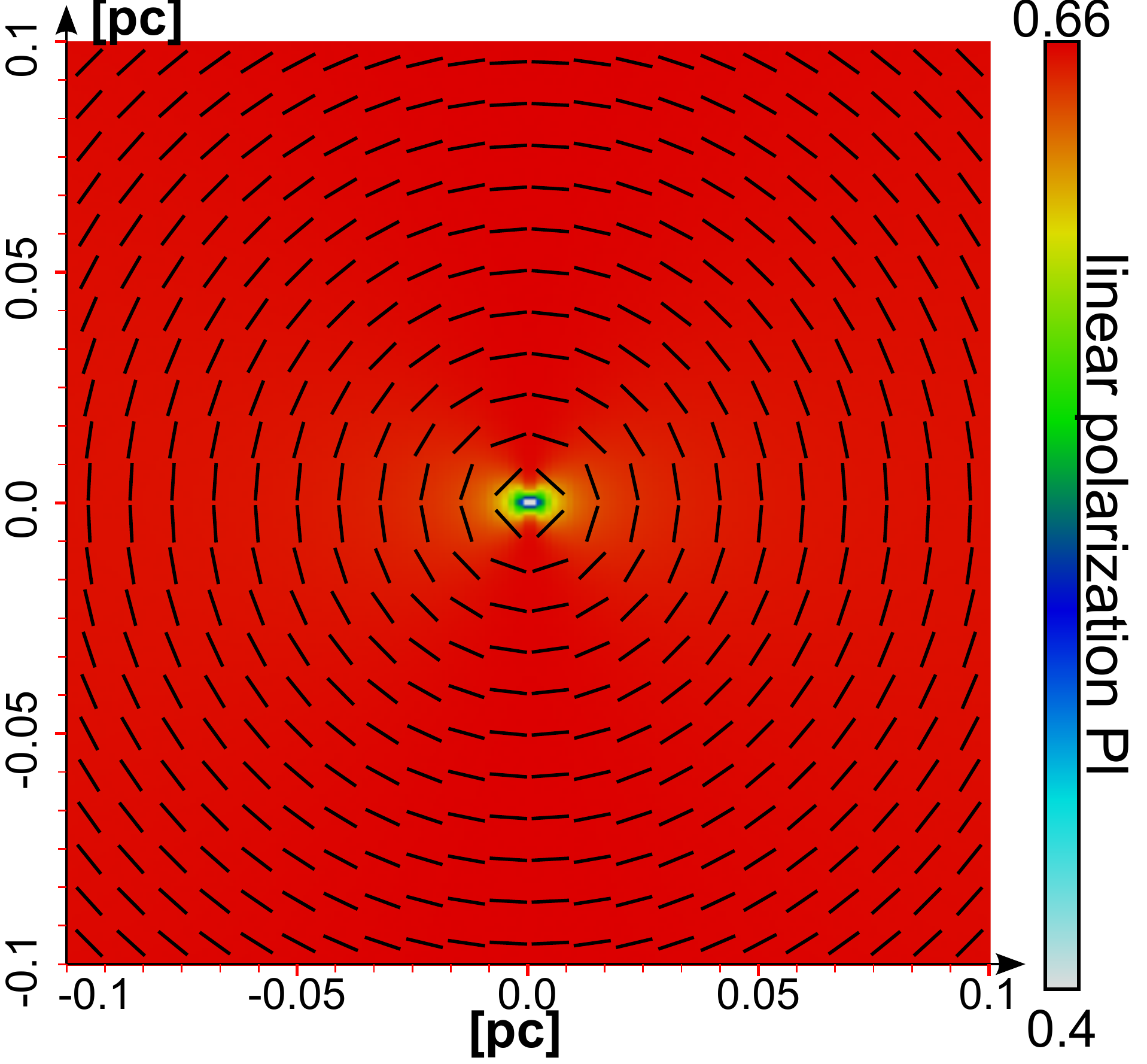}\\
				\includegraphics[width=1.0\textwidth]{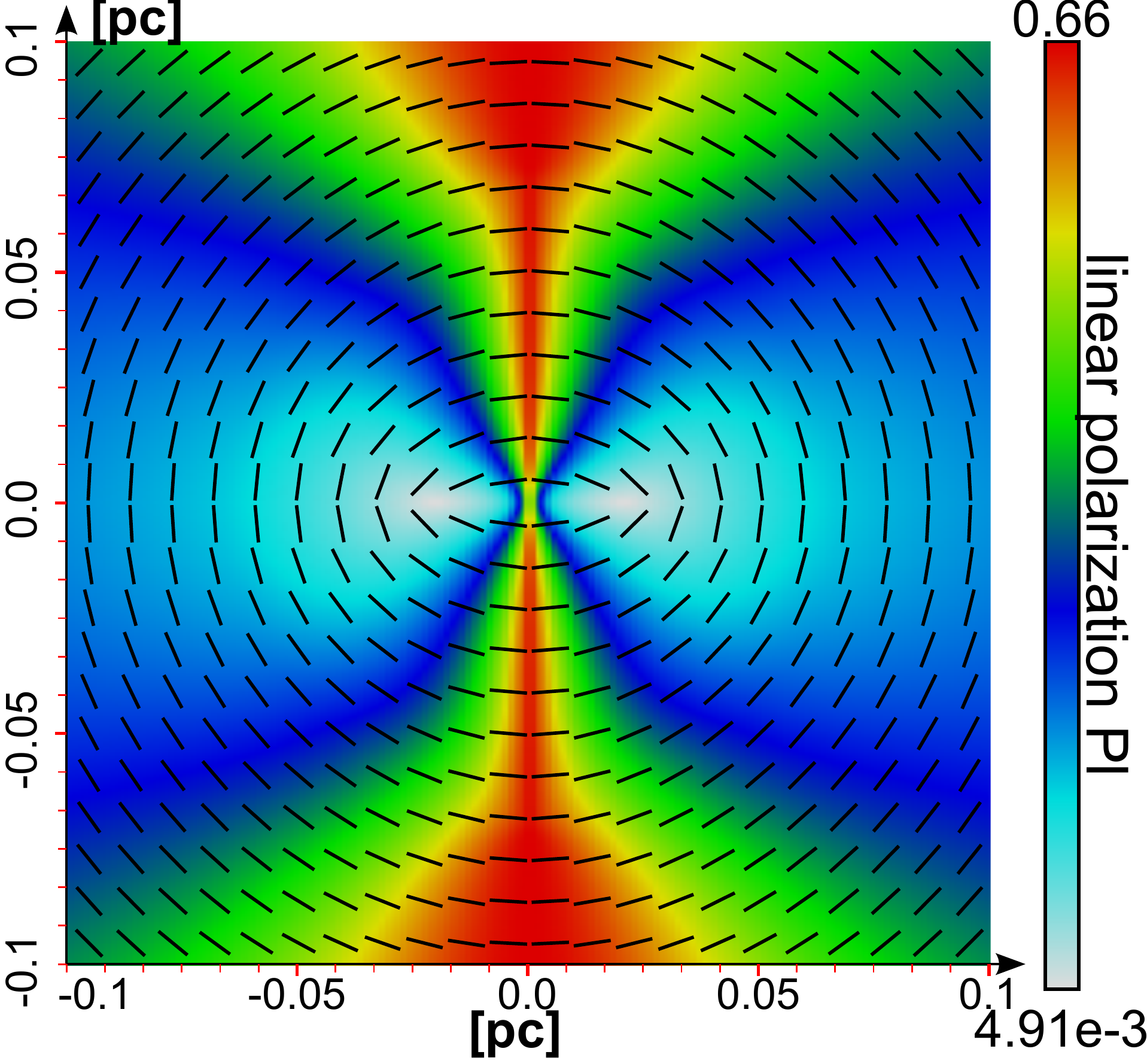}\\
				\includegraphics[width=1.0\textwidth]{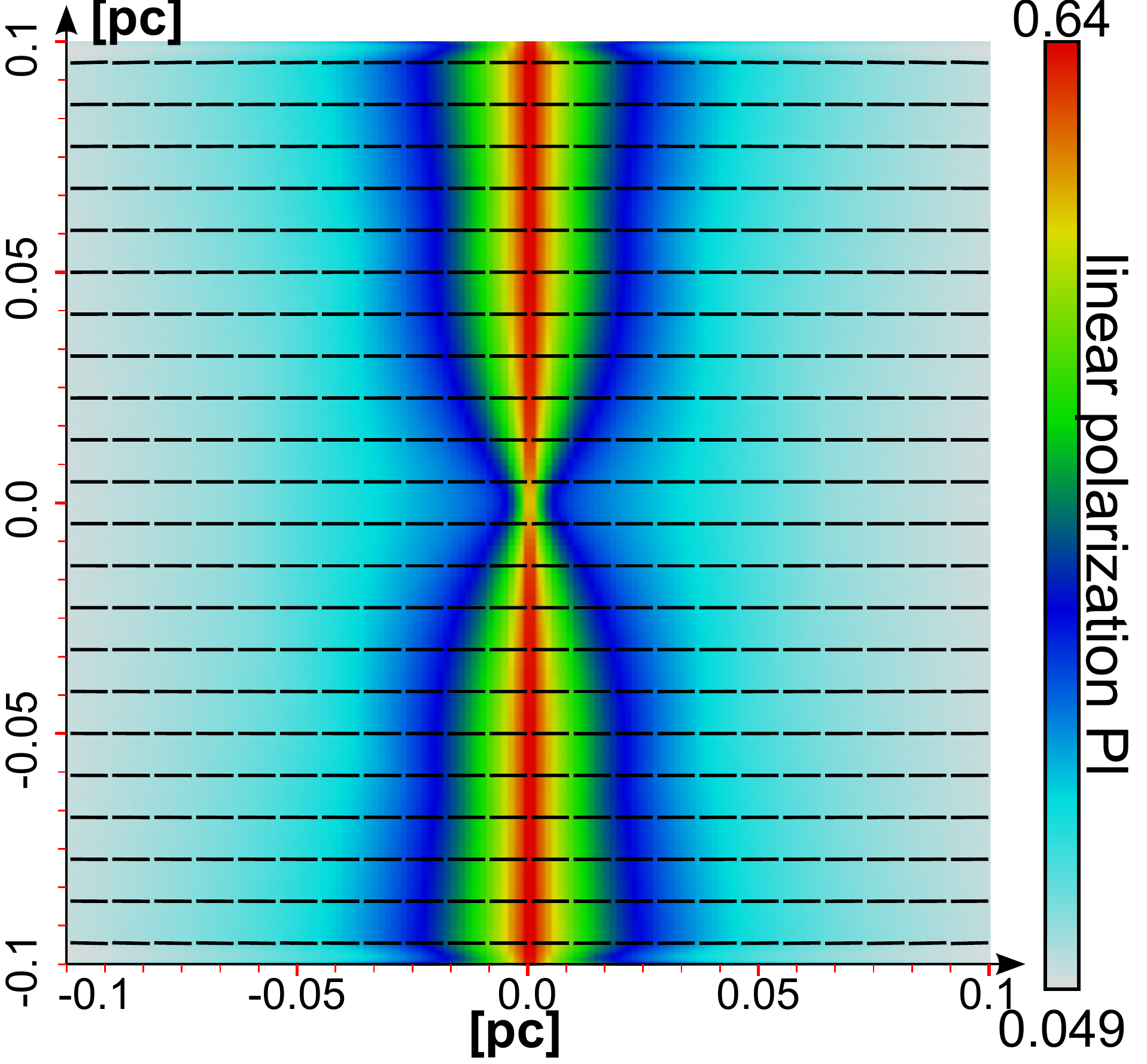}
			\end{center}
		\end{minipage}
		\begin{minipage}[c]{0.49\linewidth}
			\begin{center}
				\includegraphics[width=1.0\textwidth]{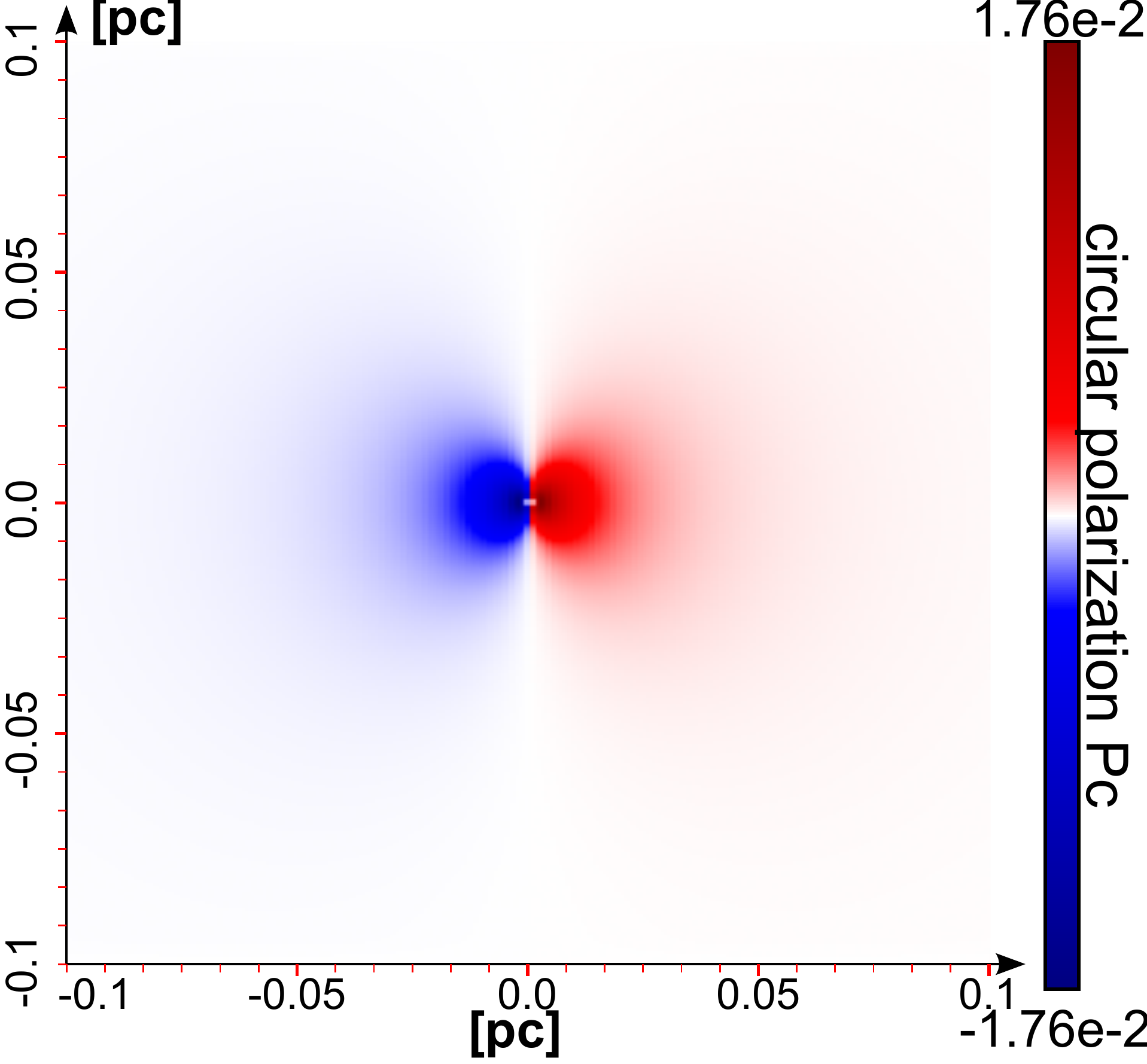}\\
				\includegraphics[width=1.0\textwidth]{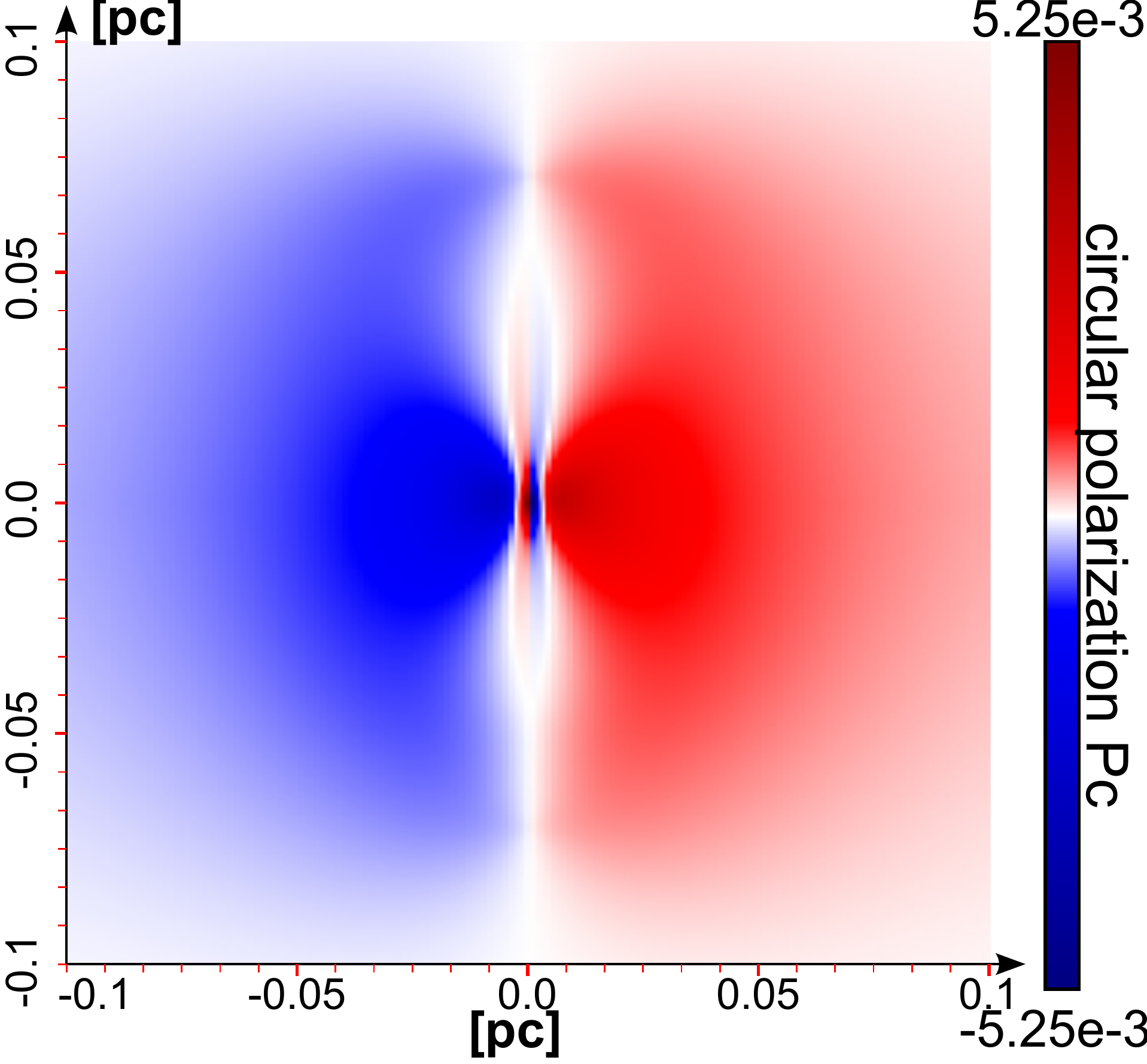}\\
				\includegraphics[width=1.0\textwidth]{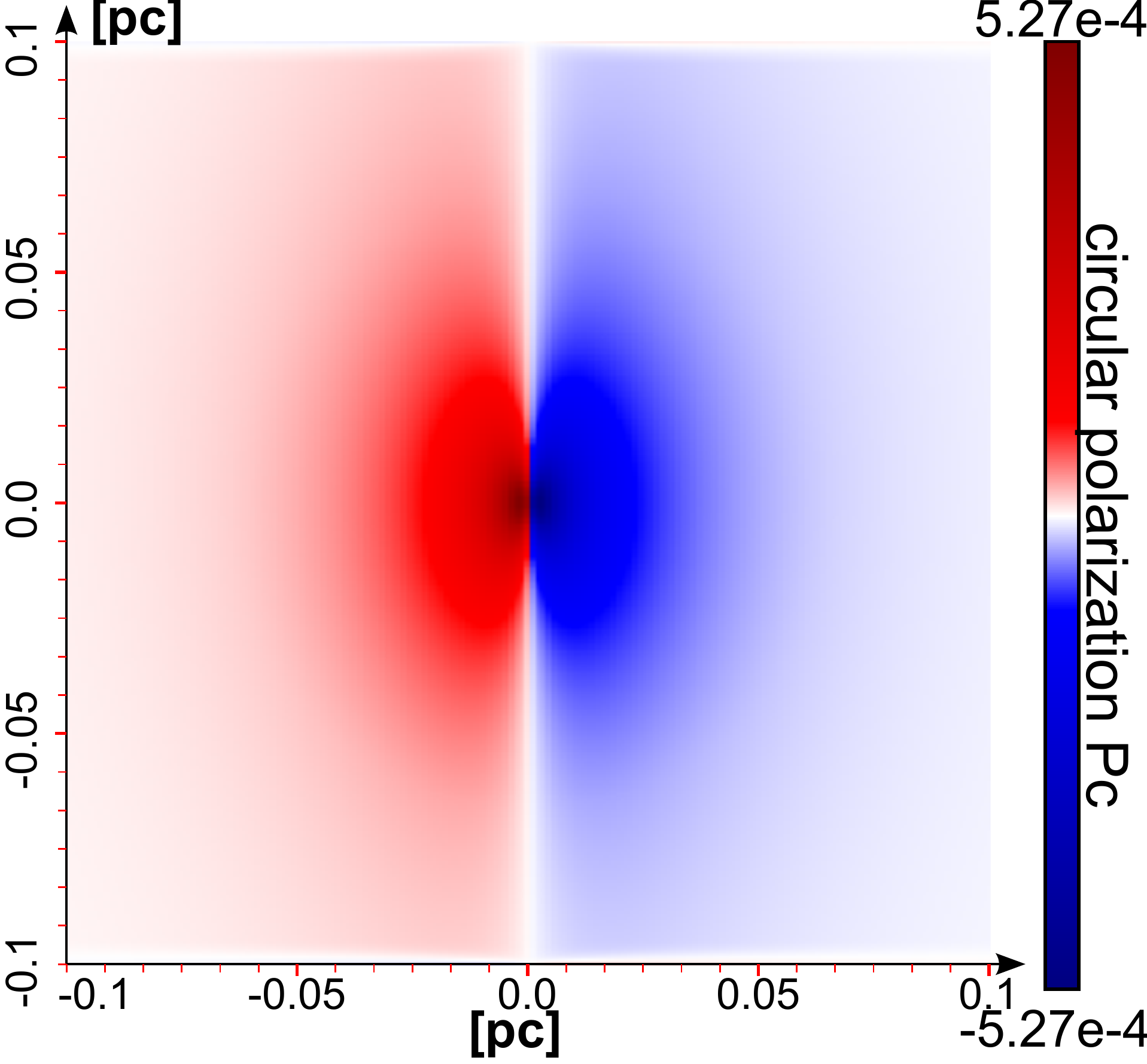}
			\end{center}
		\end{minipage}			
	
	\caption{\small Linear polarization (left) and circular polarization (right) of ideal setup $\rm{BE_{toro}}$ with inclination angles of $3^{\circ}$ (top), $45^{\circ}$ (middle) and $87^{\circ}$ (bottom), a constant dust temperature $T_{\rm{d}} = 15\ \rm{K}$, a characteristic radius $R_c= 1100\ \rm{AU}$ and a density $n_0=10^{13}\ \rm{m^{-3}}$ in the center. We added an offset angle of $90^{\circ}$ to the vectors of linear polarization to match the projected magnetic toroidal field. The dust particles are perfectly aligned.}
\label{IDToro}
\end{figure}

\begin{figure}[]
	\begin{minipage}[c]{0.49\linewidth}
			\begin{center}
				\includegraphics[width=1.0\textwidth]{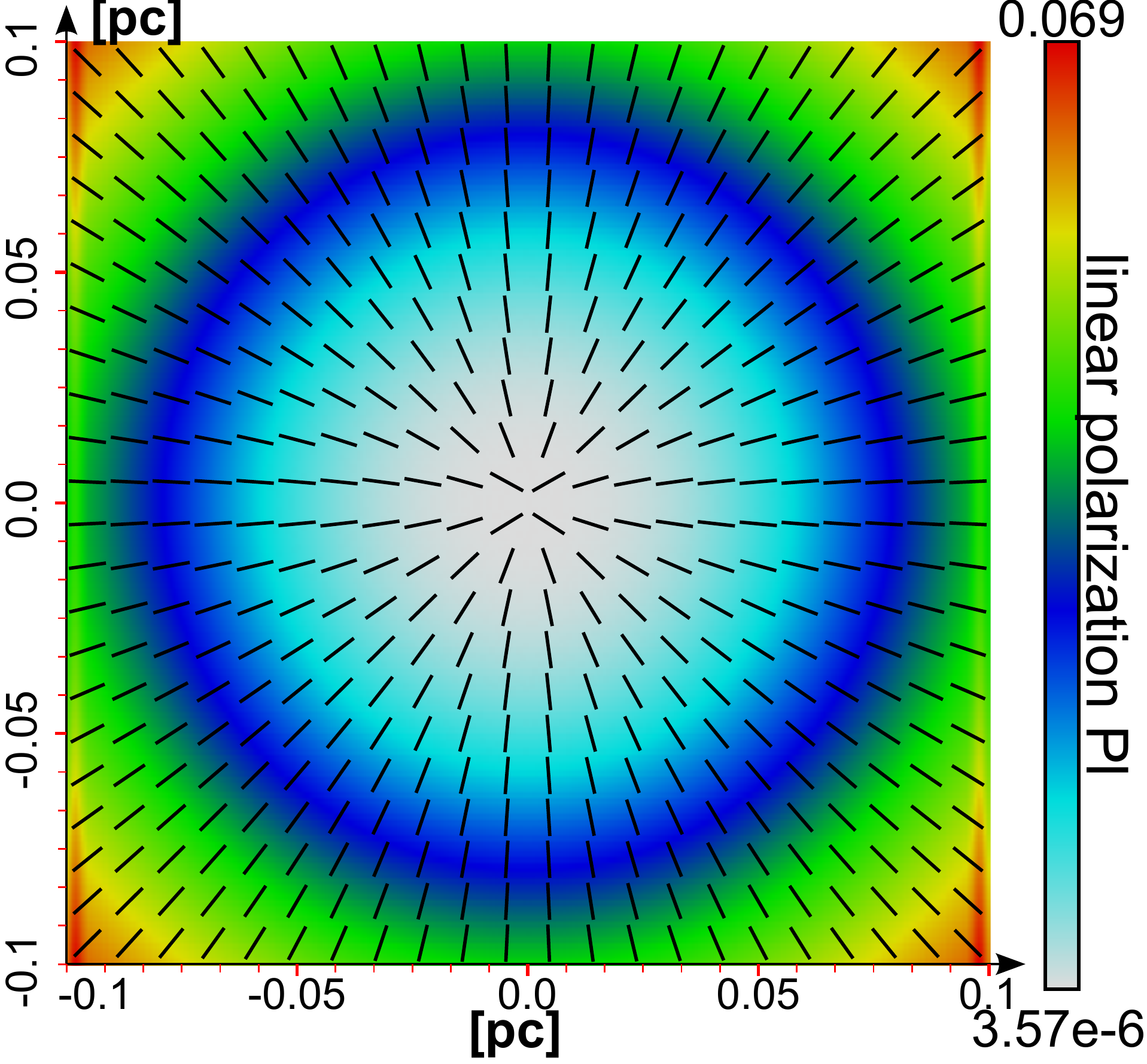}\\
				\includegraphics[width=1.0\textwidth]{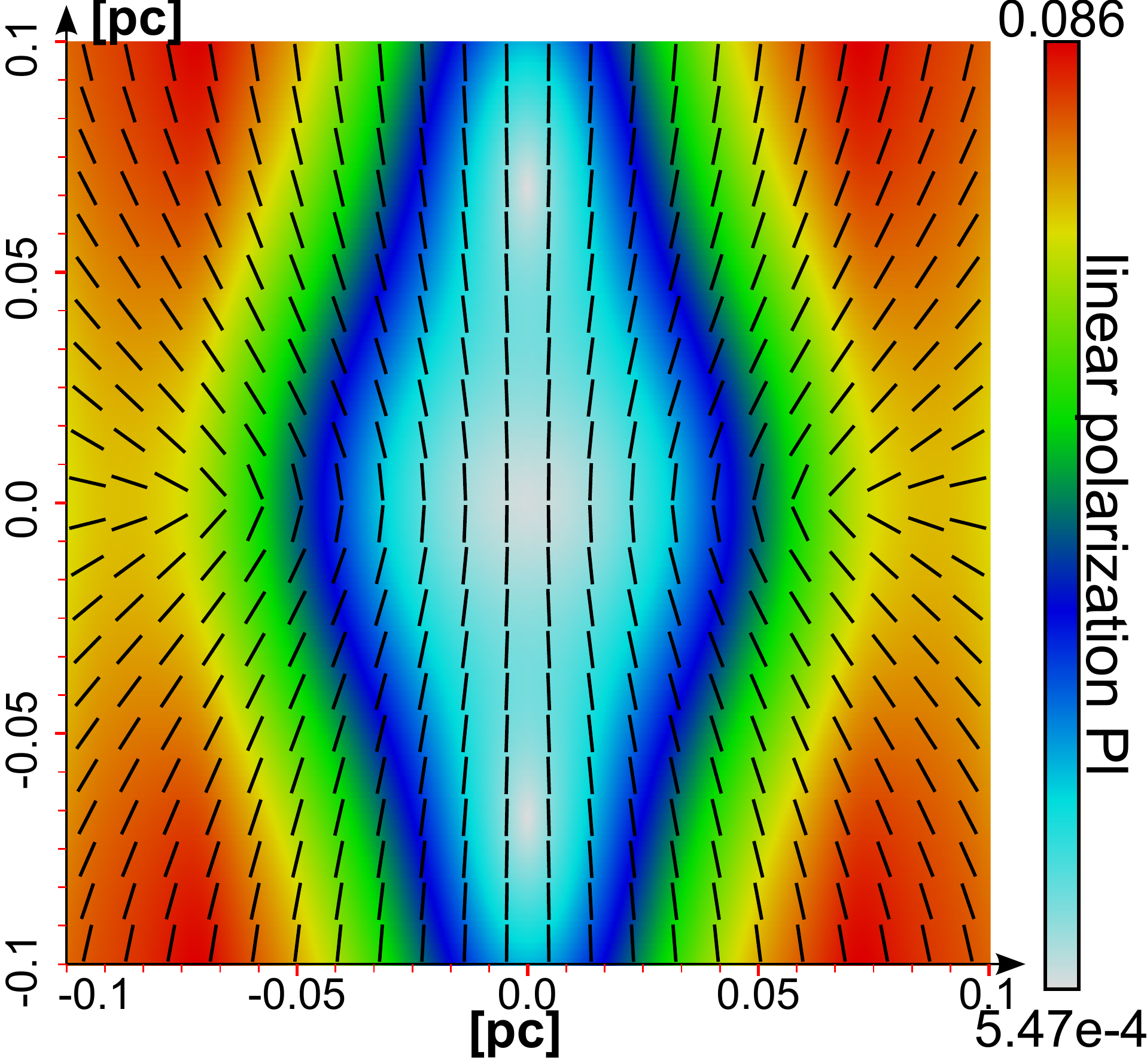}\\
				\includegraphics[width=1.0\textwidth]{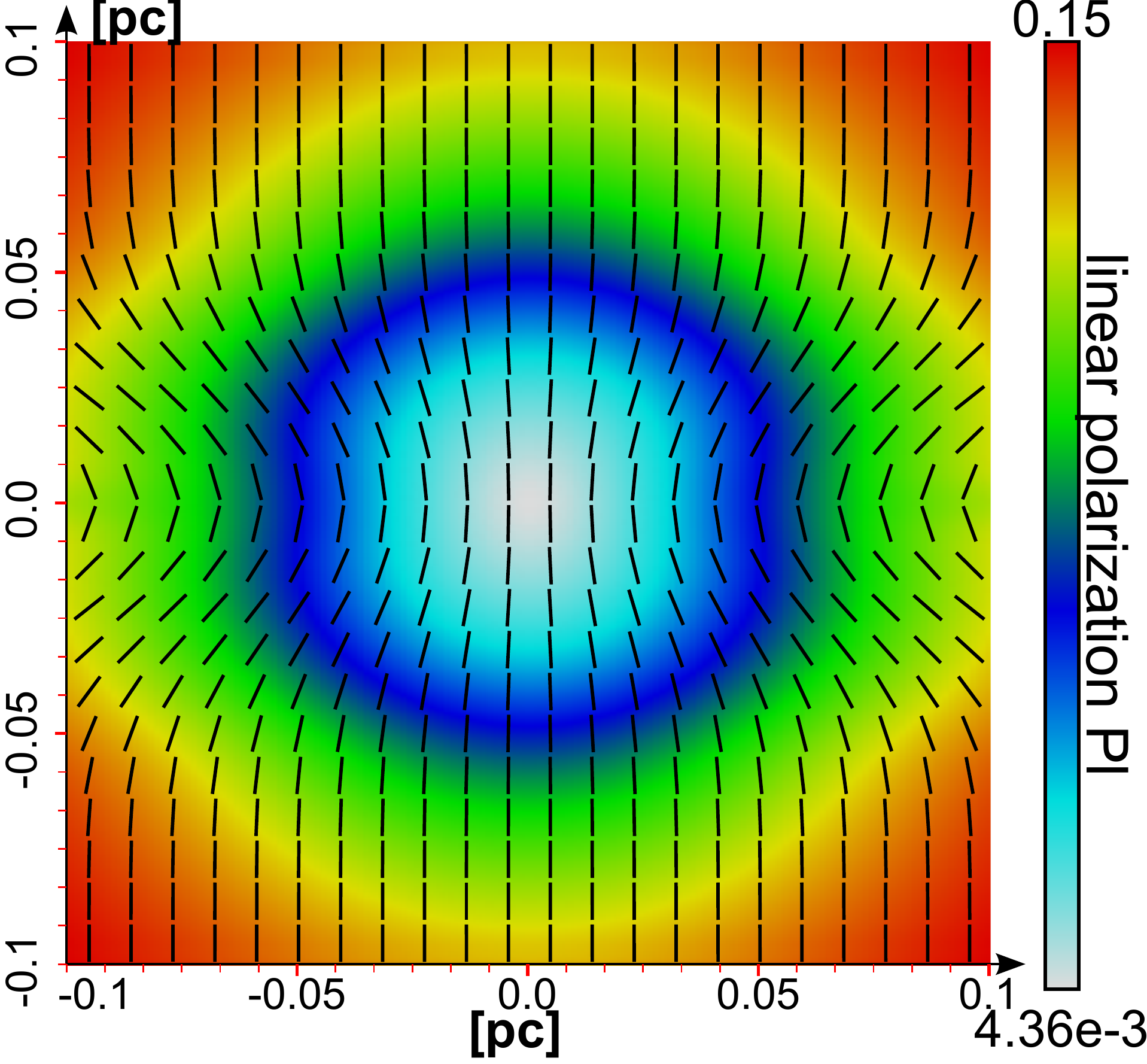}
			\end{center}
		\end{minipage}
		\begin{minipage}[c]{0.49\linewidth}
			\begin{center}
				\includegraphics[width=1.0\textwidth]{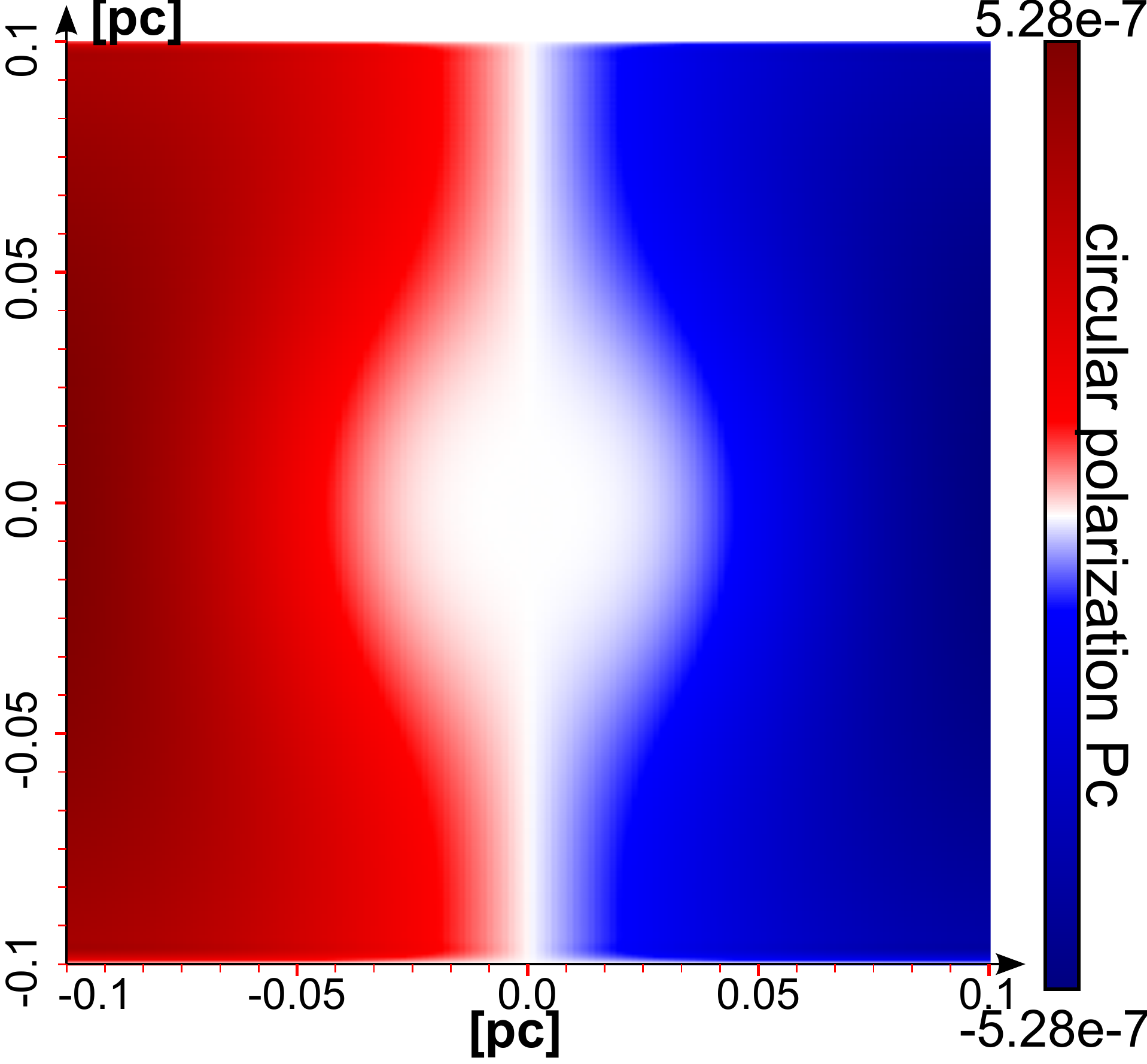}\\
				\includegraphics[width=1.0\textwidth]{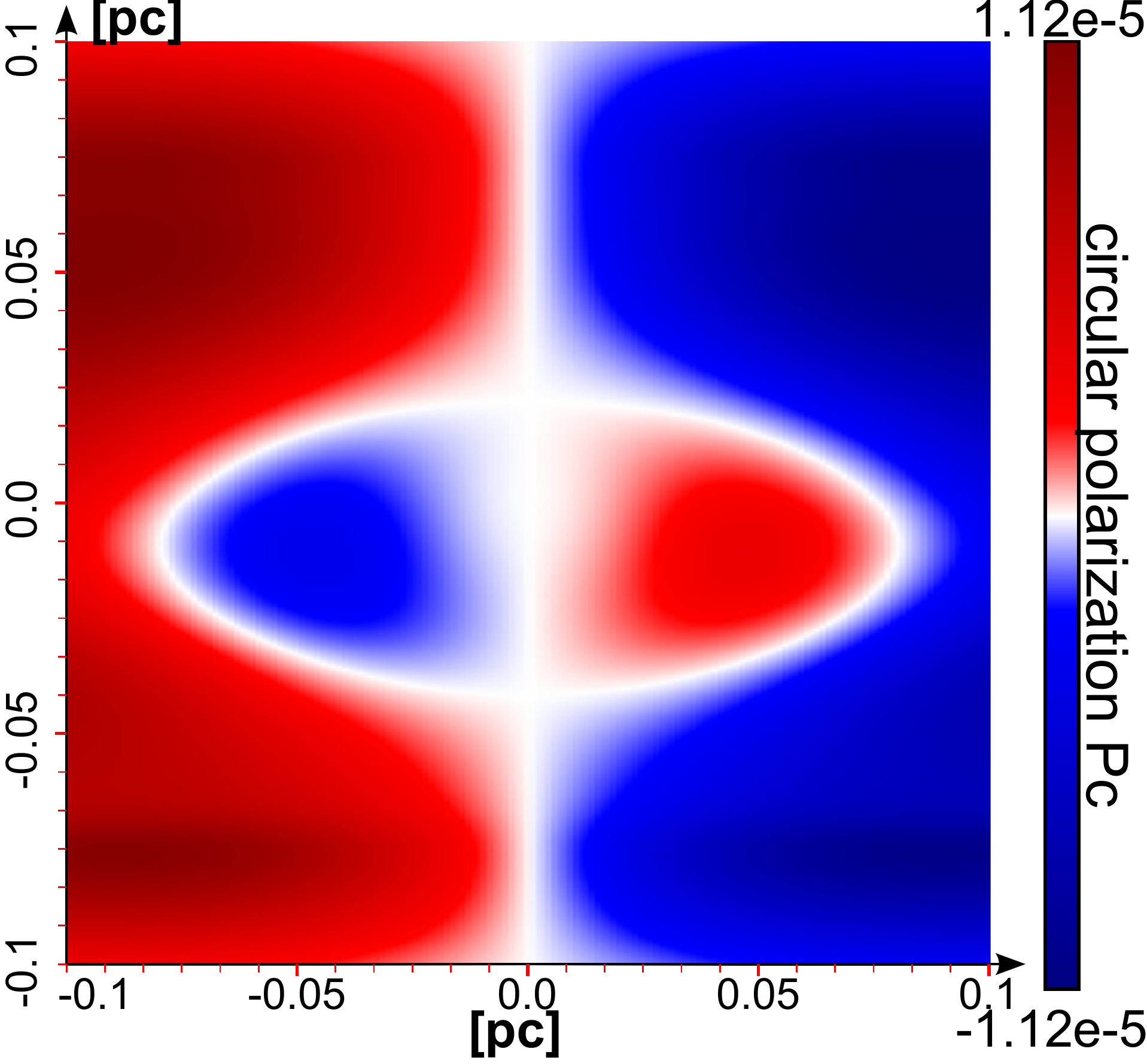}\\
				\includegraphics[width=1.0\textwidth]{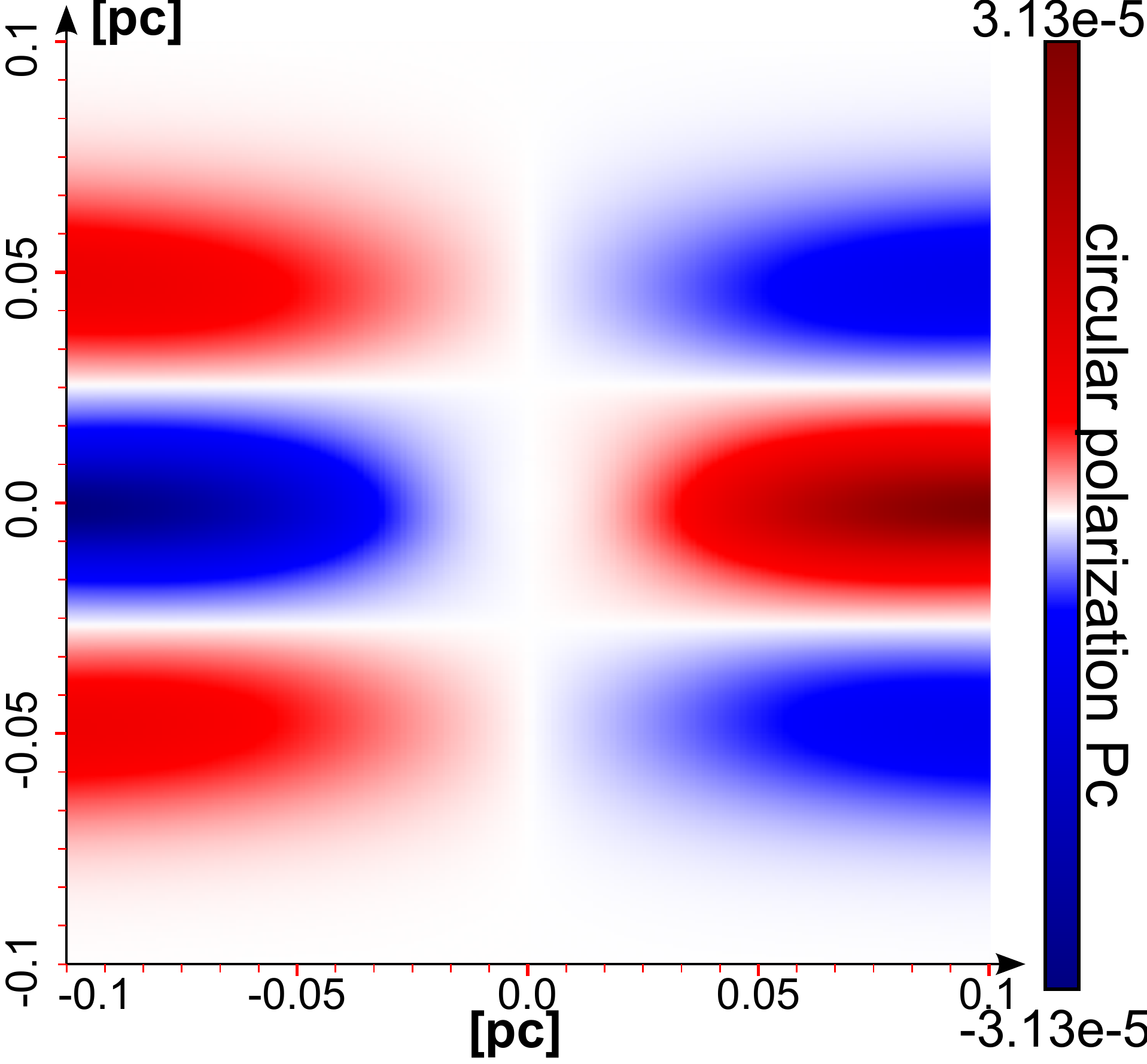}
			\end{center}
		\end{minipage}

	\caption{\small Same as in Fig. \ref{IDToro} for model $\rm{BE_{hour2}}$.	The characteristic radii for the temperature distribution are adjusted to keep the gas temperature and dust temperature in a range of $10\ \rm{K} - 25\ \rm{K}$ and $5\ \rm{K} - 15\ \rm{K}$, respectively. The dust particles are imperfectly aligned.}
						\label{IDHour}
\end{figure}

\begin{figure}[]
	\begin{minipage}[c]{0.49\linewidth}
			\begin{center}
				\includegraphics[width=1.0\textwidth]{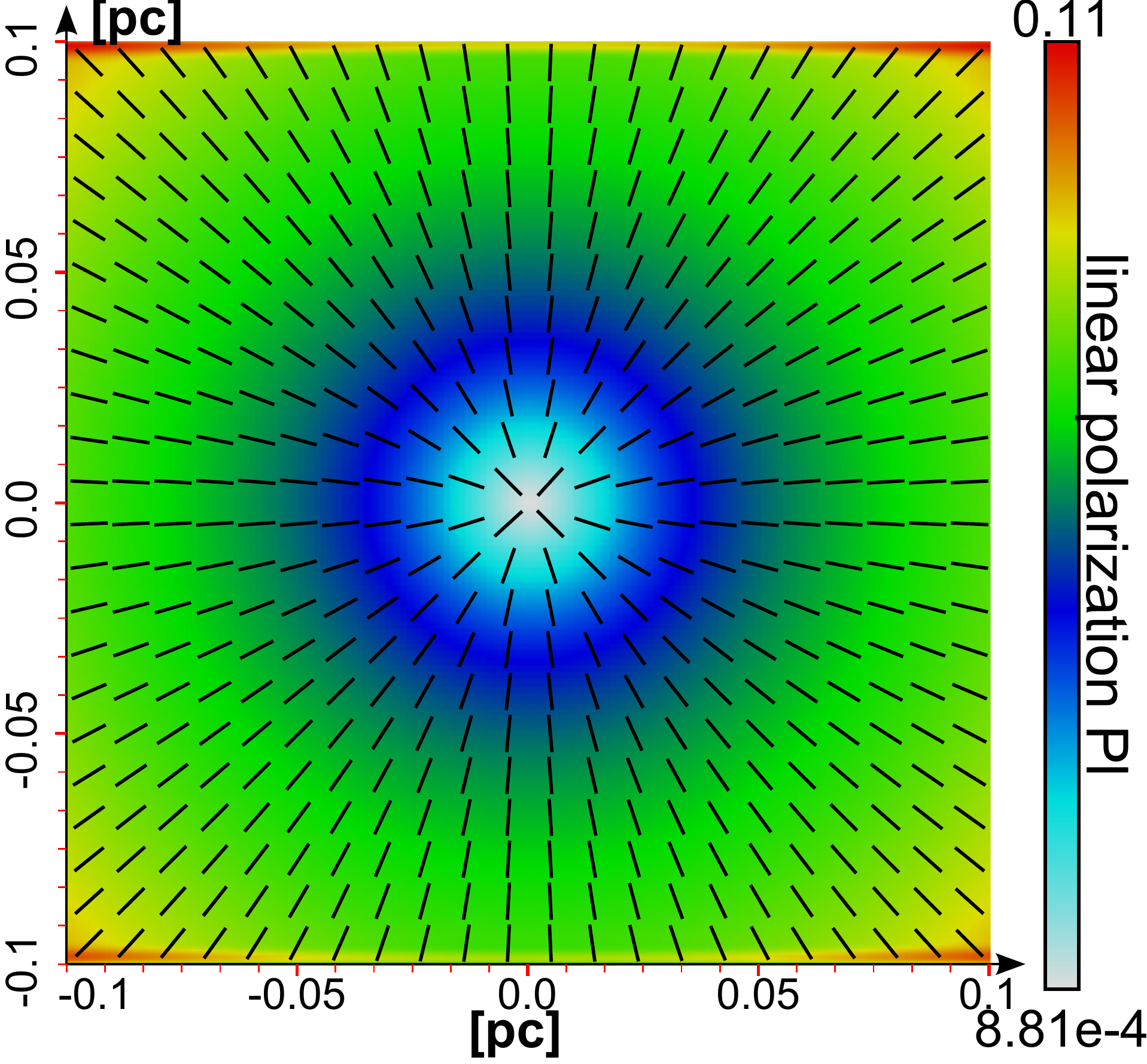}\\
				\includegraphics[width=1.0\textwidth]{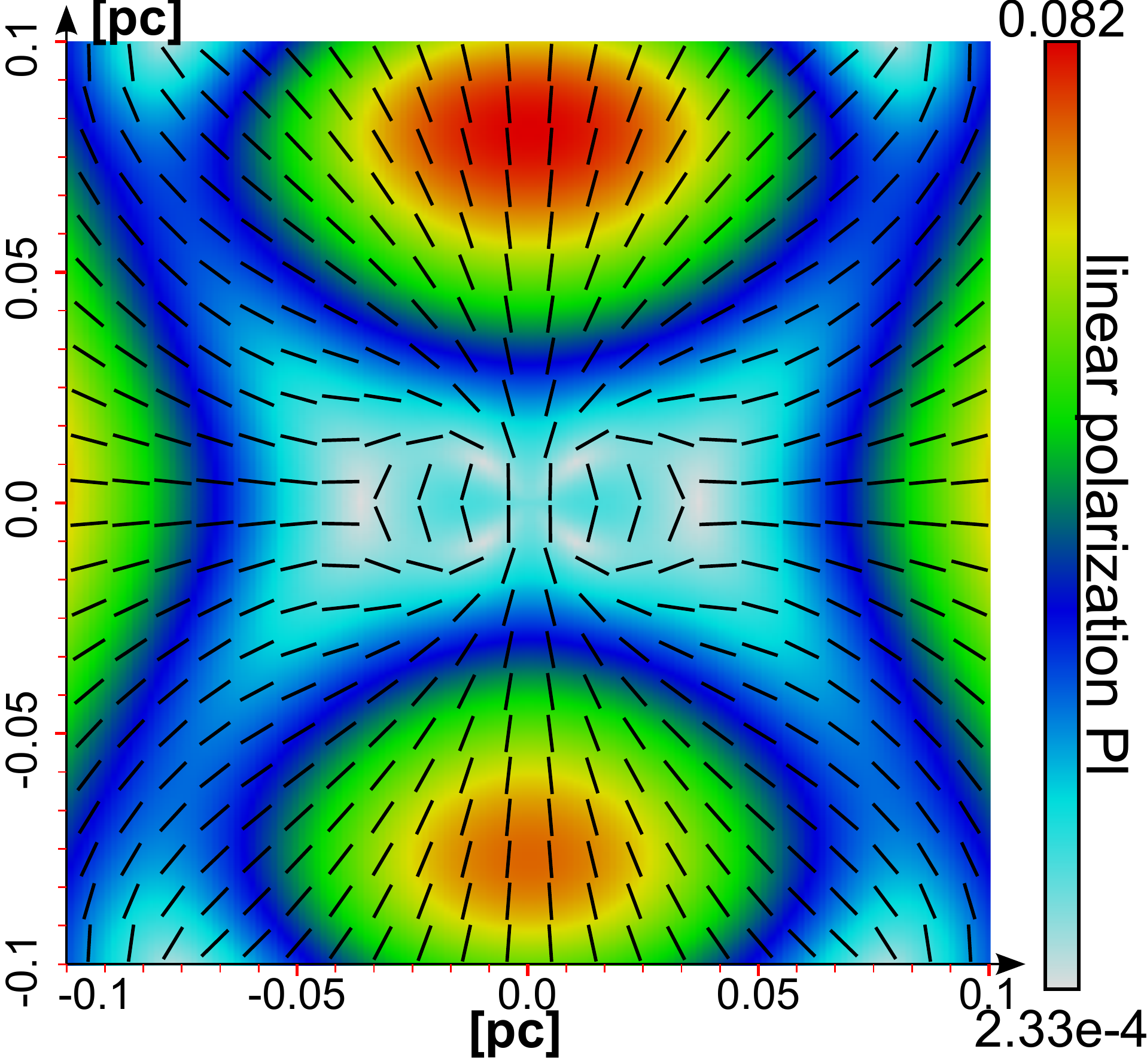}\\
				\includegraphics[width=1.0\textwidth]{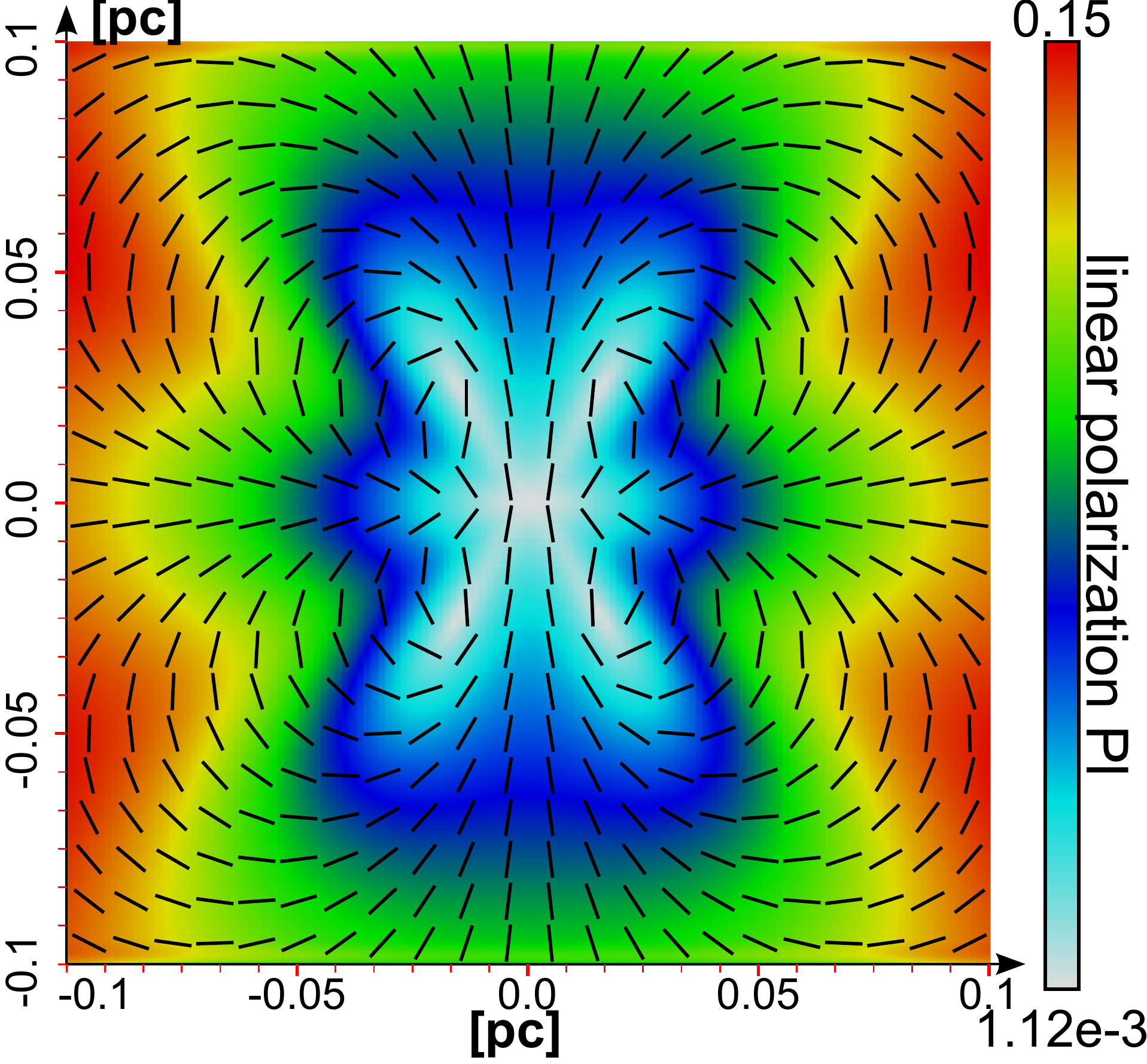}
			\end{center}
		\end{minipage}
		\begin{minipage}[c]{0.49\linewidth}
			\begin{center}
				\includegraphics[width=1.0\textwidth]{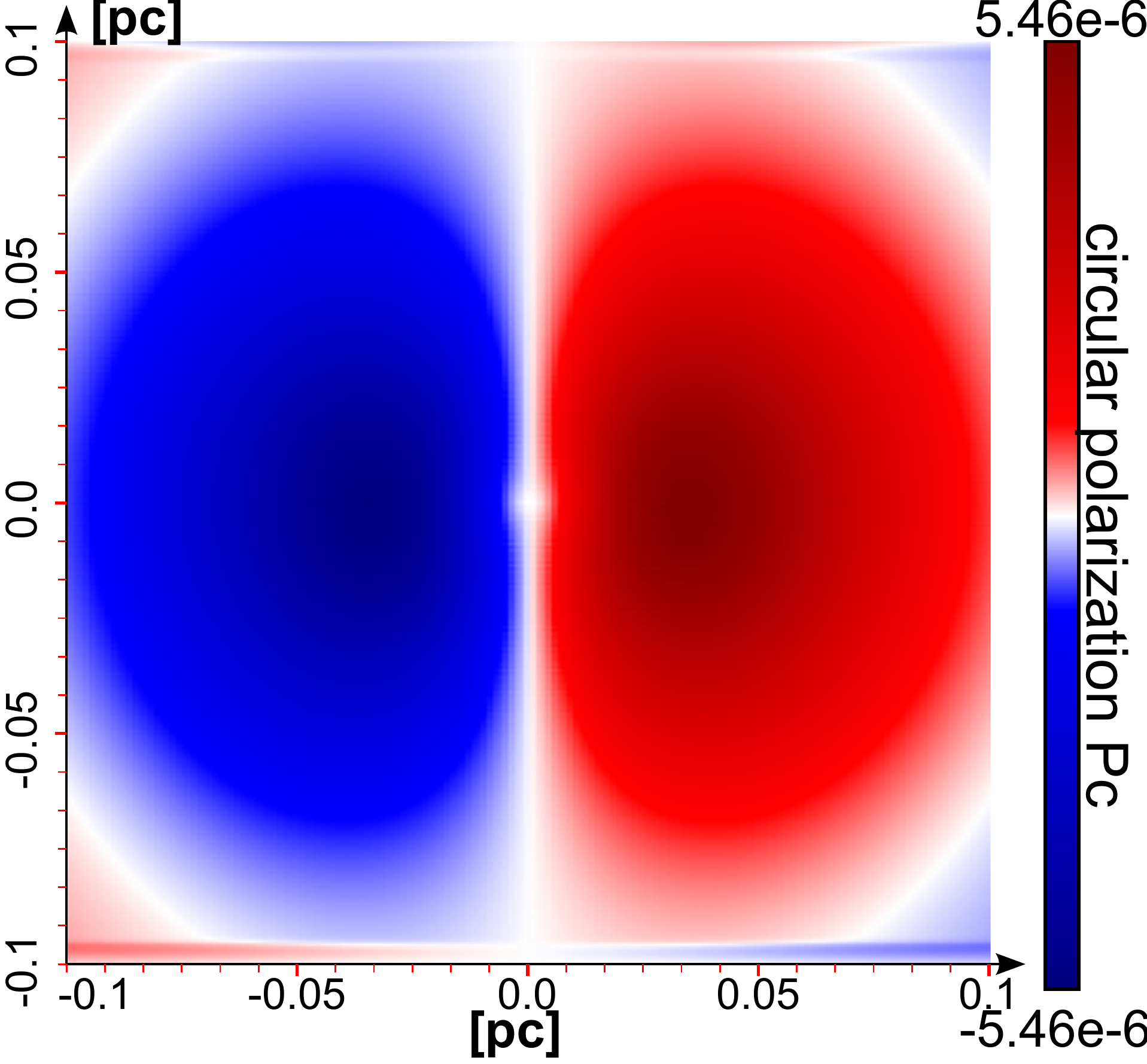}\\
				\includegraphics[width=1.0\textwidth]{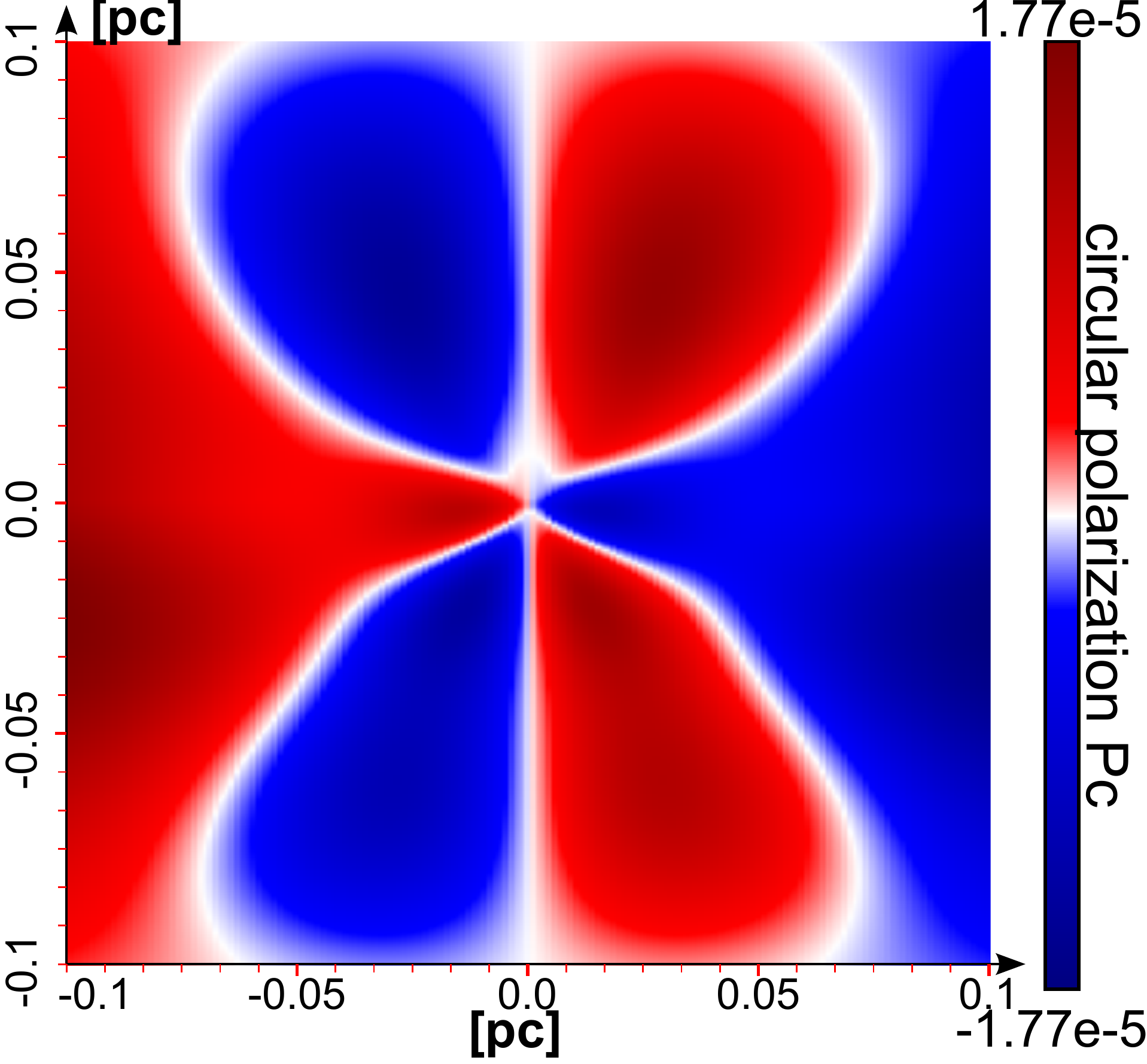}\\
				\includegraphics[width=1.0\textwidth]{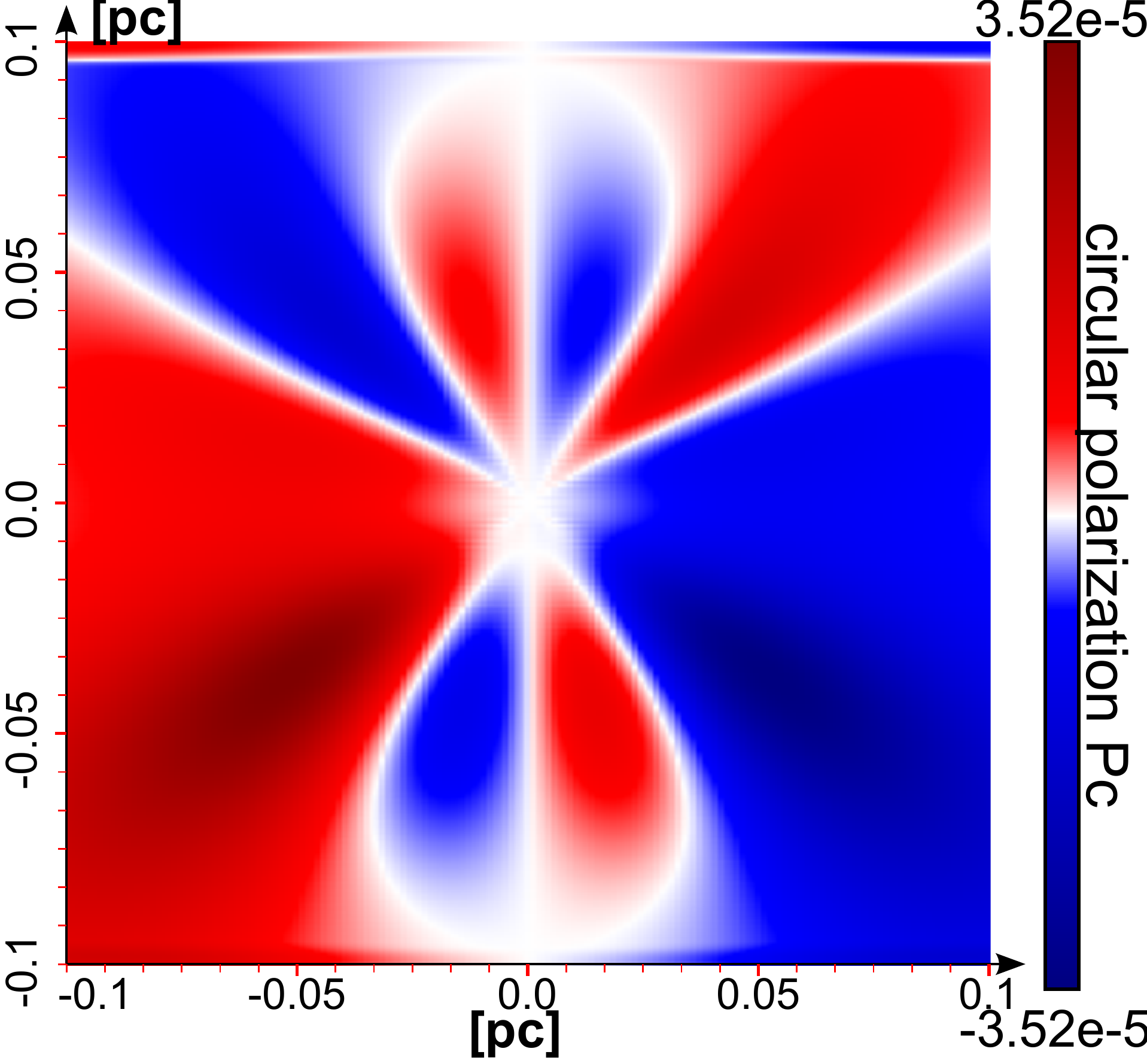}
			\end{center}
		\end{minipage}			
	
	\caption{\small Same as in Fig. \ref{IDQuad} for model $\rm{BE_{quad2}}$.}
						\label{IDQuad}
\end{figure}

\begin{figure}[]
	\begin{minipage}[c]{0.49\linewidth}
			\begin{center}
				\includegraphics[width=1.0\textwidth]{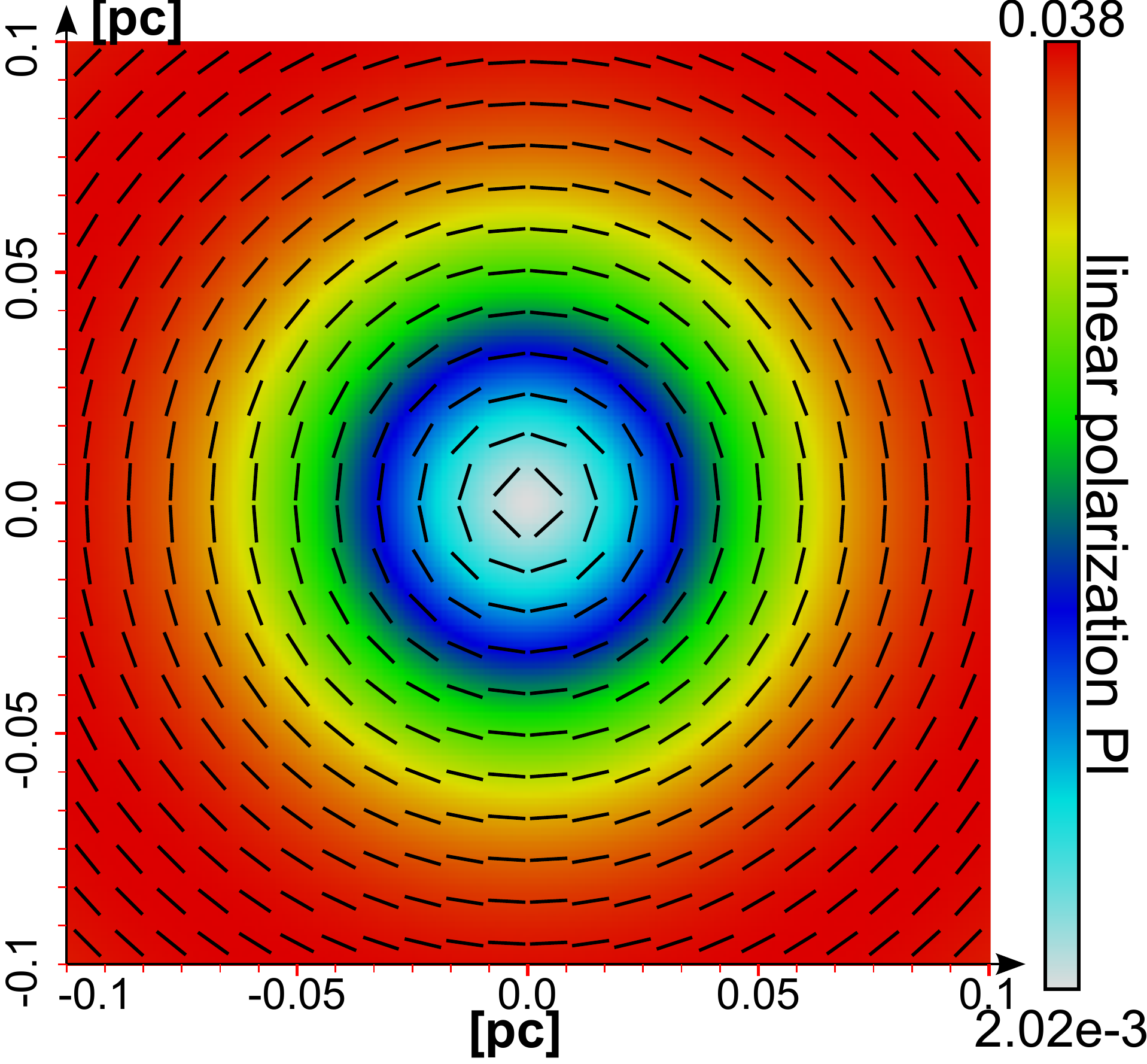}\\
				\includegraphics[width=1.0\textwidth]{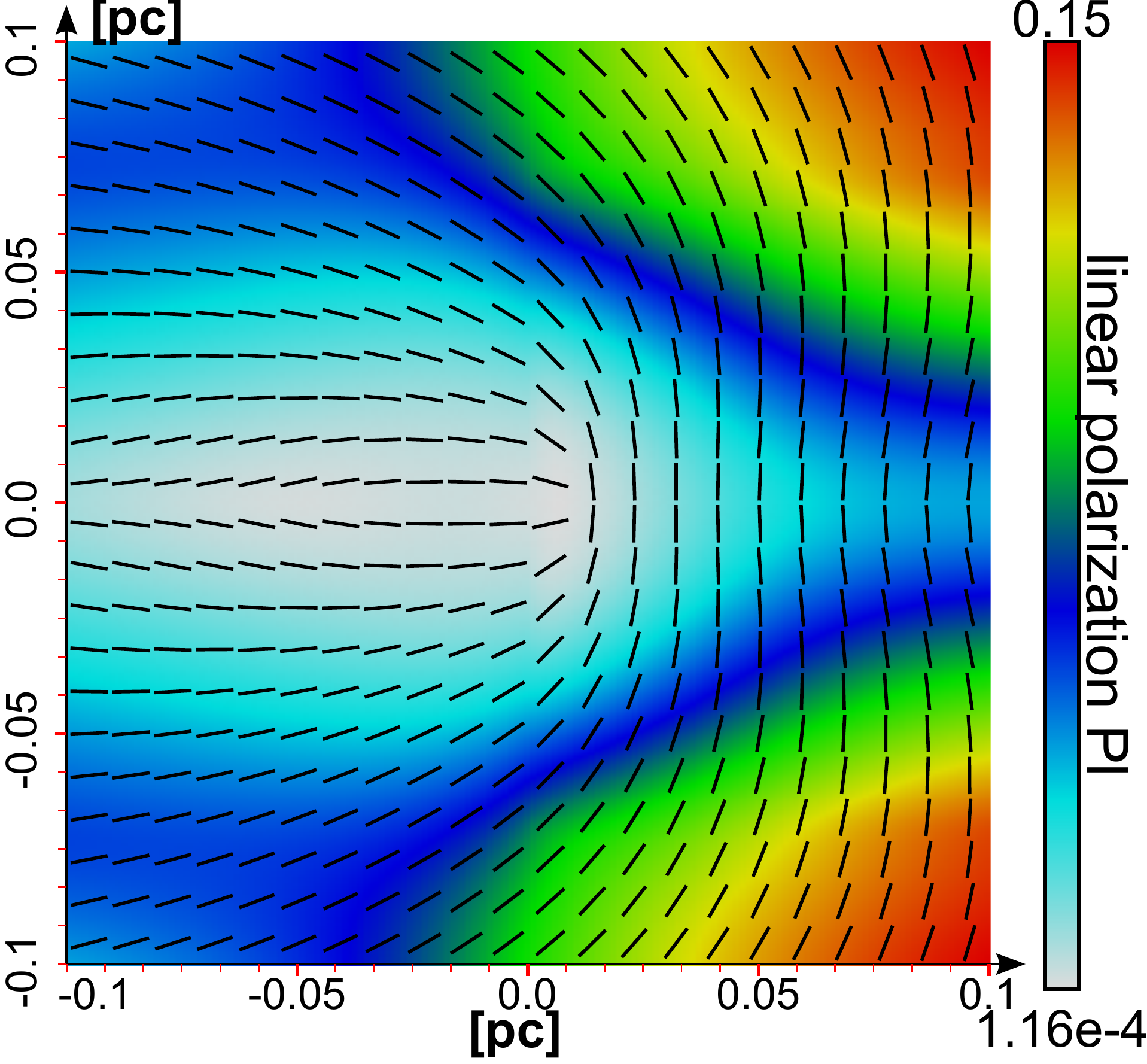}\\
				\includegraphics[width=1.0\textwidth]{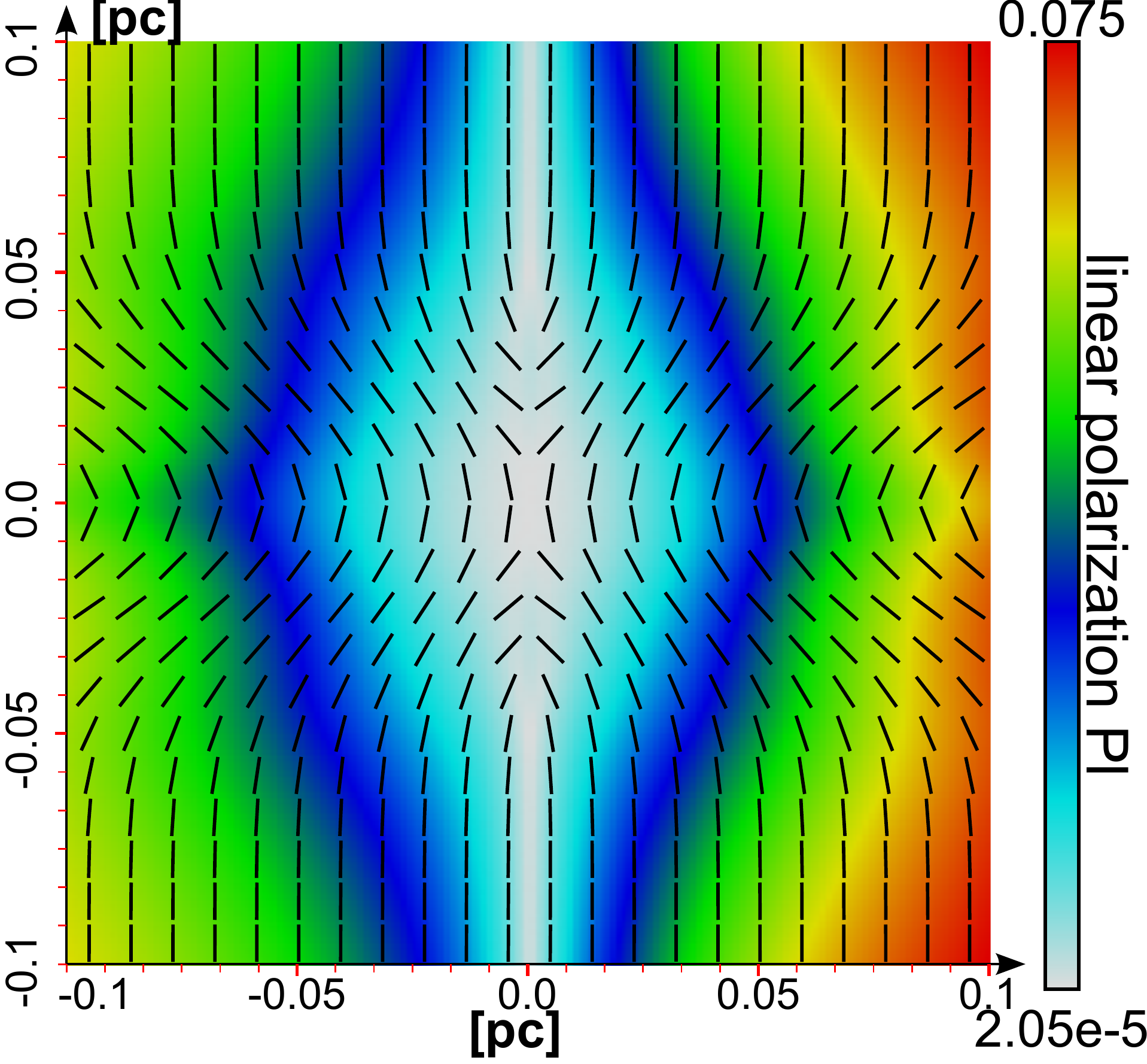}
			\end{center}
		\end{minipage}
		\begin{minipage}[c]{0.49\linewidth}
			\begin{center}
				\includegraphics[width=1.0\textwidth]{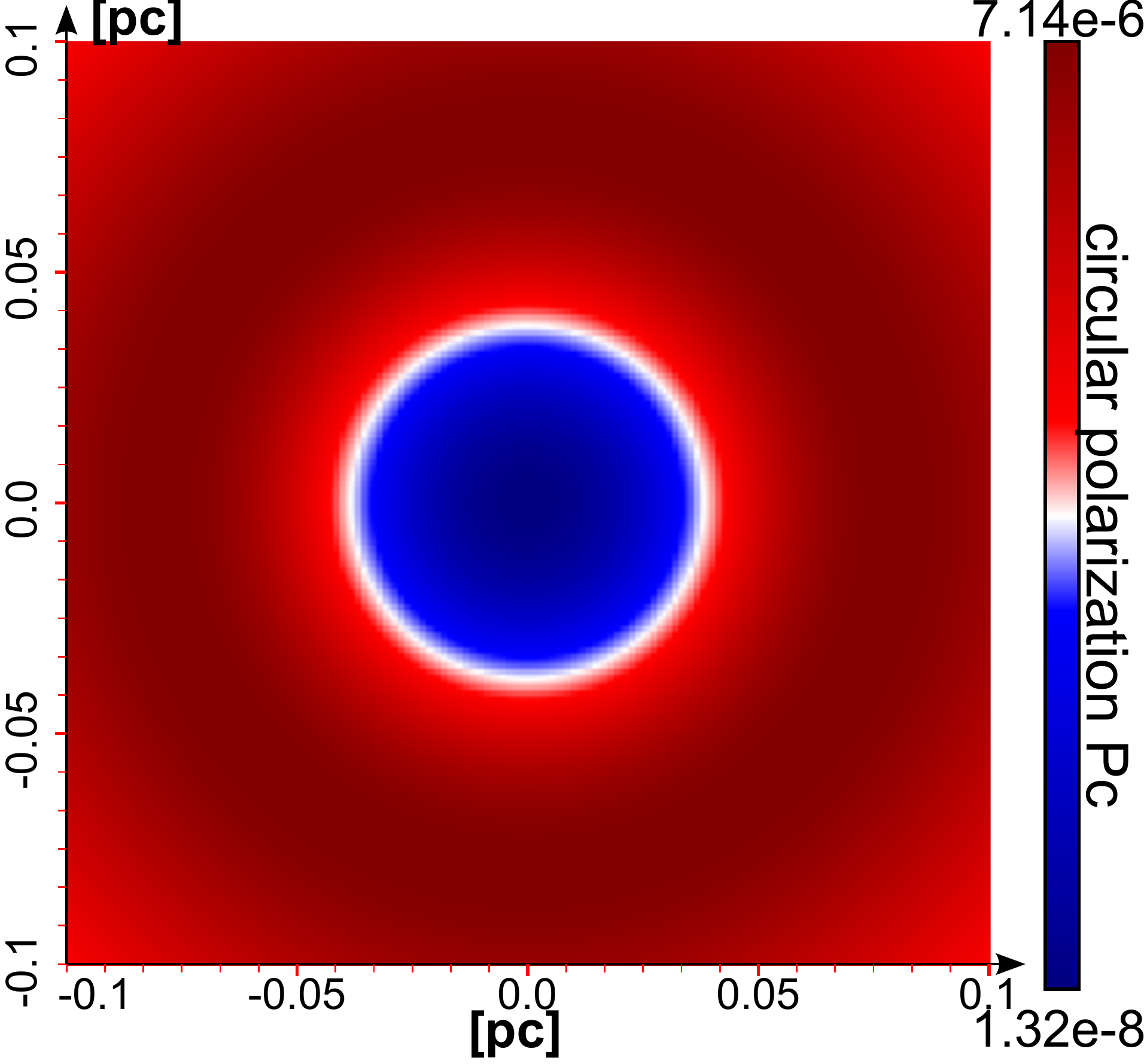}\\
				\includegraphics[width=1.0\textwidth]{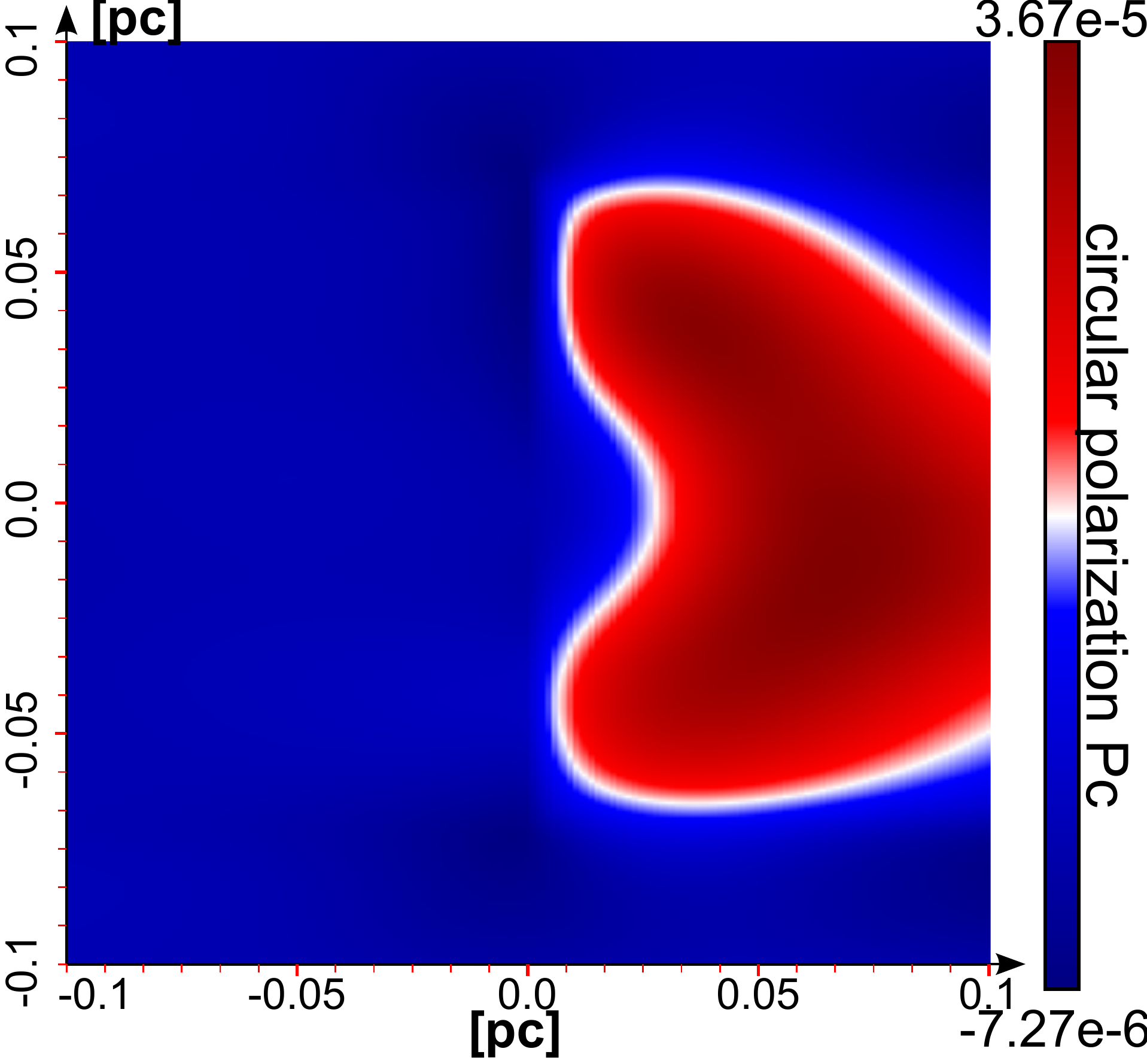}\\
				\includegraphics[width=1.0\textwidth]{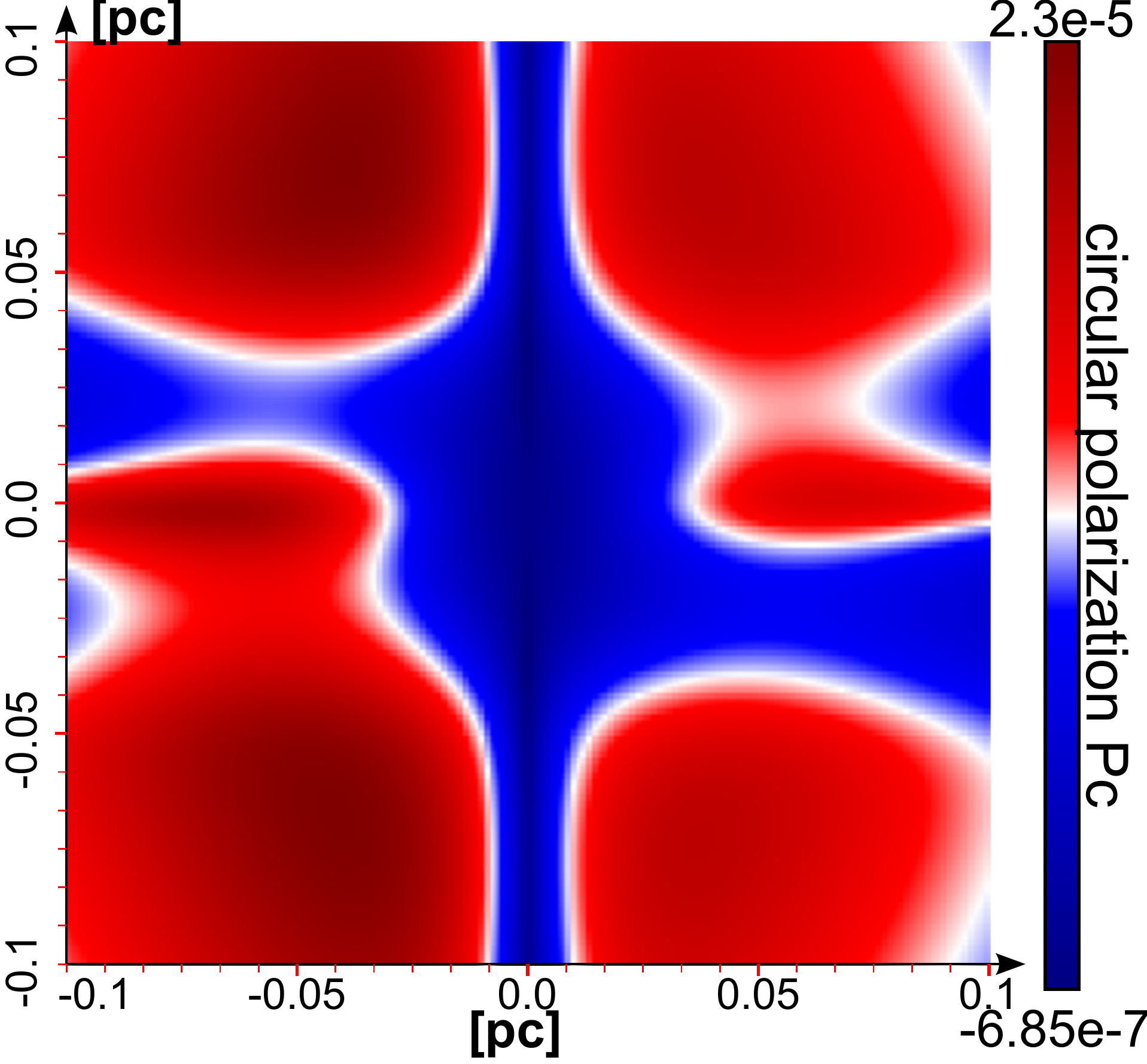}
			\end{center}
		\end{minipage}			
	
	\caption{\small Same as in Fig. \ref{IDQuad} for model $\rm{BE_{helic2}}$ with inclination angles of $0^{\circ}$ (top), $45^{\circ}$ (middle) and $90^{\circ}$ (bottom).}
	\label{IDHelic}
\end{figure}

\subsection{Ideal test - setups}
\label{setupIDEAL}
We began with analytical models assuming ideal conditions and included our dust model as presented in Sect.\ref{sect:dust}. The density profile is spherical and follows a Bonnor-Ebert profile \citep{ 1956MNRAS.116..351B,1955ZA.....37..217E} with a characteristic radius of $R_c = 1100\ \rm{AU}$ and a central number density of $n_0=10^{13}\ \rm{m^{-3}}$ typical of protostellar cores \citep{1997A&A...326..329L,2002A&A...394..275G,2011A&A...535A..49S}. The side length of the model space is $0.2\ \rm{pc}$. All models and relevant physical parameter are listed in Tab. \ref{tab:1}. A 3D plot of all considered magnetic field morphologies is provided in Fig. \ref{fig:morph}.\\
In the first setup ($\rm{BE_{toro}}$) we assumed a constant temperature of $T_{\rm{d}} = 15\ \rm{K}$ for the dust component and applied a toroidal magnetic field morphology

\begin{equation}
\vec{B}(x,y,z) =  \frac{1}{\sqrt{ x^2+y^2}}  \begin{pmatrix} y \\ -x \\ 0 \end{pmatrix}.
\label{eq:Btor}
\end{equation}
The dust particles were considered to be perfectly aligned (PA).\\
In the following, we include the effects of imperfectly aligned dust grains as described in Sect. \ref{sq:IDG}. Here, the additional parameter of dust temperature $T_{\rm{g}}$ and field strength $B_0$ are required. We chose the parameter $B_0$ for a Bonnor-Ebert sphere with an average temperature of $T_{\rm{g}} = 20\ \rm{K}$ to keep the mass-to-flux ratio consistent with recent observations \citep{2004Ap&SS.292..239W,2008A&A...487..247F,2009Sci...324.1408G,2010ApJ...724L.113B}. We varied the temperature such that $T_{\rm{g}}$ was in the range $10\ \rm{K} - 25\ \rm{K}$ and $T_{\rm{d}}$ is in the range $5\ \rm{K} - 15\ \rm{K}$ with the higher temperature in the center for later research on the impact of IDG grain alignment. This range of temperatures agrees well with theoretical models \citep[e.g.][]{2002A&A...394..275G,2005ApJ...635.1151K,2011A&A...535A..49S} and observational data \citep[e.g.][]{1997A&A...326..329L}.\\
For the magnetically supported model of star formation a characteristic hourglass field is predicted by theory \citep[e.g.][]{1993ApJ...417..243G} and MHD simulations \citep[e.g.][]{2006MNRAS.373.1091B}. An originally parallel magnetic field morphology turns into an hourglass morphology since the tension of the bent magnetic field lines provides support. Therefore, a preserved orientation in the polarization maps on multiple scales should be a strong indication for a significant contribution of the magnetic field to the star formation process. Recent polarization measurements \citep{2011AA...535A..44F,2013ApJ...769L..15S} strongly indicate the existence of such hourglass structures. However, for a definitive conclusion measurements of such magnetic field structure are still too rare.\\
The problem may be resolved by exploring the influence of inclination angle and grain alignment on synthetic polarization maps for different field morphologies. In the next setups we compare the pattern of linear polarization and circular polarization for an hourglass and a quadrupole field morphology dependent on inclination angle.\\
In contrast to exact physical solutions \citep[e.g.][]{1993ApJ...417..243G,1996ApJ...472..211L} for the models $\rm{BE_{hour1}}$ and $\rm{BE_{hour2}}$ we approximated the hourglass morphology of the magnetic field by an analytical function (see Fig. \ref{fig:morph} b)

\begin{equation}
\vec{B}(x,y,z) =  \frac{B_0}{\sqrt{ 1 + (\alpha x z)^2 e^{-2 \alpha z^2}  + (\alpha y z)^2 e^{-2 \alpha z^2}}}  \begin{pmatrix}  \alpha xz e^{-\alpha z^2} \\ \alpha yz e^{-\alpha z^2} \\ 1 \end{pmatrix}.
\label{eq:Bhour}
\end{equation}
The scaling parameter $B_0$ is for the field strength and is listed in Tab. \ref{tab:1} for each model. Here, the quantity $\alpha$ is a shape parameter to adjust the field geometry. Higher values of $\alpha$ will lead to denser magnetic field lines toward the center region. This geometry allows for an accurate and expedient treatment of ambipolar diffusion to demonstrate the effects of different inclination angles to the maps of linear and circular polarization. As shape parameter for the magnetic field we used $\alpha = 10$. The quadrupole field for the models $\rm{BE_{quad1}}$ and $\rm{BE_{quad2}}$ is defined as follows (see Fig. \ref{fig:morph} c):
\begin{equation}
\vec{B}(x,y,z) = \frac{B_0}{2 (x^2 + y^2 + z^2)^{7/2}}  \begin{pmatrix}  -3 x (x^2 + y^2 - 4 z^2) \\ -3 y (x^2 + y^2 - 4 z^2) \\ 3 z (-3 x^2 - 3 y^2 + 2 z^2) \end{pmatrix}.
\label{eq:Bquad}
\end{equation}
Cloud cores are observed to rotate \citep[e.g.][]{1993ApJ...406..528G}. It is expected for a magnetic field in a rotating cloud to have a toroidal component as well. We assumed the axis of rotation to be parallel to the predominant direction of the hourglass-shaped magnetic field lines and modeled the morphology as a normalized superposition of equations \ref{eq:Btor} and \ref{eq:Bhour}. The resulting magnetic field morphology for models $\rm{BE_{helic1}}$ and $\rm{BE_{helic2}}$ is plotted in Fig. \ref{fig:morph} d.\\
We calculated the inclination-angle-dependent polarization maps of linear and circular polarization resulting from thermal re-emission and dichroic extinction. Simultaneously, we obtained the corresponding intensity maps of the optical depth and net polarization.
		
\subsection{MHD - setups}
\label{setupMHD}
To test the results of our code against realistic scenarios we used our code to post-process an MHD simulation ($\rm{MHD_{sim1}}$ and $\rm{MHD_{sim2}}$) of a collapsing magnetized molecular cloud presented in \cite{2013MNRAS.432.3320S}. We took a snapshot of a run with a core mass of $300\ \rm{M_{\odot}}$ that has a diameter of $0.24\ \rm{pc}$, and a magnetic field initially parallel to the z – axis \citep[for more details we refer to][]{2013MNRAS.432.3320S}. The snapshot is taken $11.8\ \rm{kyr}$ after the first protostar has formed. The magnetic field is in a range of $B \in [8.8\times 10^{-10}\ \rm{T}; 9.4\times 10^{-5}\ \rm{T}]$. However, the dust temperature is not calculated from the radiation field of the newly born stars but is calculated in the MHD collapse simulation itself \citep[for details about the heating and cooling processes see][]{2006MNRAS.373.1091B}. We just probed the inner core with our ray-tracing algorithm, leading to a side length of $7734\ \rm{AU}$. The highest temperature in this area is below the sublimation temperature of the silicate grains and graphite grains, respectively, so that each cell contributes fully to the resulting maps of optical depth, intensity, linear polarization, and circular polarization. We adjusted the adaptive grid to maintain the original data structure of the simulation.\\

\section{Discussion}
\label{disc}		
\subsection{Impact of different inclination angles} Fig. \ref{IDToro} shows the degree of circular and linear polarization overlaid with the normalized orientation vectors for three different inclination angles for the model setup $\rm{BE_{toro}}$. For inclination angles of $0^{\circ}$ and $90^{\circ}$, respectively, the toroidal magnetic field has no crossing field-lines along the line of sight. The same holds for the hourglass and quadrupole morphology, respectively (see Fig. \ref{fig:morph} a - c). As expected, the resulting degree of circular polarization is zero (see Eq. \ref{eq:sol1}). For an inclination angle of $3^{\circ}$ the toroidal magnetic field is still radially symmetric. However, we find a recognizable pattern of circular polarization. The underlying toroidal morphology is not yet apparent. In model setup $\rm{BE_{toro}}$, the temperature is constant and the dust grains are perfectly aligned. For a low inclination angle near zero the angle of aligned dust grains with respect to the line of sight remains constant, leading to an expected pattern of almost homogenous linear polarization. In this case deviations in the pattern of linear polarization are just a result of the 3D density distribution and inclination angle. With increasing inclination angle, the angle of grain orientation with respect to the line of sight remains constant along the projected symmetry axis of the toroidal field, whereas in the outer regions, the dust grain orientation reaches $90^{\circ}$ and $\Delta C_{\rm{abs}}$  decreases. As expected, this results in a low degree of linear polarization in the outer regions for an inclination angle of $87^{\circ}$, demonstrating the accuracy of the applied radiative transfer code. The behavior of the degree of linear and circular polarization dependent on grain alignment and inclination angle is presented in Fig. \ref{fig:grIDEAL}. This Figure is below discussed in detail.\\
Figs. \ref{IDQuad} and \ref{IDHour} show the degrees of linear and circular polarization for three different inclination angles for the model setup $\rm{BE_{quad2}}$ and $\rm{BE_{hour2}}$. Here, the degree of polarization also depends on dust temperature $T_{\rm{d}}$, gas temperature $T_{\rm{g}}$, and alignment mechanism, leading to an inhomogeneous polarization pattern. With increasing inclination angle the hourglass morphology and quadrupole morphology (see Fig. \ref{fig:morph} b \& \ref{fig:morph} c), respectively, becomes apparent. However, for an inclination angle as low as $3^{\circ}$, the underlying field morphology remains ambiguous. In both configurations the vector field of linear polarization shows a similar radial symmetric pattern. Both setups $\rm{BE_{hour2}}$ and $\rm{BE_{quad2}}$ would be indistinguishable by measurements of linear polarization alone. Most important, the same pattern would be observable for a variety of field morphologies (e.g. a dipole field). In contrast to the ambiguities in the pattern of linear polarization, the maps of circular polarization remain unique for each field morphology, as we show in the right columns in Figs. \ref{IDQuad} and \ref{IDHour}. Here, we demonstrate that the hourglass and quadrupole fields are  distinguishable even for observations with low inclination angels because of the unique symmetry axis of each morphology which results in the characteristic pattern of circular polarization.\\
In Fig. \ref{IDHelic} we present the degree of circular polarization and degree of linear polarization overlaid with the normalized position vectors for the model setup $\rm{BE_{helic2}}$. Since $\rm{BE_{helic2}}$ has no rotation symmetry, crossing magnetic field lines along any line of sight and pattern of circular polarization are present even for inclination angles of $0^{\circ}$ and $90^{\circ}$. Note that the presence of crossing field lines does not necessarily lead to a low degree of linear polarization. As long as the directions of the field lines are not perpendicular to each other, they can still contribute to the net polarization due to dichroic extinction. Furthermore, nearby hot but optically thin regions can also rapidly build up linear polarization because of thermal remission.\\
Because of its helical symmetry, the initial symmetric pattern of $\rm{BE_{helic2}}$ in the synthetic linear and circular polarization maps becomes distorted while rotating the model space toward higher inclination angles. Here, we see toroidal patterns in the top rows of Fig. \ref{IDToro} similar to those in Fig. \ref{IDHelic} for low inclination-angles. The same would be observable if the dust grains in model $\rm{BE_{toro}}$ were imperfectly aligned. A similar behavior can be found for high-inclination angles in the model setups $\rm{BE_{hour2}}$ and $\rm{BE_{helic2}}$, as shown in the bottom rows of \ref{IDHelic} and Fig. \ref{IDHour}. Although the patterns of linear polarization deviate in both cases, the orientation vectors match quite well, which provide the possibility for misinterpretations under realistic observational conditions. Again, unique patterns in circular polarization measurements allow an unambiguous identification of the predicted magnetic field morphology independent of inclination angle.

\subsection{Influence of alignment mechanism} 
\begin{figure*}[]
		
		\begin{minipage}{0.49\linewidth}
		\begin{center}
                \includegraphics[width=1.0\textwidth]{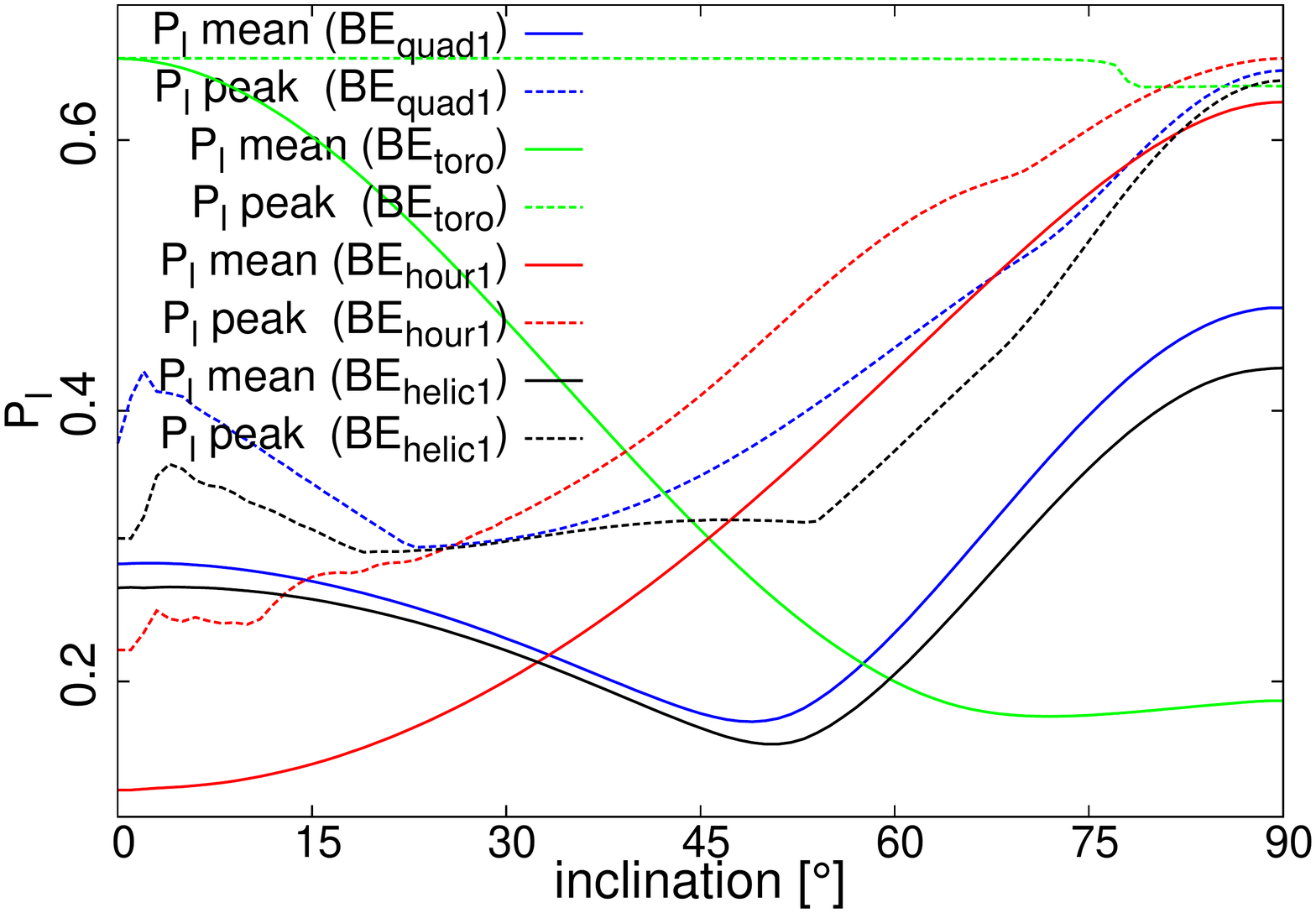}\\
								\includegraphics[width=1.0\textwidth]{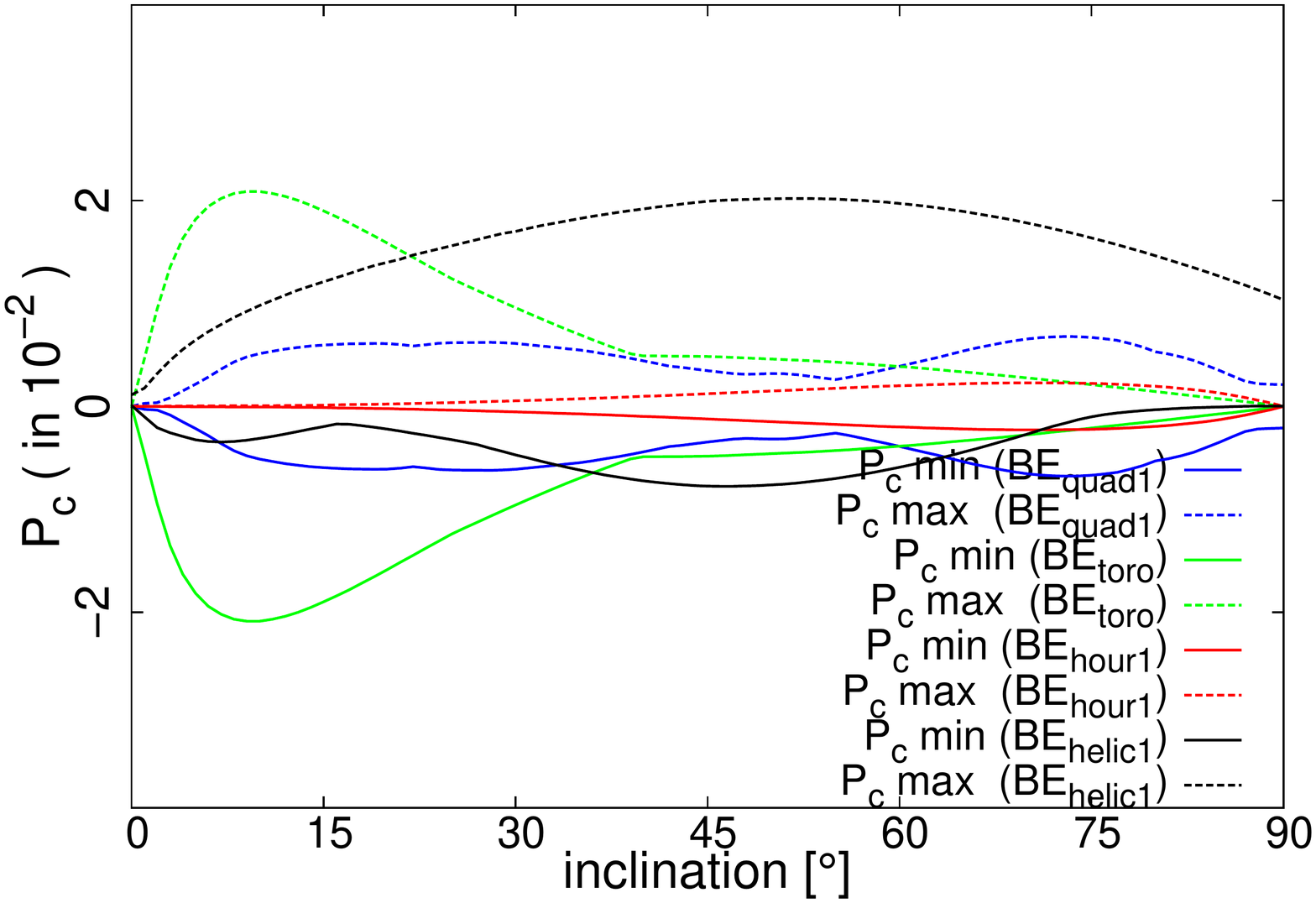}
\end{center}
            \end{minipage}
            \begin{minipage}[c]{0.49\linewidth}
						
\begin{center}
								\includegraphics[width=1.0\textwidth]{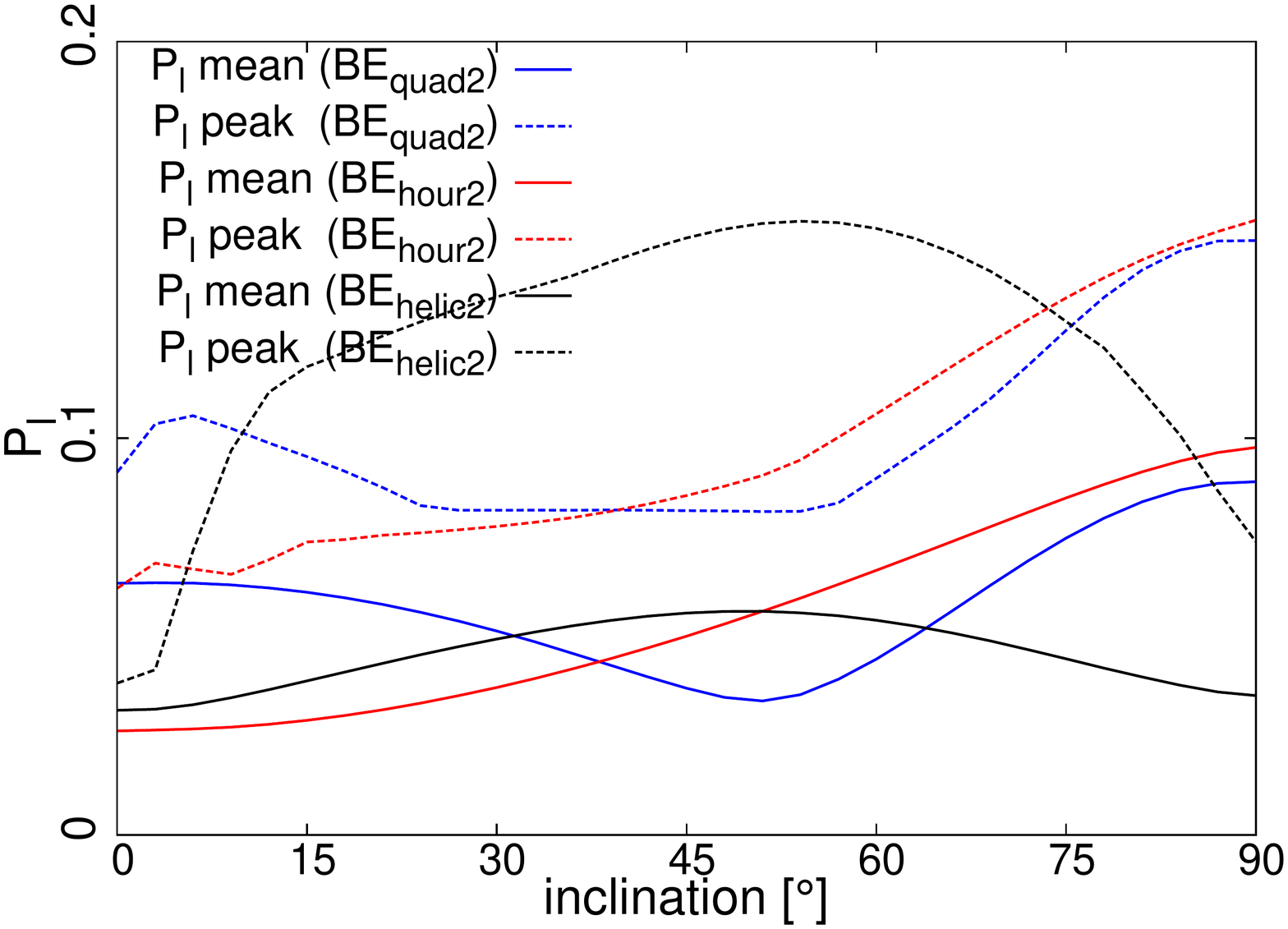}\\
								\includegraphics[width=1.0\textwidth]{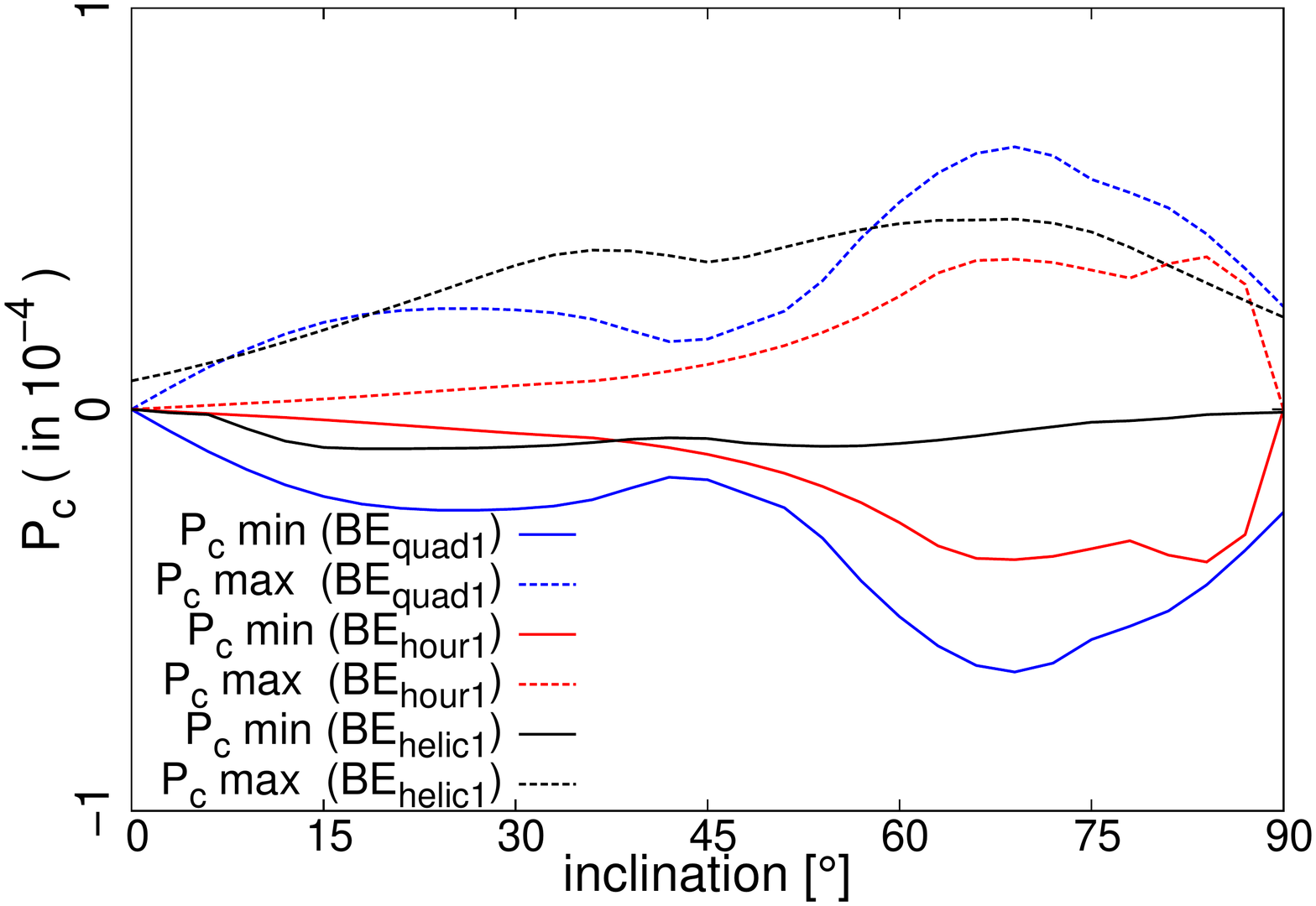}
\end{center}
		          \end{minipage}	
						\caption{\small Resulting degrees of linear polarization $P_{\rm{l}}$ (top) and circular polarization $P_{\rm{c}}$ (bottom) for perfectly (left) and imperfectly (right) aligned dust grains depending on inclination angle at a wavelength of $811\ \rm{\mu m}$ for the ideal Bonnor – Ebert sphere setups as presented in Sect. \ref{setupIDEAL}. For the degree of linear polarization $P_{\rm{l}}$ the mean values are plotted as solid lines and the peak values are in dashed lines. For circular polarization $P_{\rm{c}}$ the lowest values are plotted as solid lines, the highest values as dashed lines. The details of all model setups are listed in in Tab. \ref{tab:1}.}
						\label{fig:grIDEAL}
    \end{figure*}

\begin{figure*}[]
        		\begin{minipage}{0.49\linewidth}
		\begin{center}
                \includegraphics[width=1.0\textwidth]{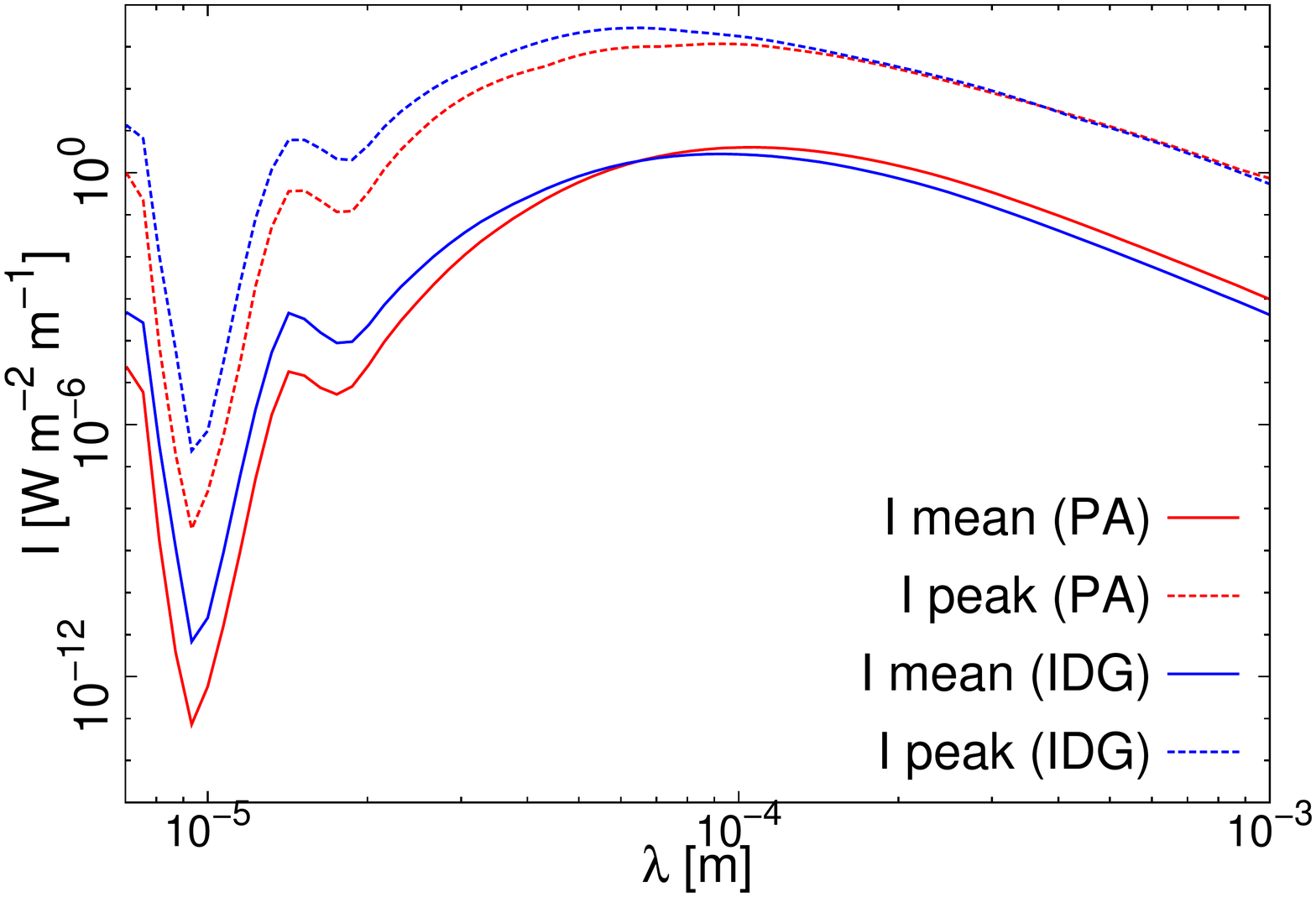}\\
								\includegraphics[width=1.0\textwidth]{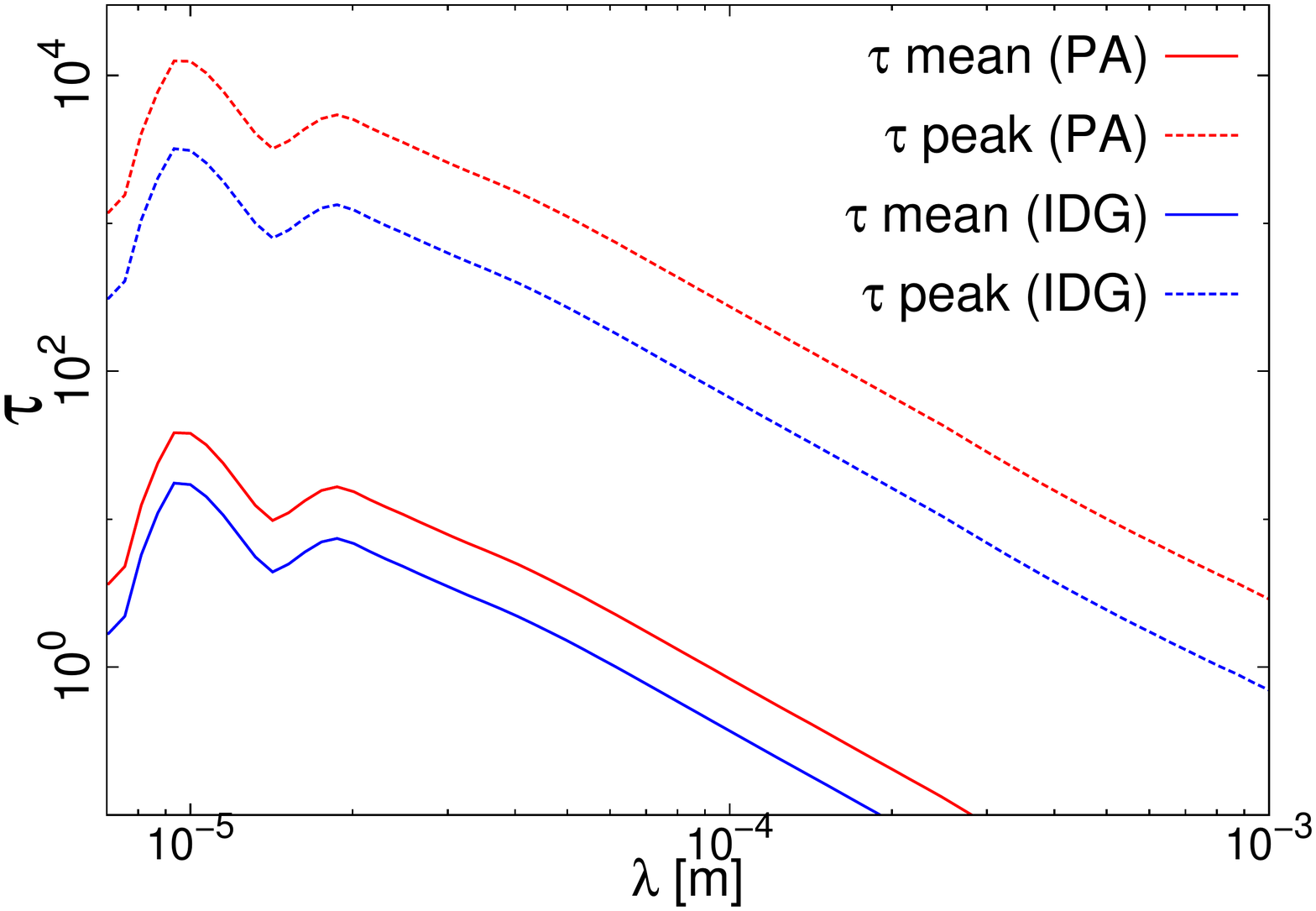}
\end{center}
            \end{minipage}
            \begin{minipage}[c]{0.49\linewidth}
						
\begin{center}
								\includegraphics[width=1.0\textwidth]{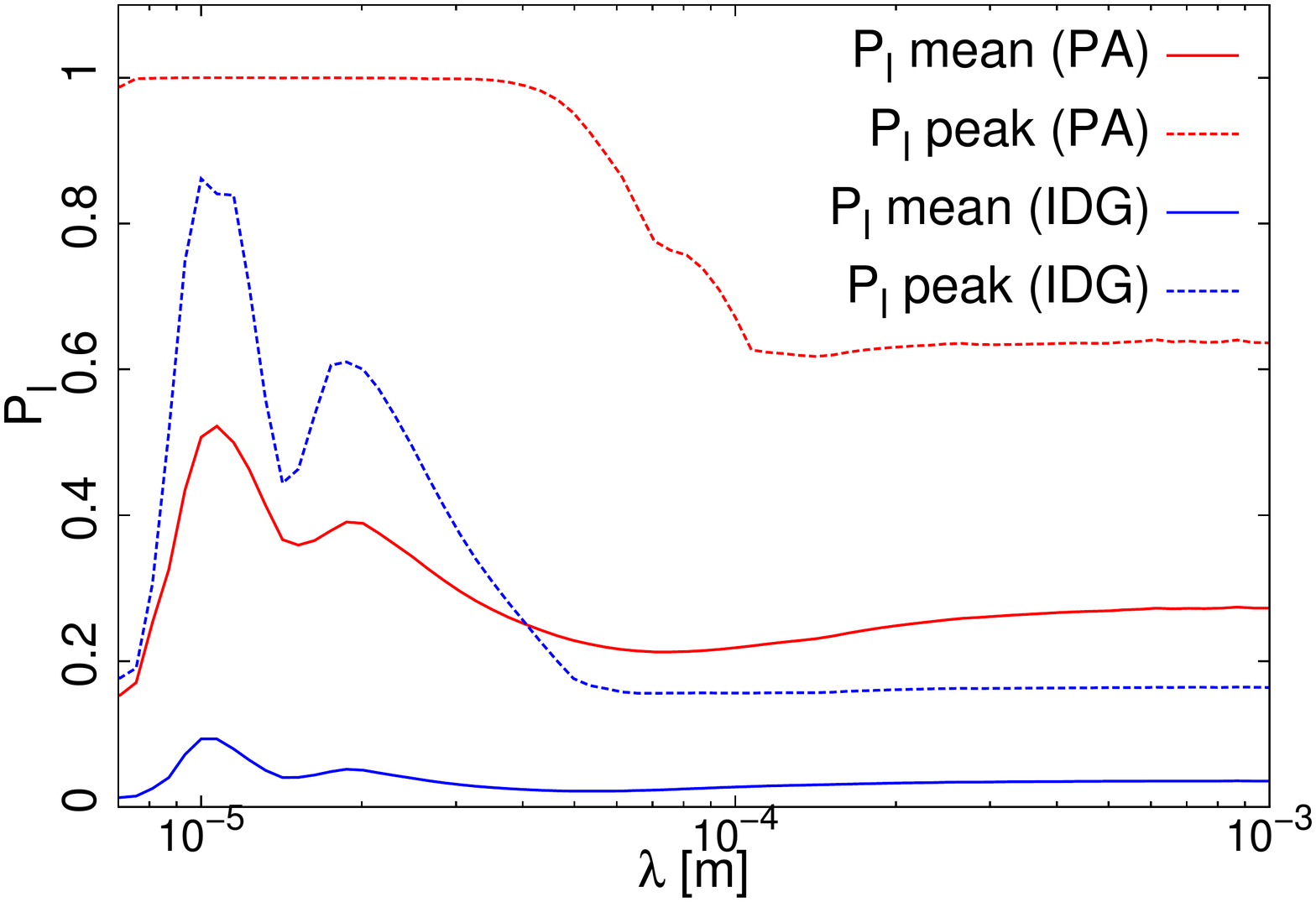}\\
								\includegraphics[width=1.0\textwidth]{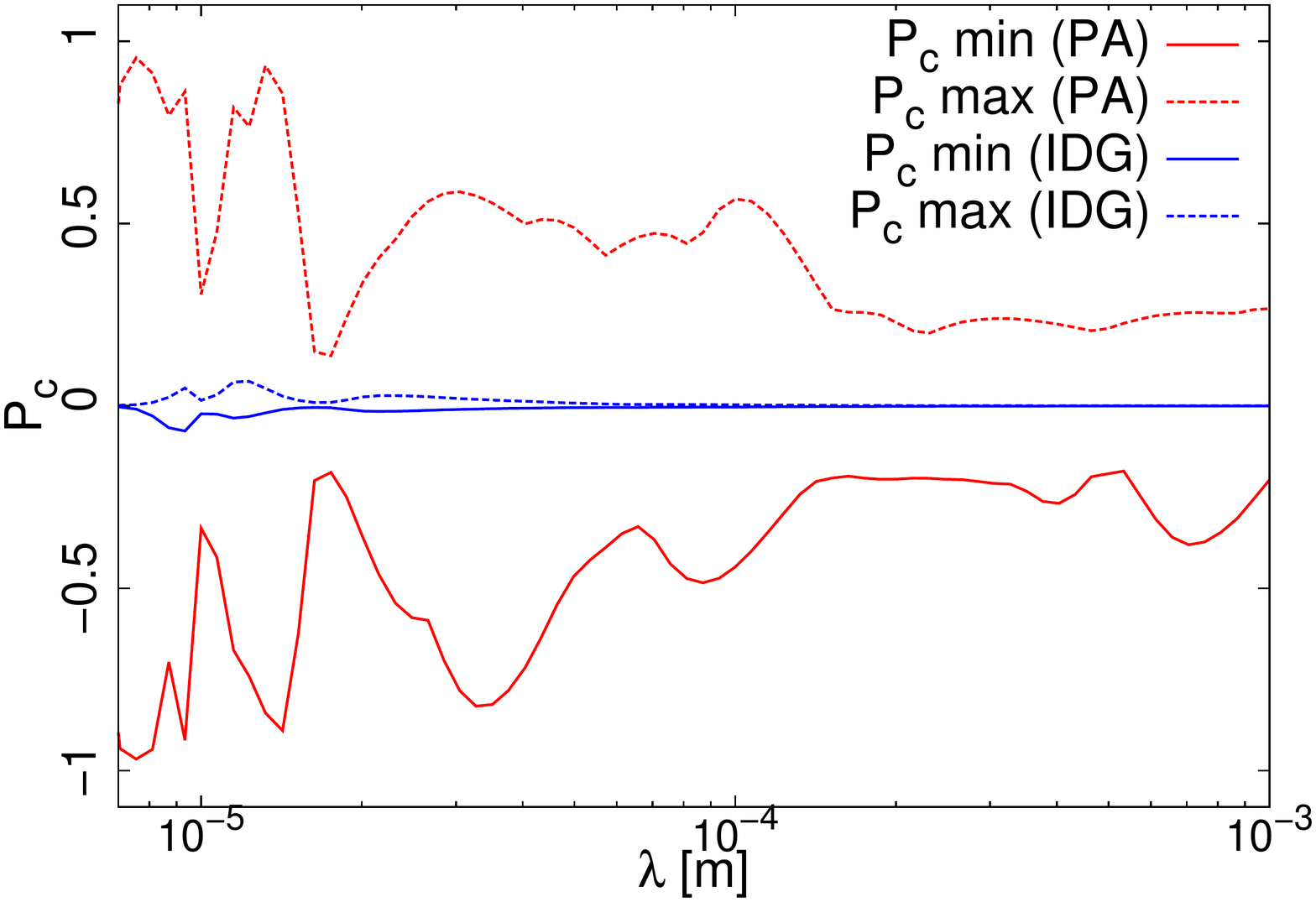}
\end{center}
		          \end{minipage}							
						\caption{\small Resulting peak values (dashed lines) and mean values (solid lines) of intensity $I$ , optical depth $\tau$, and degree of linear polarization $P_{\rm{l}}$ for the MHD – simulation as presented in Sect. \ref{setupMHD}. For circular polarization polarization $P_{\rm{c}}$ the lowest values are plotted as solid lines and the highest values as dashed lines. The red lines correspond to the setup with perfectly aligned dust grains  ($\rm{MHD_{sim1}}$) and the blue lines to the setup with imperfectly aligned grains ($\rm{MHD_{sim2}}$).}
						\label{fig:grMHD}
    \end{figure*}

\begin{figure*}[]
	\begin{minipage}{0.49\linewidth}
		\begin{center}
				\includegraphics[width=0.69\textwidth]{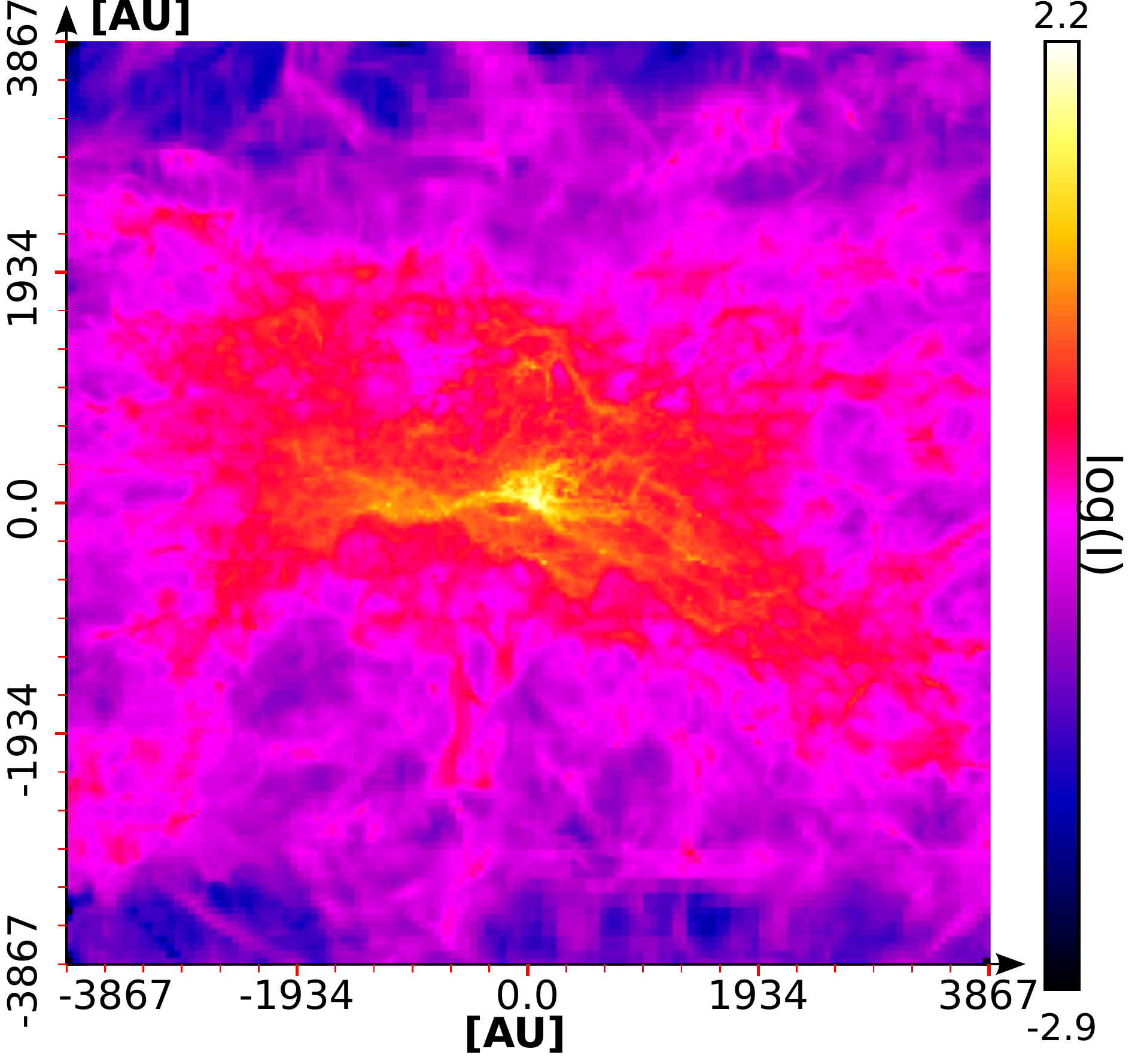}
				\includegraphics[width=0.69\textwidth]{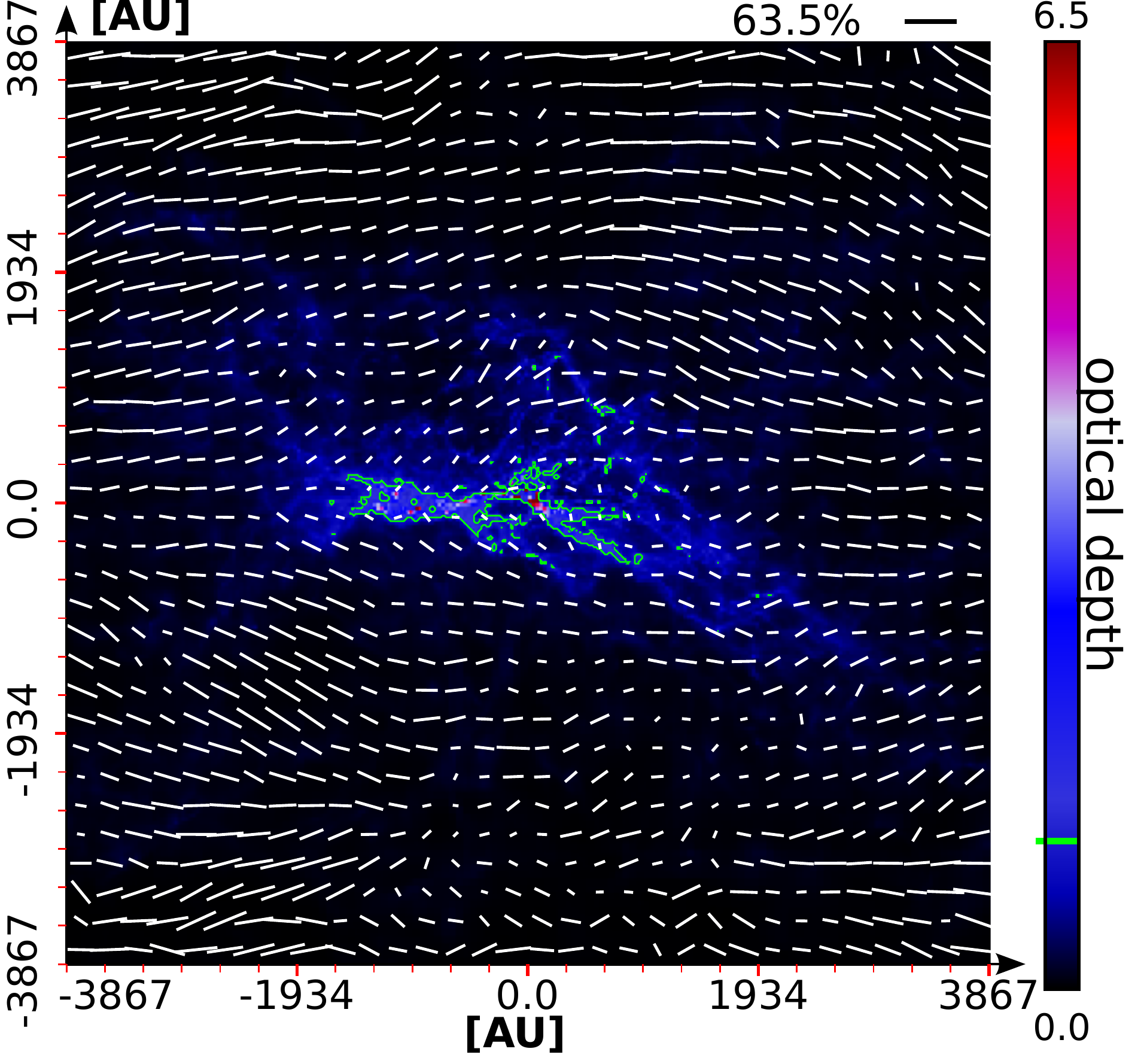}
				\includegraphics[width=0.69\textwidth]{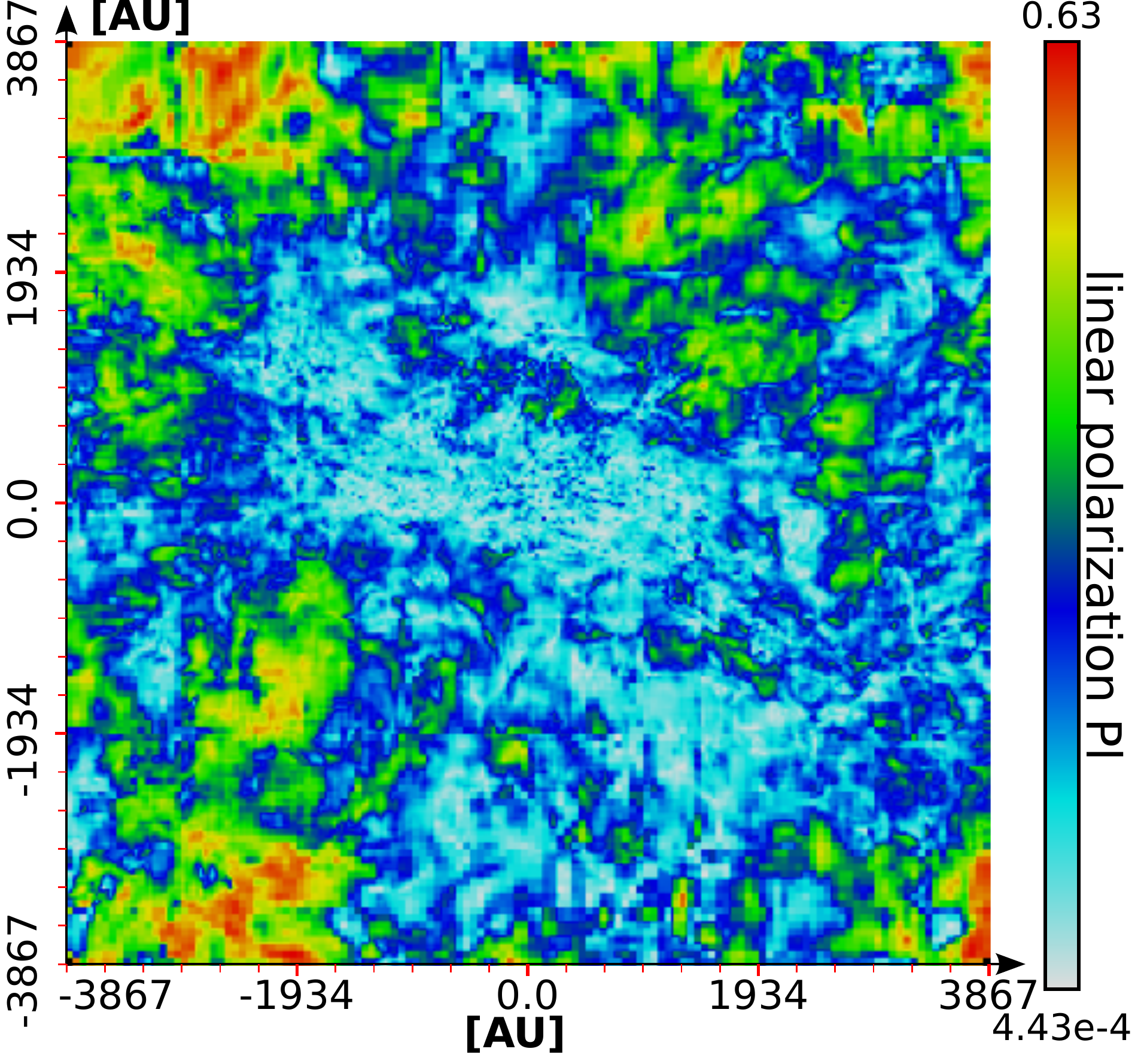}
				\includegraphics[width=0.69\textwidth]{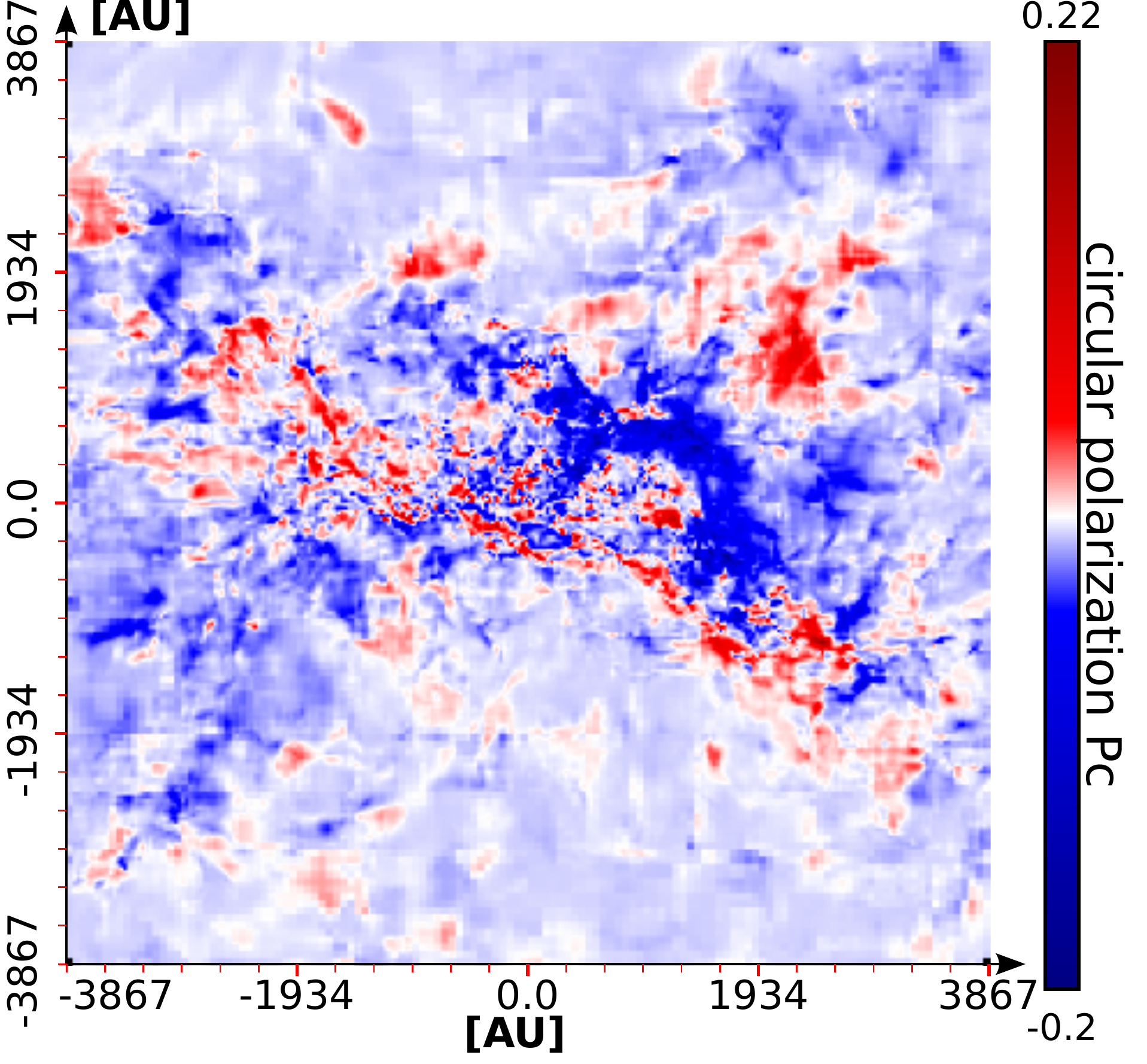}
		\end{center}
		\end{minipage}
		\begin{minipage}{0.49\linewidth}
			\begin{center}
				\includegraphics[width=0.69\textwidth]{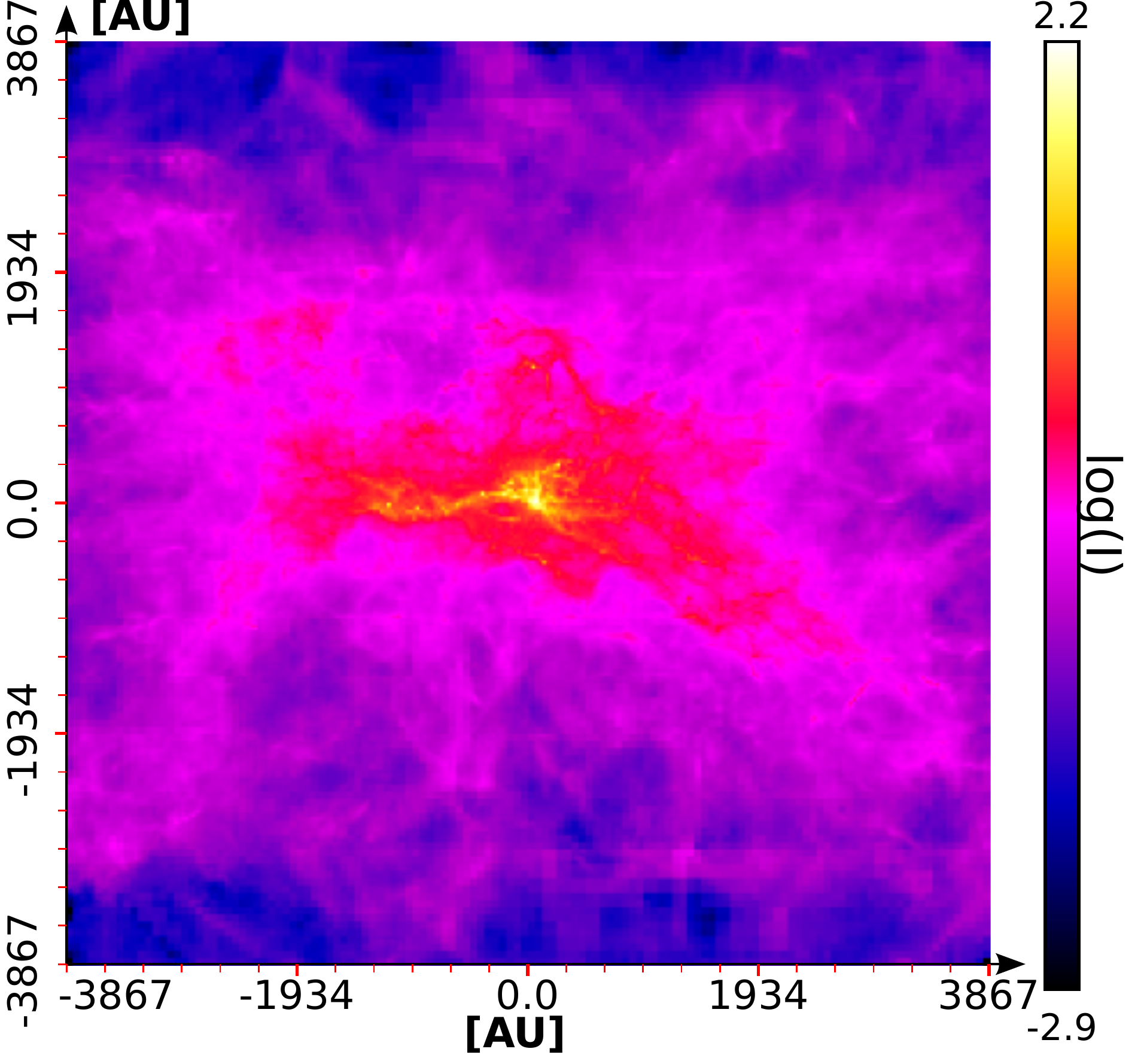}
				\includegraphics[width=0.69\textwidth]{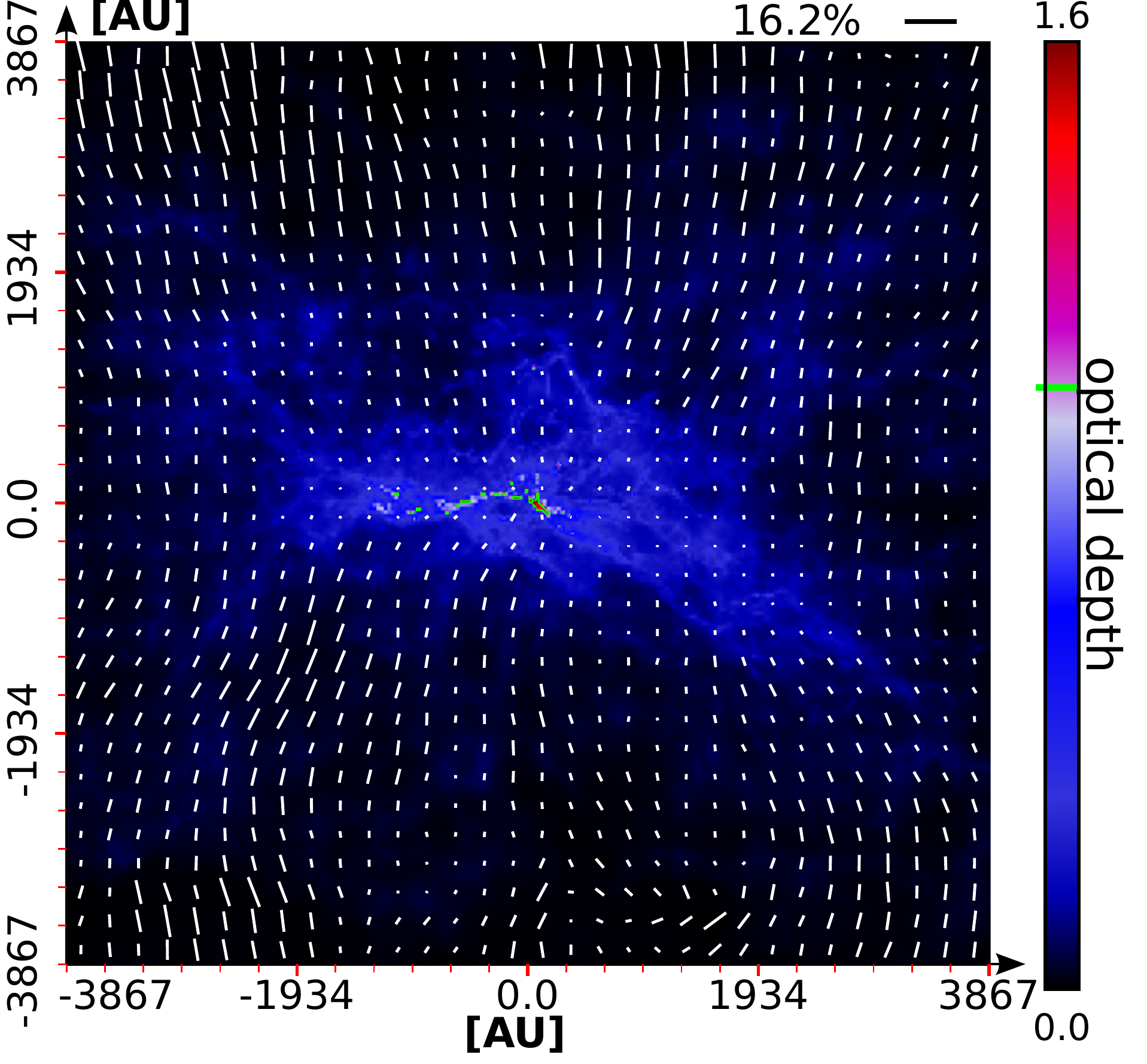}
				\includegraphics[width=0.69\textwidth]{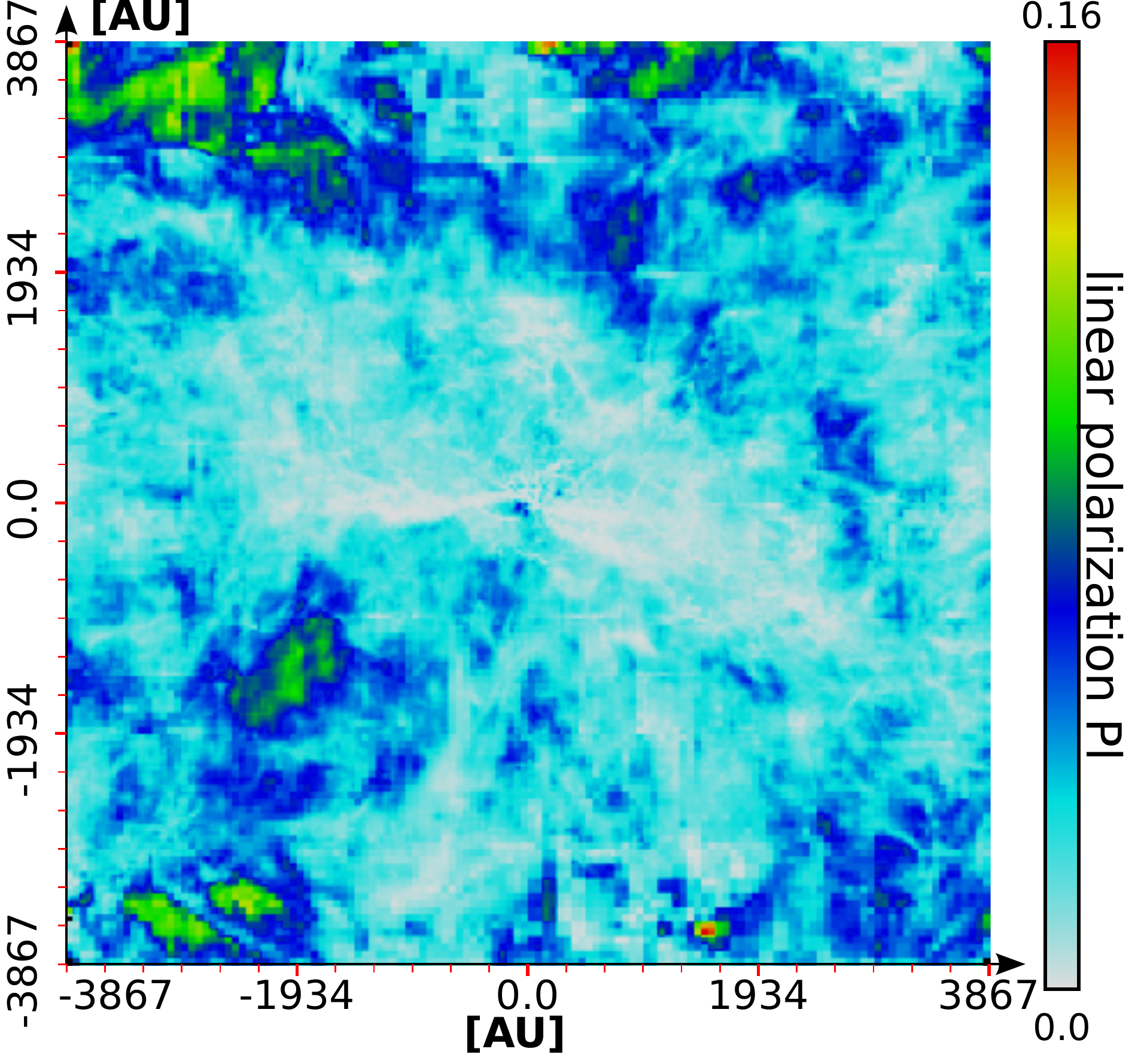}
				\includegraphics[width=0.69\textwidth]{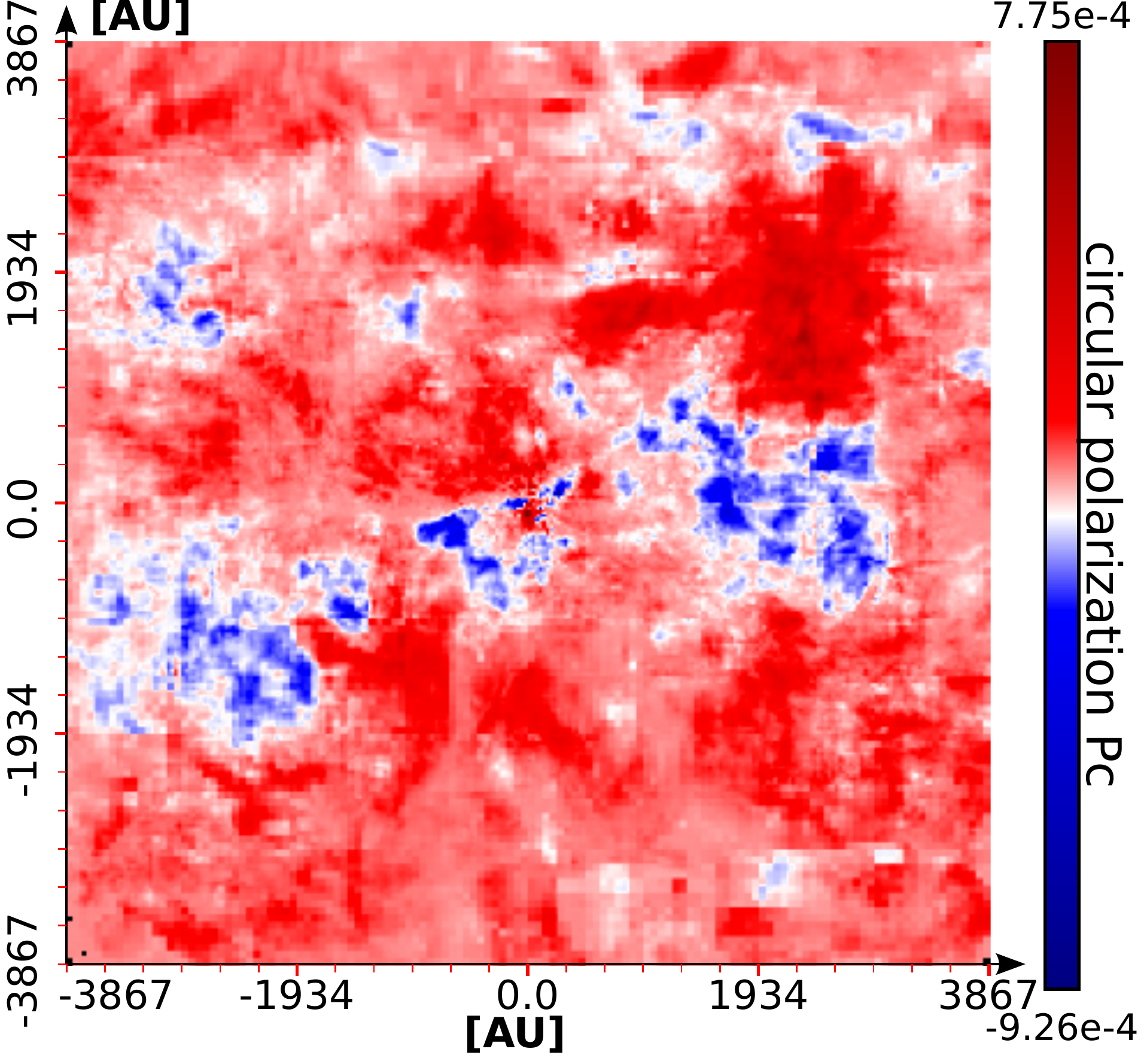}
				\end{center}
		\end{minipage}	
\caption{\small Resulting intensity maps (\rm{top row}), optical depth maps (\rm{second top row}), maps of linear (\rm{third top row}) and circular (\rm{bottom row}) polarization of setup $\rm{MHD_{sim1}}$ (\rm{left column}) and $\rm{MHD_{sim2}}$ (\rm{right column}) at a wavelength of $\lambda =248\ \rm{\mu m}$. The intensity is in $\rm{W m^{-2} m^{-1}}$ and the vectors of linear polarization have an offset angle of $90^{\circ}$. The color bar of the intensity map is on a logarithmic scale, and for the optical depth we apply an upper cut-off at $15 \%$ for illustrative purposes. The transition between optically thin and optically thick regions is indicated by the green contour lines and a marker in the color bar.}
\label{fig:MHDIDG}
\end{figure*}

In Fig. \ref{fig:grIDEAL} we present plots of the degree of linear (top) and circular (bottom) polarization as a function of the inclination for models with perfect alignment ($\rm{BE_{toro}}$, $\rm{BE_{quad1}}$, $\rm{BE_{hour1}}$, and $\rm{BE_{helic1}}$ on the left) and imperfect alignment ($\rm{BE_{quad2}}$, $\rm{BE_{hour2}}$, and $\rm{BE_{helic2}}$ on the right). In spherical models the degree of circular polarization is expected to be symmetrical with respect to the x-axis. But for the models $\rm{BE_{helic1}}$ and $\rm{BE_{helic2}}$ the inherent helical symmetry of the magnetic field morphology becomes apparent and the mirror symmetry in the plot is broken (see black lines and bottom row in Fig. \ref{fig:grIDEAL}). Perfectly aligned grains lead to an unrealistically high degree of linear polarization. One approach to solve this problem is provided in \cite{2000ApJ...544..830F} where the cross sections of the dust and the alignment efficiencies were modeled by a single parameter to adjust the degree of linear polarization to observational data. However, this is without any physical motivation. As we demonstrate here, lower degrees for the highest linear polarization of $\approx 15 \%$ as required by observational data \citep{1999ApJ...525L.109G,2005AA...430..979G,2011ApJ...732...97D} can be achieved by applying a realistic dust grain model and alignment mechanism to radiative transfer calculations. For the exemplary case with a wavelength of $811\ \rm{\mu m}$, as we show in Fig. \ref{fig:grIDEAL} on the right-hand side, the peak values and  mean values of linear polarization behave very well in the observed limitations. So far, there is no observational data available for the limitations of circular polarization caused by dichroism.\\
Beside the reduction in the maximum degree of linear and circular polarization the usage of the IDG mechanism leads to some additional effects. Compared with Fig. \ref{IDToro}, toward the center region in figures \ref{IDQuad}, \ref{IDHour} and \ref{IDHelic}, the IDG mechanism results in depolarization for the linear polarization. The same behavior can be found  for the circular polarization because of the adjustment of the parameter $\delta_{\rm{0}}$ with respect to the higher density and and temperatures (see Eq. \ref{eq:delta}).\\
However, even with circular polarization measurements the interpretation of linear polarization data remains difficult in a complex environment because of other effects of ambiguity. In each cell of the model space the linear polarization is determined by dichroic extinction and thermal re-emission perpendicular to each other. Depending on the cross sections of extinction and absorption, the process of polarization can be dominated either by extinction or thermal re-emission (see Appendix \ref{apB}). Since the cross sections are functions of a distinct set of physical parameters (wavelength, alignment efficiency, orientation and strength of the magnetic field) in each region inside the ISM, this can lead to a flip of the orientation angle by $90^{\circ}$.\\
In Fig. \ref{fig:MHDIDG} we compare the effects of grain alignment in a complex environment provided by MHD simulations (see Sect. \ref{setupMHD}) for an exemplarily chosen wavelength of $\lambda =248\ \rm{\mu m}$. With perfectly aligned dust grains ($\rm{MHD_{sim1}}$) the resulting intensity and linear polarization pattern is completely dominated by the outer regions with a thin and hot gas phase. With the exception of the dense core in both cases ($\rm{MHD_{sim1}}$ \& $\rm{MHD_{sim2}}$), the model space is almost completely optically thin at $248\ \rm{\mu m}$. With the IDG mechanism, acting on the dust grains, a high number density $n_{\rm{d}}$ and the resulting lower value of the parameter $\delta_{\rm{0}}$ leads to a depolarization in the central region. Due to the averaging of the cross sections according to the IDG mechanism the cross sections decrease and the dust grains behave more like spherical particles independent of the angle with respect to the incident light-ray, and the degree of linear polarization is reduced. As a result of this, the intensity and optical depth is lower and the optical depth is smoother than for PA aligned grains (see Figs. \ref{fig:grMHD} and \ref{fig:MHDIDG}). The effect of the $90^{\circ}$ flip can be observed depending on grain alignment in the pattern of the linear polarization vectors. However, the density, temperature, and the magnetic field structure are far too complex in both cases to derive any meaningful interpretation about the underlying magnetic field morphology from the pattern of linear and circular polarization. We plan to investigate this problem in a forthcoming paper. 

\subsection{Effect on multi-wavelength polarization maps} 
\begin{figure*}[]
	\begin{minipage}{0.49\linewidth}
		\begin{center}
				\includegraphics[width=0.82\textwidth]{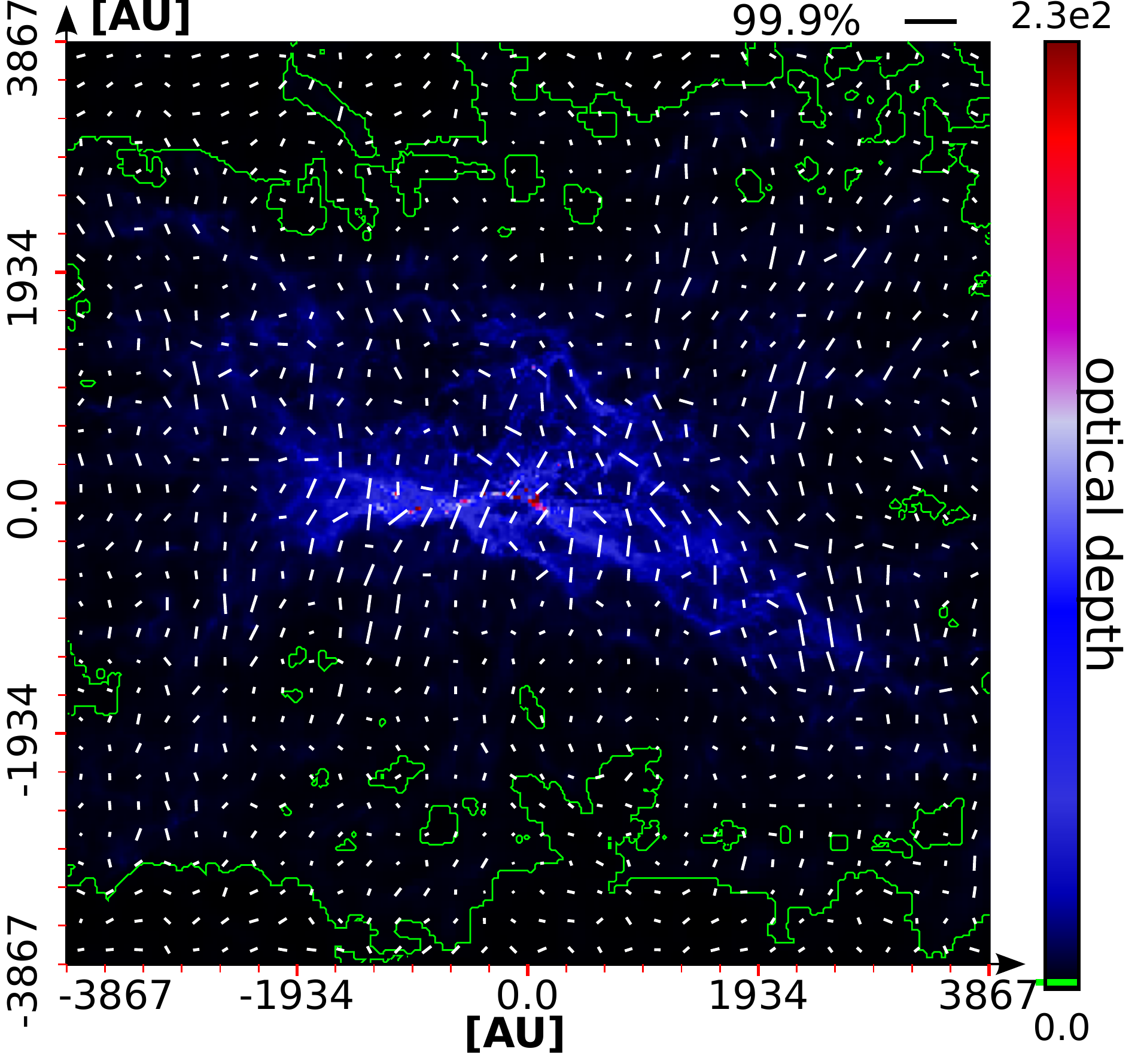}
					\caption*{ $\lambda = 7.1\ \rm{\mu m}$}
				\includegraphics[width=0.82\textwidth]{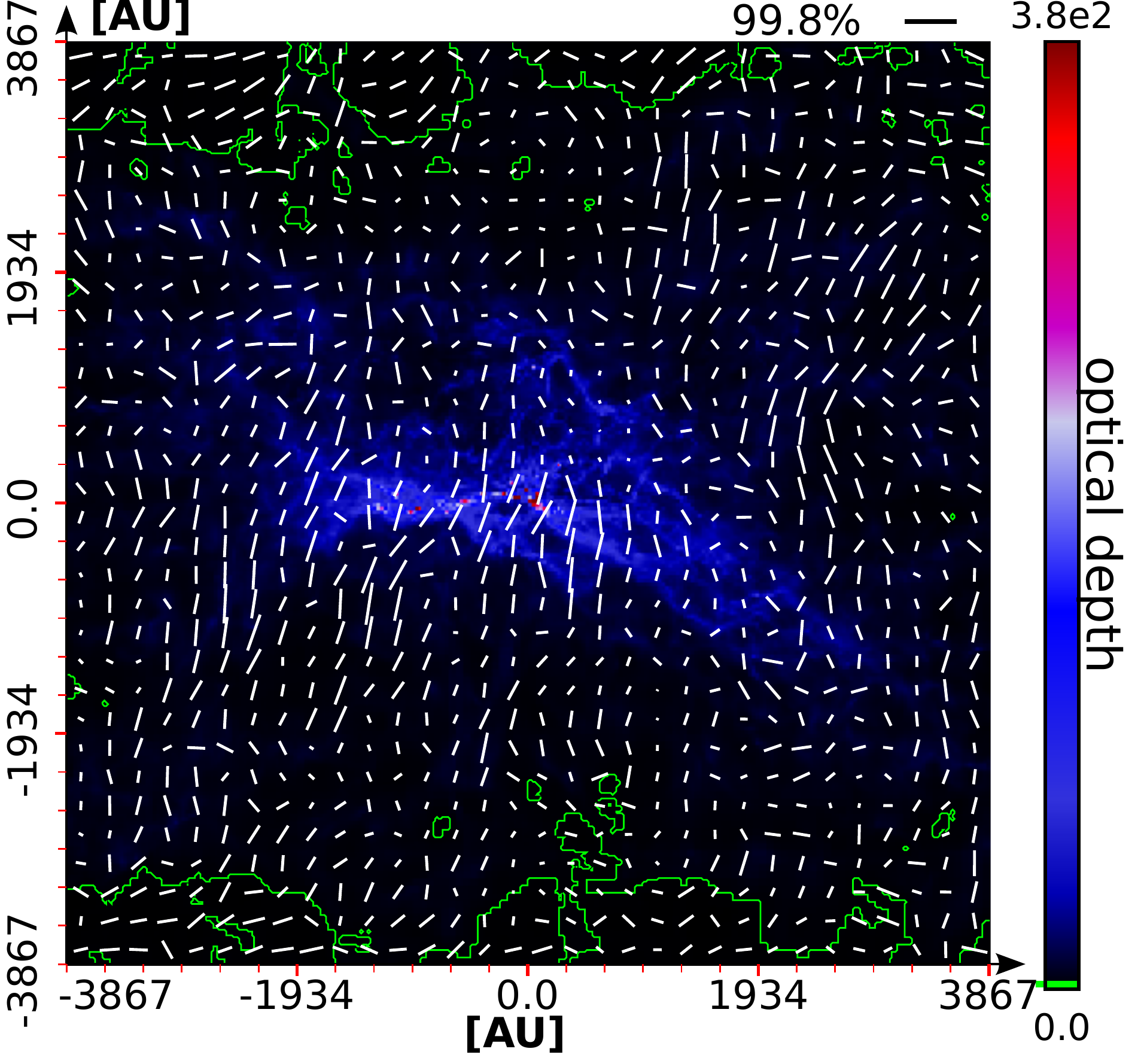}
					\caption*{ $\lambda = 30.5\ \rm{\mu m}$}
				\includegraphics[width=0.82\textwidth]{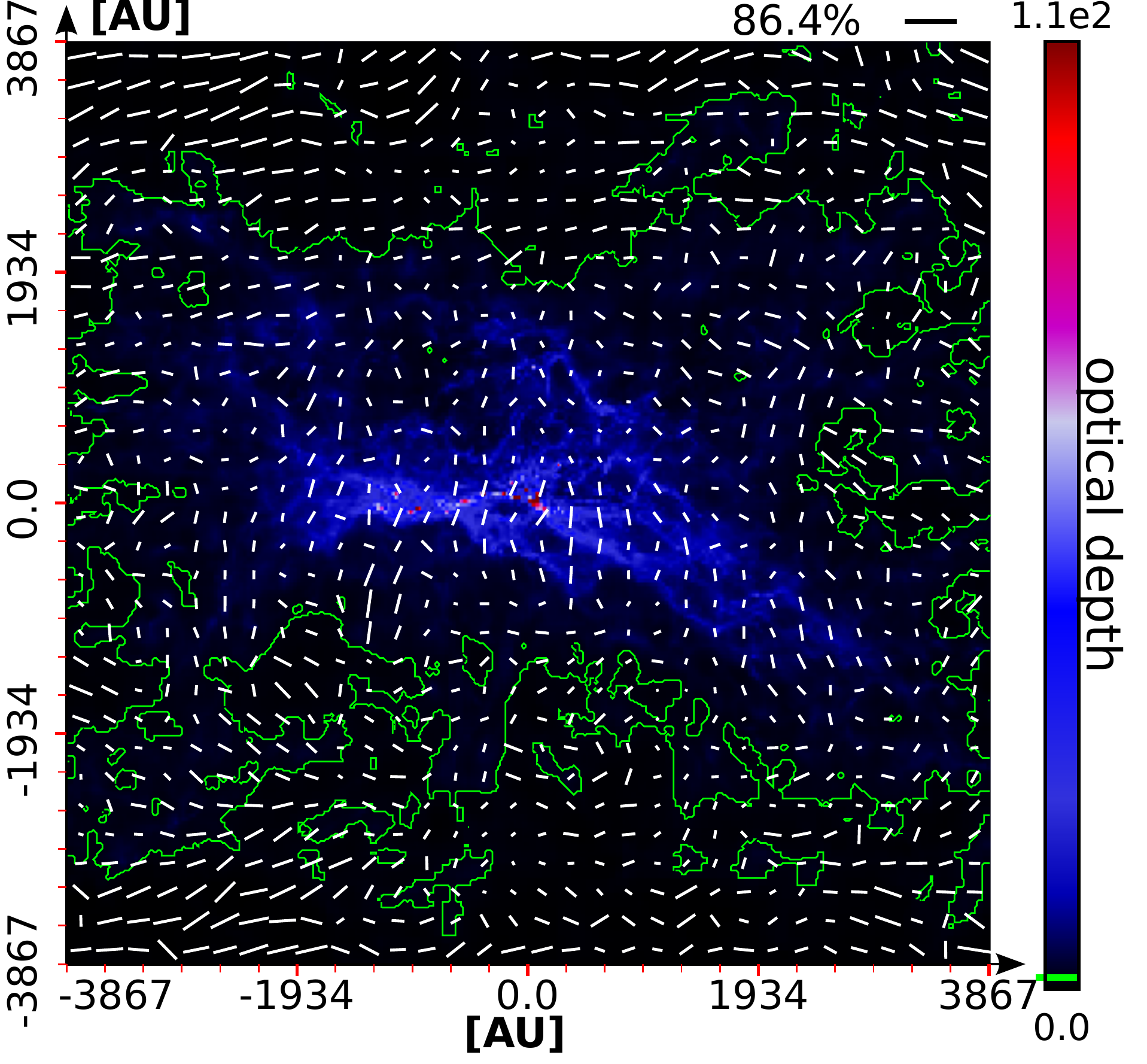}
					\caption*{ $\lambda = 61.4\ \rm{\mu m}$}
		\end{center}
		\end{minipage}
		\begin{minipage}{0.49\linewidth}
			\begin{center}
				\includegraphics[width=0.82\textwidth]{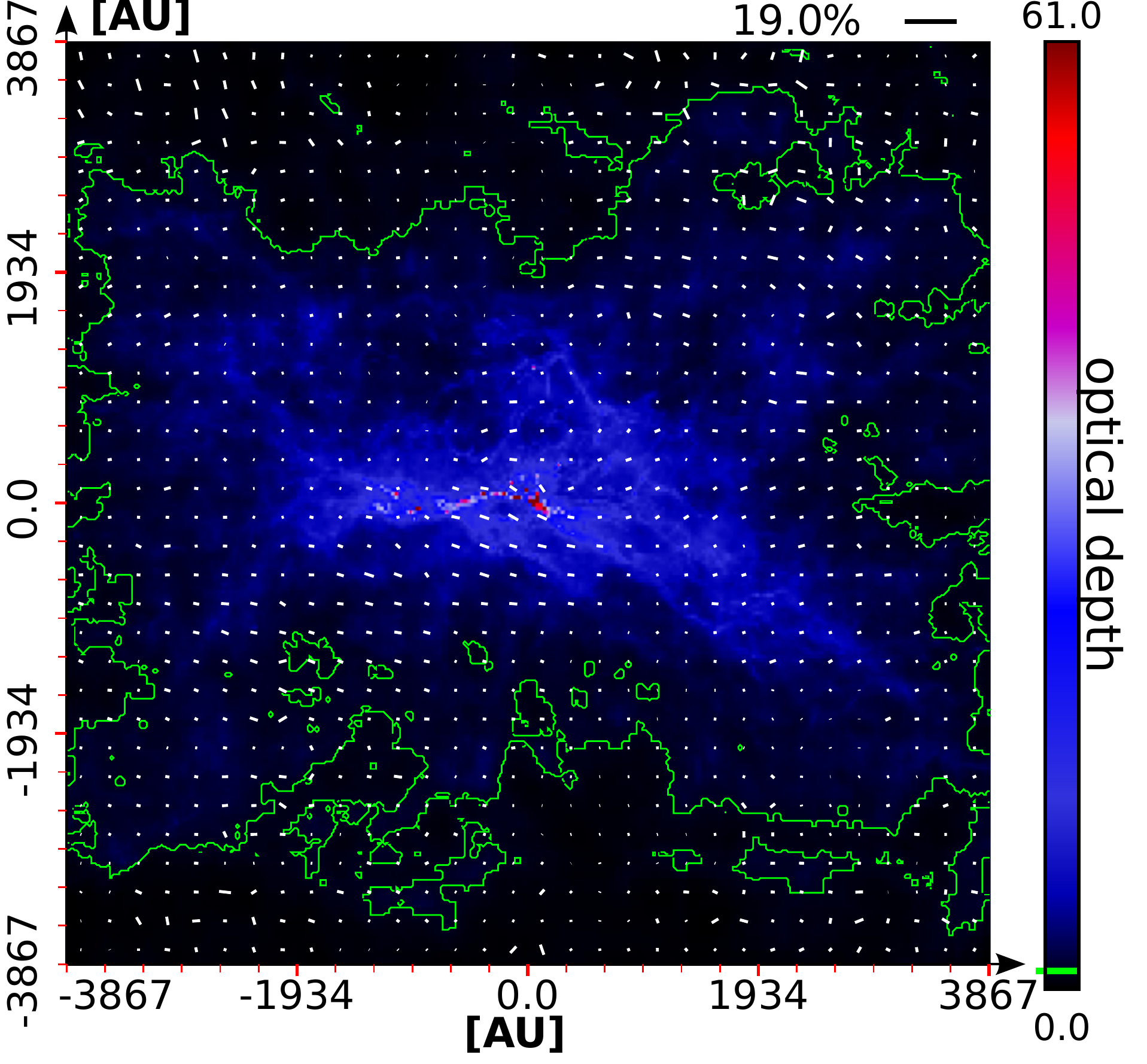}
					\caption*{ $\lambda = 7.1\ \rm{\mu m}$}
				\includegraphics[width=0.82\textwidth]{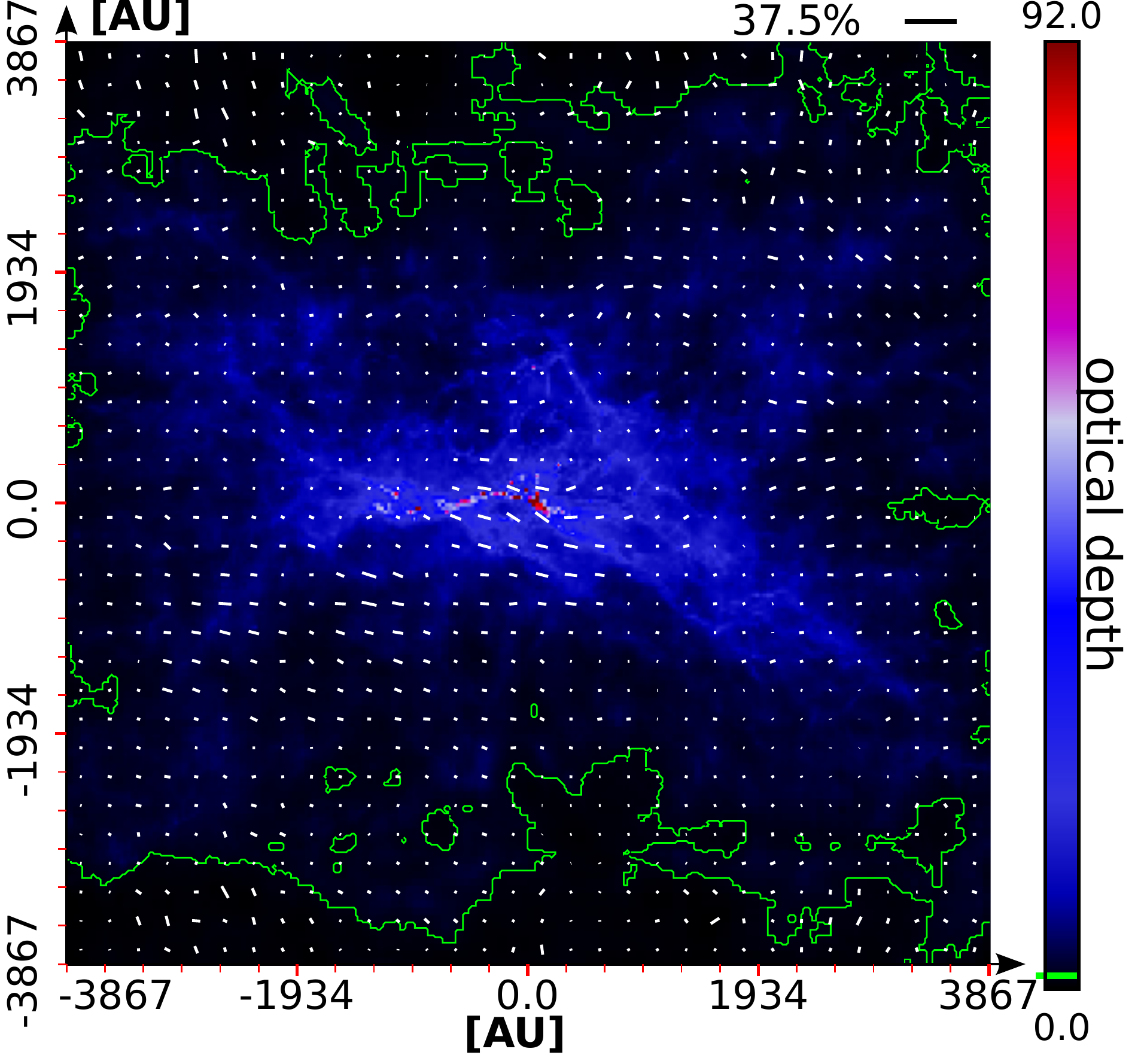}
					\caption*{ $\lambda = 30.5\ \rm{\mu m}$}
				\includegraphics[width=0.82\textwidth]{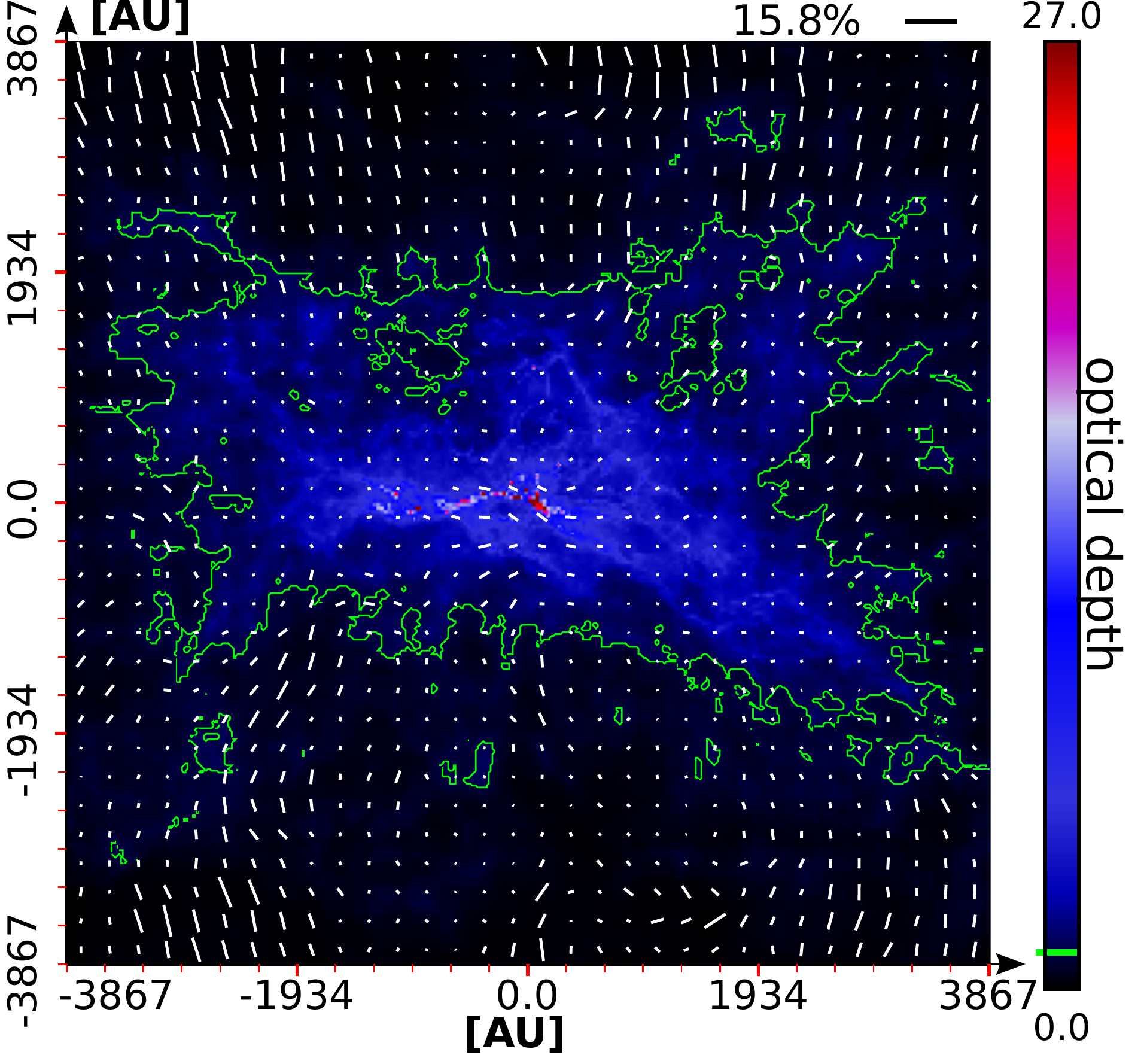}
					\caption*{ $\lambda = 61.4\ \rm{\mu m}$}
				\end{center}
		\end{minipage}	
	\caption{\small Resulting optical depth maps of setup $\rm{MHD_{sim1}}$ (\rm{left column})  in comparison with setup $\rm{MHD_{sim2}}$ (\rm{right column}) at three distinct wavelengths. The vectors of linear polarization have an offset angle of $90^{\circ}$ to match the projected magnetic field. We adjust the color bars of the optical depth with an upper cut-off at $15 \%$ for illustrative purposes. The transition between optically thin and optically thick regions is indicated by the green contour lines and a marker in the color bar.}
	\label{fig:MHDrot}
\end{figure*}

\begin{figure}[]
	\begin{minipage}[c]{0.49\linewidth}
			\begin{center}
				\includegraphics[width=1.0\textwidth]{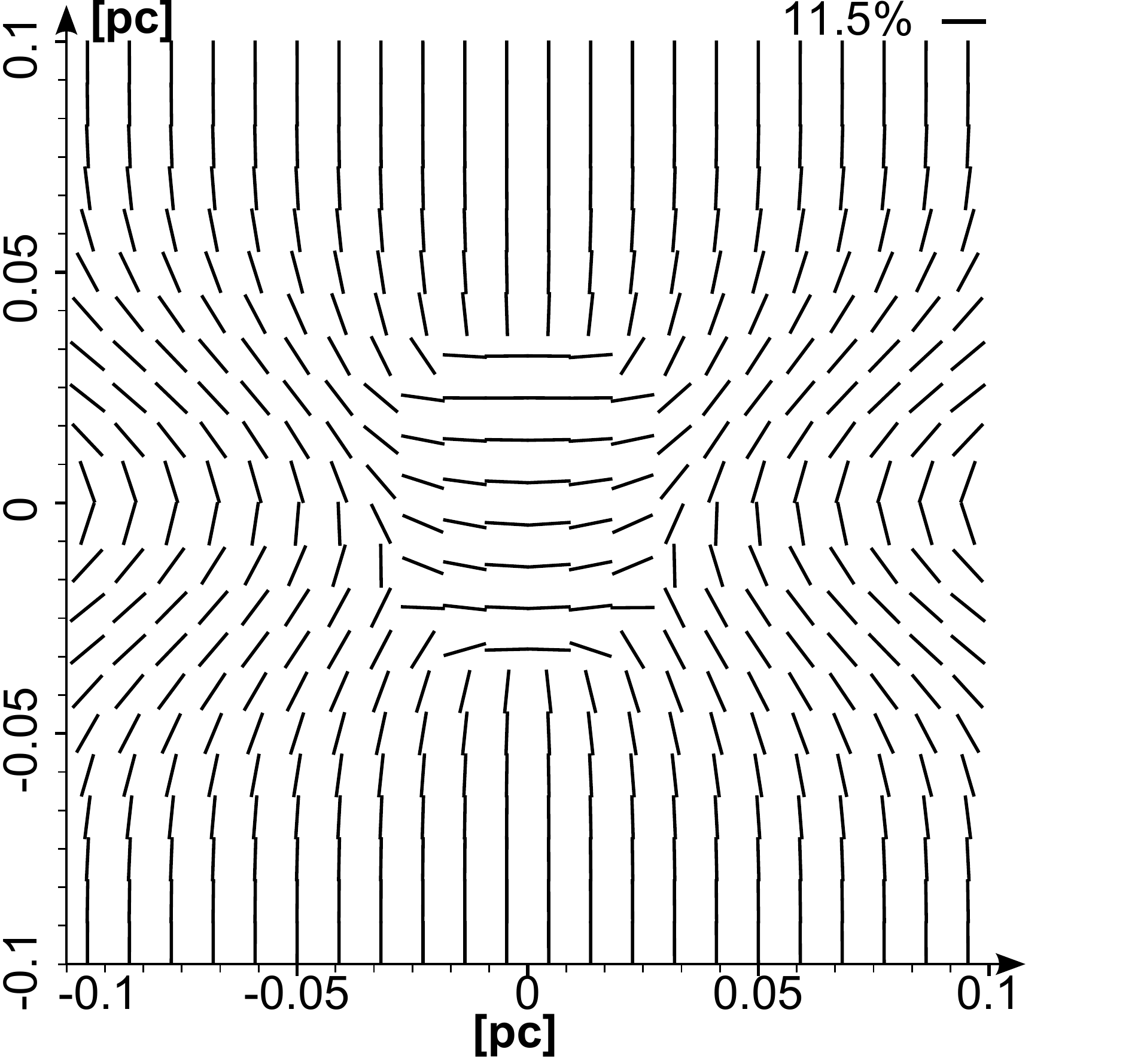}
			\end{center}
		\end{minipage}
		\begin{minipage}[c]{0.49\linewidth}
			\begin{center}
				\includegraphics[width=1.0\textwidth]{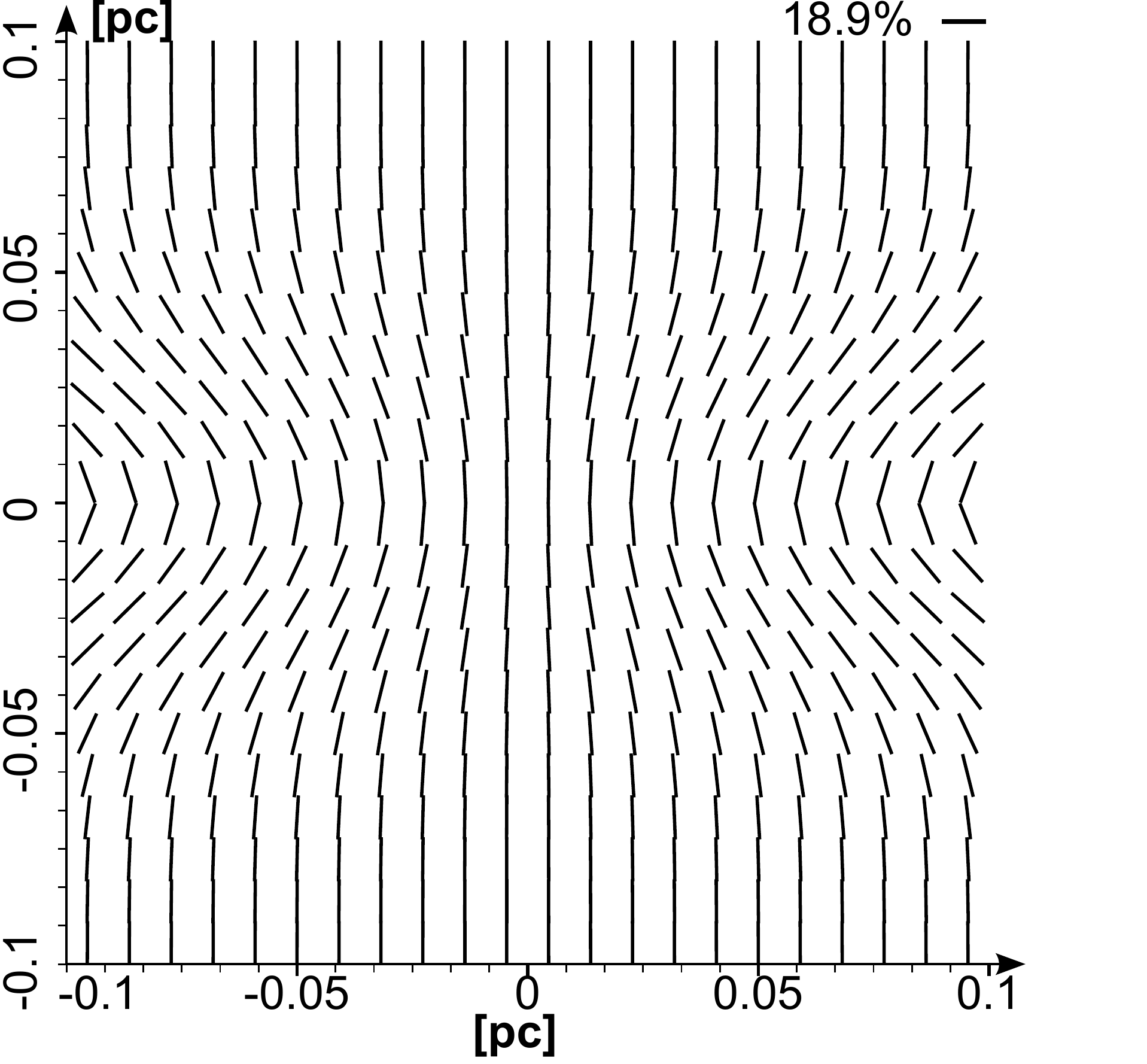}
			\end{center}
		\end{minipage}			
	
	\caption{\small Orientation vectors of linear polarization with an inclination angle of $75^{\circ}$ for a wavelength of $\lambda = 46\ \rm{\mu m}$ (left) and $\lambda = 811\ \rm{\mu m}$ (right). The parameters are identical to those of $\rm{BE_{hour2}}$. However, the density is $n_0=10^{12}\ \rm{m^{-3}}$ in the center and the field strength is $6.0\times 10^{-8} T$. The linear polarization vectors have an offset angle of $90^{\circ}$ to match the projected magnetic hourglass field. The orientation angles of linear polarization in the center regions differ from each other because of the $90^{\circ}$ flip (see Appendix \ref{apB})}.
\label{IDlPolFlip}
\end{figure}

Fig. \ref{fig:grMHD} shows the intensity, optical depth, polarization, and circular polarization of the post-processed MHD collapse simulation ($\rm{MHD_{sim2}}$) as a function of wavelength for PA and IDG alignment. The intensity and optical depth (left panels) follow the dependence determined by the characteristic features of the applied silicate - graphite dust model. The characteristic mid-IR silicate bump at $10\ \rm{\mu m}$ leads to an increased optical depth and a diminished intensity due to extinction. This behavior is also recognizable in the curves of linear and circular polarization (right panels). The degree of circular polarization also depends on $\sin$ and $\cos$ terms (see Eq. \ref{eq:sol2}). This results in an oscillation in the minimum and maximum degree of circular polarization as a function of wavelength. As mentioned in the previous discussion, the degree of linear polarization in model $\rm{MHD_{sim1}}$ amounts to $100 \%$, which is unrealistically high. In the case of IDG alignment ($\rm{MHD_{sim2}}$) the trend in the curves of intensity and optical depth remain unaffected. However, as demonstrated before, the IDG alignment mechanism changes the degree of linear and circular polarization dramatically. We did not smooth our synthetic data over beam size or angular resolution to mimic realistic observational conditions or existing  equipment. Nevertheless, for the mean value, the degree of linear polarization corresponds very well to the limitations derived from observation \citep{1999ApJ...525L.109G,2005AA...430..979G,2011ApJ...732...97D}.\\
In Fig. \ref{fig:MHDrot}, we show maps of the optical depth overlaid with vectors of linear polarization for three distinct wavelengths with perfectly aligned grains ($\rm{MHD_{sim1}}$) in comparison with imperfectly aligned grains ($\rm{MHD_{sim2}}$). The border of optically thin and optically thick regions is indicated by contour lines. With increasingly longer wavelengths, the transition between the regimes of wavelength-dominated by dichroic extinction and thermal re-emission manifests itself in a flip of the orientation angles of $90^{\circ}$, as described before. However, a detailed comparison of maps at different wavelengths reveals several polarization vectors with a continuous rotation. The effect of thermal re-emission depends on both dust temperature and number density. Regions with different temperatures and densities can dominate the thermal re-emission at different wavelengths. A region with low density and hot dust becomes optically thick at shorter wavelength, blocking the contributions of colder and denser regions appearing first along the line of sight. For longer wavelengths, this region becomes optical thin and its contribution to thermal remission decreases because of the behavior of the Planck function. The thermal re-emission process is now dominated by the cold and dense regions. Since along the line of sight the magnetic field direction varies, a continuous rotation of the observed orientation vectors of linear polarization can occur. This is consistent with the effect of the $90^{\circ}$ flip (see Appendix \ref{apB})  since the cross sections of extinction of absorption depend on wavelength and the applied alignment mechanism.\\ To demonstrate this effect, we calculateed additional linear polarization maps similar to the parameter of setup $\rm{BE_{hour2}}$ for two different wavelengths. We lowered the central density of the Bonnor - Ebert sphere to $n_0=10^{12}\ \rm{m^{-3}}$. In this case, the magnetic field strength scales with $\sqrt{10} \approx 3$. The result is shown in Fig. \ref{IDlPolFlip}. For the longer wavelength (left), the hourglass-field morphology is apparent. However, for the shorter wavelength (right), the effect of the $90^{\circ}$ flip dominates the central region. A careful comparison along the way from the center to the outer regions also reveals polarization vector of angles between $0^{\circ}$ and $90^{\circ}$.

\section{Summary and conclusions}
We demonstrated the influence of grain alignment mechanisms and magnetic field configurations by a first case study of the potential of multi - wavelength polarization measurements. For this reason we developed an extended version of the radiative transfer code MC3D \citep{2003CoPhC.150...99W} and demonstrateed its accuracy. We included a dust grain model consistent with observations and studied the effects of non-spherical dust particles with perfect alignment and imperfect alignment with respect to the direction of the magnetic field. The effects of dichroic polarization and thermal re-emission allowed us to calculate synthetic polarization maps. To do this we used ideal Bonnor - Ebert sphere setups and a sophisticated MHD collapse simulation as test models (see Tab. \ref{tab:1}.). We showed that measurements of linear polarization in combination with measurements of circular polarization as an additional source of information allow one to reveal the underlying morphology of the magnetic field for different inclination angles. The limiting factors for interpreting continuum polarization measurements are the current uncertainty of different physical parameters concerning alignment theory and dust composition. We demonstrated that the influence of imperfect grain alignment changes the behavior for circular and linear polarization radically. We conclude that possible ambiguities in the interpretation of observational data can be resolved by including additional measurements of circular polarization in future observation missions. However, it remains unclear whether the low values of circular polarization in the order of $10^{-4} - 10^{-6}$ as predicted by our models are accessible to observational equipment in the near future.\\
An additional source of ambiguity is inherent in the dichroic polarization mechanism itself. We identified two effects of rotation in the orientation angles of the vectors of linear polarization projected on the plane sky. The orientation of linear polarization is determined by a critical value of the cross sections for absorption and extinction (see Appendix \ref{apB}). Since the cross sections depend on wavelength and alignment this effect leads to a flip in the orientation of linear polarization by $90^{\circ}$. Additionally, thermal re-emission depends on dust temperature and number density.  
This results in a continuous rotation of the orientation angle of linear polarization as a function of wavelength. This might easily veil the underlying field morphology even for a symmetrical density distribution and field morphology. Therefore, measurements for linear polarization at a distinct wavelength alone are insufficient to identify the underlying magnetic field morphology because of effects of the wavelength-dependent rotations on the polarization vectors. \\
Resolving complex density and magnetic field structures, however, is currently ambiguous because of open questions about interstellar dust composition and the absence of a consistent theory of grain alignment mechanisms \citep{2007AAS...21113807H,2010MNRAS.404..265D}. A reliable theory about the dominating grain alignment mechanisms inside the ISM \citep[e.g.][]{2011ASPC..449..116L} and a sophisticated understanding of the dust composition is essential to resolve these uncertainties. 
\appendix
\section{Radiative transfer equation}
\label{apA}
Here, we provide a brief description of the equations we used to treat the radiative transfer and dichroic polarization in dusty environments. Under the assumption that no light is scattered into the line of sight, the radiative transfer equation can be written in the Stokes vector formalism as follows \citep{1974ApJ...187..461M}:

\begin{equation}
\frac{\rm{d}}{dl}\begin{pmatrix} I\\ Q \\ U \\ V \end{pmatrix}=- n_{\rm{d}}(s)\begin{pmatrix} C_{\rm{ext}} & \Delta C_{\rm{ext}} & 0 & 0 \\ \Delta C_{\rm{ext}} & C_{\rm{ext}} & 0 & 0\\ 0 & 0 & C_{\rm{ext}} & -\Delta C_{\rm{circ}} \\ 0 & 0 & C_{\rm{circ}} & \Delta C_{\rm{ext}} \end{pmatrix}\begin{pmatrix} I\\ Q \\ U \\ V \end{pmatrix}.
\label{eq:radtrans}
\end{equation} The mechanisms of dichroic polarization are straightforward to calculate by analytical functions since we consider the number density $n_{\rm{d}}$ of the dust to be constant in each cell. This condition is given since each cell in the model space of MC3D operates with a set of constant physical parameter. The matrix elements $C_{\rm{ext}}$ and $\Delta C_{\rm{ext}}$ are the cross section for extinction and linear polarization, $\Delta C_{\rm{circ}}$ for circular polarization due to birefringence.\\
With the constant number density of the dust ($n_{\rm{d}}$) the system of equations decays into two uncoupled systems \citep{2002ApJ...574..205W}.  The first system of equations solely describes the change of the $I$ and $Q$ parameter due to dichroic extinction. It can be solved by simple substitution and integration:

\begin{equation}
\begin{pmatrix} I_{i+1}+Q_{i+1}\\ I_{i+1}-Q_{i+1} \end{pmatrix} = \begin{pmatrix} (I_i+Q_i) \times exp(-n_i \times l_i\left[C_{\rm{ext},i}+\Delta C_{\rm{ext},i}\right])   \\ (I_i-Q_i) \times exp(-n_i \times l_i\left[C_{\rm{ext},i}-\Delta C_{\rm{ext},i}\right]) \end{pmatrix}.
\label{eq:sol1}
\end{equation}
The second system of equations can be handled as a complex eigenvalue problem. This leads to additional $\rm{cosine}$ and $\rm{sine}$ terms and a transfer between linear and circular polarization
\begin{equation}
\begin{pmatrix} U_{i+1}\\ V_{i+1} \end{pmatrix} =  e^{-n_i \times l_i C_{\rm{ext},i}}\begin{pmatrix}  U_i \cos\left(n_i \times l_i \Delta C_{\rm{circ},i}\right) - V_i \sin\left(n_i \times l_i \Delta C_{\rm{circ},i}\right) \\ U_i \sin\left(n_i \times l_i \Delta C_{\rm{circ},i}\right) - V_i \cos\left(n_i \times l_i \Delta C_{\rm{circ},i}\right)  \end{pmatrix}.
\label{eq:sol2}
\end{equation}
Linear polarization arises from linear dichroism alone, while circular polarization depends on both a non-zero value in the $U$ parameter and birefringence. Subsequently, circular polarization can occur in the case of non-parallel magnetic field lines along the line of sight.

\section{Orientation of polarization}
\label{apB}
It is possible to determine the exact conditions for the $90^{\circ}$ flip in a single cell of the model space. In general, a threshold for this effect does not exist along the entire line of sight. In each cell of our model space we have the two opposing effects of dichroic extinction and thermal re-emission adding to the linear polarization perpendicular to each other. In the reference frame of the magnetic field the dichroic extinction provides a negative contribution to the Q Parameter while thermal re-emission contributes positively to Q. In this orientation the U and V parameter remain zero. If we solve equation \ref{eq:sol1} for the $Q_{i+1}$ parameter, we can calculate the conditions when the two effects cancel each other out:

\begin{equation}
Q_{i+1} = e^{-n_i \times l_i C_{\rm{ext},i}}\left[ Q_i \cosh(n_i  l_i \Delta C_{\rm{ext},i}) - I_i \sinh(n_i  l_i \Delta C_{\rm{ext},i}) \right] = 0
\label{eq:Qnext}.
\end{equation}
The contribution of thermal re-emission is determined by the temperature of the dust $T_{\rm{d}}$, the number density $n_{\rm{d}}$, the cross sections for absorption $\Delta C_{\rm{abs}},\ C_{\rm{abs}}$ and the path length $l$. Inside each cell all the parameters and functions $n_{\rm{d}}$,\ $l$,\ $\Delta C_{\rm{abs}}$,\ $C_{\rm{abs}}$,\ $C_{\rm{ext}}$,\ $B_{\rm{\lambda}}(T)$ are positive and constant, so one can solve equation \ref{eq:Qnext}. As we can derive from equations \ref{eq:reemissionI} and \ref{eq:reemissionQ}, the contributions of $ I_i$ and $ Q_i$ in a single cell are as follows:

\begin{equation}
I_i = n_i \times l_i  C_{\rm{abs}, i} B_{\rm{\lambda}}(T_i),
\end{equation}

\begin{equation}
Q_i = n_i \times l_i \Delta C_{\rm{abs},i} B_{\rm{\lambda}}(T_i).
\end{equation} 
For a wavelength of $\lambda > 7\ \rm{\mu m}$ one can approximate $C_{\rm{abs},i} \approx C_{\rm{ext},i}$. By introducing the optical depths for extinction

\begin{equation}
\tau_{\rm{ext}, i} = n_i \times l_i  C_{\rm{ext}, i},
\end{equation} and polarization

\begin{equation}
\tau_{\rm{pol}, i} = n_i \times l_i \Delta C_{\rm{ext},i}, 
\end{equation} we can derive an inequality for the Q parameter to change its sign as a function of the inverse hyperbolic tangent:

\begin{equation}
1 \lessgtr  \frac{1}{\tau_{\rm{ext}}} tanh^{-1}\left(\frac{\tau_{\rm{pol}}}{\tau_{\rm{ext}}}\right).
\label{eq:crit}
\end{equation}
If the right-hand side is larger than 1, the polarization process is dominated by thermal re-emission and, in the reverse case, by dichroic extinction. However, in the calculated synthetic polarization maps the observed flip of $90^{\circ}$ for the orientation of linear polarization depends on all the physical quantities along the entire line of sight and cannot be determined with this inequality.

\begin{acknowledgements}
We wish to thank the referee for a careful review and for the comments that helped improve our paper. We also thank Robi Banerjee, Gesa Bertrang and Peter Scicluna for fruitful discussions and Florian Kirchschlager for the introduction into DDSCAT. For this project the authors S.R. and S.W. acknowledge the support of the DFG: WO 857/11-1. D.S. acknowledges funding by the Deutsche Forschungsgemeinschaft via grant BA 3706/3-1 within the SPP \textit{The interstellar medium}.
\end{acknowledgements}

\bibliographystyle{aa}
\bibliography{./bibtex}
\end{document}